\documentclass[11pt]{article}
\usepackage{jheppub}
\usepackage[utf8]{inputenc}
\usepackage{amsfonts, amsmath, amsthm, amssymb}     
\usepackage{bbold}
\usepackage{bm}
\usepackage{mathtools, cuted}      
\usepackage{commath}
\usepackage[mathscr]{euscript}
\usepackage{color}
\usepackage{physics}
\usepackage{tensor}
\usepackage{hyperref}
\usepackage{multirow}
\usepackage[dvipsnames]{xcolor}
\usepackage{bclogo}
\usepackage{comment}
\usepackage{dsfont}
\usepackage{cancel}
\usepackage[normalem]{ulem} 
\usepackage{nicefrac}
\usepackage{appendix}
\usepackage{microtype}
\usepackage{etoolbox}

        \hypersetup{
           breaklinks=true,   
           colorlinks=true,   
        }

\usepackage{tikz}
    \usetikzlibrary{decorations.pathreplacing}
    \usetikzlibrary{decorations.pathmorphing} 
    \usetikzlibrary{decorations.markings} 
    \usetikzlibrary{arrows.meta,bending}
    \usetikzlibrary{decorations}
    \usetikzlibrary{shapes.geometric}
    \usetikzlibrary{shapes}
    \usetikzlibrary{positioning}
    \usetikzlibrary{intersections}
    \usetikzlibrary{external}
    \usetikzlibrary{arrows,patterns, calc,through,backgrounds}

\numberwithin{equation}{section}

\newcommand\numberthis{\addtocounter{equation}{1}\tag{\theequation}} 

\newcommand{\p}{\partial}
\newcommand{\Disc}{\mathbb{D}}

\renewcommand{\i}{\operatorname{i}\!}

\renewcommand{\d}{\operatorname{d}\!}

\newcommand{\hbb}[0]{\mathbb{h}}
\newcommand{\Deltabb}[0]{\mathbb{\Delta}}

\def\centerarc[#1](#2)(#3:#4:#5)
    { \draw[#1] ($(#2)+({#5*cos(#3)},{#5*sin(#3)})$) arc (#3:#4:#5); } 

\begin{document}

\title{\vspace*{2cm} Thermal $\bm{n}$-Point  Conformal Blocks in Four Dimensions from Oscillator Representations}

\author{Martin Ammon,}
\author{Jakob Hollweck,}
\author{Tobias Hössel,}
\author{Katharina W\"olfl}
\affiliation{Theoretisch-Physikalisches Institut, Friedrich-Schiller-Universit\"at Jena,\\
Max-Wien-Platz 1, D-07743 Jena, Germany}

\emailAdd{martin.ammon@uni-jena.de}
\emailAdd{jakob.hollweck@uni-jena.de}
\emailAdd{tobias.hoessel@uni-jena.de}
\emailAdd{katharina.woelfl@uni-jena.de}

\vspace{1cm}

\abstract{We define and compute the four-dimensional thermal $n$-point conformal block in the projection channel using oscillator representations on $\mathbb{S}^1_\beta \times \mathbb{S}^3$. This is done by evaluating a class of integrals over the homogeneous space $\mathbb{D}_4$ of the four-dimensional conformal group. We restrict ourselves to scalar external operators and scalar exchange. In the low-temperature limit, our result reduces correctly to the vacuum $(n+2)$-point block in the comb channel. The corresponding expressions can be written as a series of terminating hypergeometric functions or equivalently, a series of weighted SU(2) spin-networks. Alternatively, functions adapted to the SU(2,2) representation are introduced and some properties are discussed.} 

\setcounter{tocdepth}{2} 
\maketitle


\addtocontents{toc}{\protect\setcounter{tocdepth}{3}}

\section{Introduction}

The central objects of any conformal field theory (CFT), namely correlation functions, are largely characterised by their decomposition into conformal blocks. The scalar $n$-point correlators of a CFT on $\mathbb{R}^d$ are fixed by their symmetry until $n=3$, up to the three-point coefficients $C_{ijk}$ and the conformal weights $\Delta_i$. For higher-point conformal correlation functions, the operator product expansion (OPE) defines a natural split between the kinematical data, given by the conformal blocks, and the dynamical, theory-dependent data that enters as input to these blocks. Because the OPE is associative, there are consistency conditions called crossing equations between different expansions of the correlation functions. The bootstrap program constrains the dynamical data of conformal field theories through crossing symmetry and other consistency conditions, and has become a powerful tool in both analytic and numerical investigations, see e.\,g. \cite{Ferrara:1973yt,Polyakov:1974gs,Rattazzi:2008pe,El-Showk:2012cjh, Hartman:2022zik}.

The crossing symmetry equations for higher-point correlators do not impose fundamentally new constraints beyond those already present in the four-point functions. Rather, constraint equations for higher-point correlation functions can be interpreted as a repackaging of (infinitely many) four-point crossing equations. This reformulation is advantageous for the conformal bootstrap, as it allows access to an infinite set of four-point constraints, something which is otherwise out of reach. Indeed, actually solving for (and not only constraining) the full CFT data of a theory is intrinsically a task that makes this approach necessary, as one needs to account for four-point functions of all operators. See \cite{Antunes:2020yzv,Buric:2020zlb,Harris:2025cxx} for work on the multi-point conformal bootstrap. It has also been suggested in \cite{Rosenhaus:2018zqn} that instead of studying four-point crossing equations for external spinning operators, one can consider higher-point crossing with external scalar fields.   

This motivates the extension of conformal bootstrap methods to thermal settings, where finite-temperature effects must be taken into account. In $d=2$, the vacuum conformal data $\{\Delta_i,C_{ijk}\}$ is sufficient to even determine the finite-temperature (equivalently finite size with periodic boundary conditions) corrections to conformal correlation functions on $\mathbb{S}^1_\beta 
\times \mathbb{R}$, periodic Euclidean time with circumference $\beta = \frac{1}{T}$ accounting for the temperature. This is because one can map the (punctured) plane conformally to the cylinder $\mathbb{S}^1_\beta \times \mathbb{R}$. In higher dimensions, there is no such conformal map from $\mathbb{R}^d$ to $\mathbb{S}^1_\beta \times \mathbb{R}^{d-1}$ and hence one needs additional data, namely the one-point coefficient $b_{\mathcal{O}}$ of the scalar one-point function $\langle \mathcal{O}_\Delta\rangle_\beta = b_{\mathcal{O}} T^\Delta$ \cite{Petkou:2018ynm}. Because the thermal correlation functions on $\mathbb{S}^1 \times \mathbb{R}^{d-1}$ do not manifestly satisfy the KMS condition in general, one gets \textit{thermal crossing equations} which constrain $b_{\mathcal{O}}$ in terms of the vacuum data $\{\Delta_i,C_{ijk}\}$. For the two-point function, this has been studied for a wide range of settings in various dimensions, see for example \cite{Iliesiu:2018fao,Iliesiu:2018zlz,Karlsson:2022osn,Alday:2020eua,Dodelson:2022yvn,Huang:2022vet,Dodelson:2023vrw,Esper:2023jeq,Marchetto:2023xap,Barrat:2024aoa,Buric:2025anb,Barrat:2025wbi,Barrat:2025nvu}. 
    
Instead of working on $\mathbb{S}^1_\beta \times \mathbb{R}^{d-1}$, one can  consider $\mathbb{S}^1_\beta \times \mathbb{S}^{d-1}$, which is more general, as it depends on the scale $\beta/L$, with $\beta$ thermal circle circumference and $L$ the sphere radius. Taking the high-temperature limit $\beta \to 0$ then relates correlators on $\mathbb{S}^1_\beta \times \mathbb{S}^{d-1}$ to those on $\mathbb{S}^1_\beta \times \mathbb{R}^{d-1}$. Thermal correlation functions on $\mathbb{R} \times \mathbb{S}^{d-1}$ are, due to the KMS condition, effectively defined on $\mathbb{S}^1_\beta \times \mathbb{S}^{d-1}$. In particular, the aforementioned thermal crossing equations for two-point functions can be understood, via the high-temperature limit, as arising from crossing equations for four-point functions on $\mathbb{R}^d$ with some external fields summed over \cite{Iliesiu:2018fao}. This limit is, however, challenging to take in practice, one reason being its sensitivity to CFT data at large scaling dimensions. One can also argue that $\mathbb{S}^1_\beta \times \mathbb{S}^{d-1}$ is the more natural setting for thermodynamic phenomena as it allows for the inclusion of angular potentials. Furthermore, in contrast to $\mathbb{R}^1_\beta \times \mathbb{S}^{d-1}$, the one-point function is not fixed by symmetry and thus becomes a non-trivial observable.  See for example \cite{Buric:2024kxo,Buric:2025uqt} for this perspective. 

We expect higher-point conformal blocks in thermal CFTs to play a similar role as in vacuum CFTs: higher-point crossing equations related to  $\mathbb{S}^1_\beta \times \mathbb{S}^{d-1}$ should give rise to higher-point thermal crossing equations on $\mathbb{S}^1_\beta \times \mathbb{R}^{d-1}$ that in theory allow to determine $b_\mathcal{O}$ in terms of $\{C_{ijk},\Delta_i\}$ more effectively than with the thermal crossing equation arising from the four-point function. So far, however, there are to our knowledge no results on thermal conformal blocks besides one-point blocks (see e.\,g. \cite{Gobeil:2018fzy,  Buric:2024kxo, Alkalaev:2024jxh}) for dimensions higher than two. We address this problem by defining and computing the $n$-point conformal block on $\mathbb{S}^1_\beta \times \mathbb{S}^3$ in the so-called projection (or necklace) channel. For the known previous results in this channel in two dimensions, see for example \cite{Hadasz2010,Kraus2017,Cho:2017oxl,Alkalaev:2017bzx,Alkalaev:2020yvq,Alkalaev:2022kal,Alkalaev:2023evp,Pavlov:2023asi}.

Conformal blocks can be defined naturally through eigenvalue equations of the respective Casimir elements in addition to the conformal Ward identities. A direct approach to computing them in higher dimensions is to solve these differential equations as was first done in \cite{Dolan:2000ut,Dolan:2003hv,Dolan:2011dv} for the four-point conformal block. More recently, this method has been generalised to blocks for $n>4$ in dimensions $d > 2$ in \cite{Buric:2020dyz,Buric:2021ywo, Buric:2021ttm, Buric:2021kgy}. 

We construct the conformal blocks using the oscillator construction for four-dimensional Euclidean CFTs introduced in \cite{Ammon:2024axd}. The blocks are implemented as gluings of wavefunctions defined on the Bergman space $\mathcal{H}L^2_\Deltabb(\mathbb{D}_4)$, where $\mathbb{D}_4$ is a homogeneous space of the four-dimensional conformal group. This leads to a class of integrals of basis functions $\phi^{j,m}_{a,b}$ over $\mathbb{D}_4$ which we compute by a decomposition suited to the coset structure of $\mathbb{D}_4$ and by introducing an expansion for the pointwise product of two functions $\phi^{j,m}_{a,b}$. Both the class of integrals and the expansion generalise classic results for SU(2) Wigner-$\mathcal{D}$ functions, and the derived expressions involve the SU(2) Clebsch-Gordan coefficients. This has the advantage that there are already efficient tools adjusted to the numerical evaluation of spin-network diagrams for more specific Wigner-$\mathcal{D}$ matrices and, more generally, for contractions of tensor-network diagrams. We discuss this in more detail in the outlook.

The structure of the paper is as follows. We start in section \ref{sec:def_thermal_blocks} by defining thermal conformal blocks on $\mathbb{S}^1_\beta \times \mathbb{S}^3$ in the projection channel. In section \ref{sec:comb_channel_blocks}, we review the oscillator construction for Euclidean four-dimensional CFTs and compute the vacuum $n$-point block in the comb channel. Our result involves the Clebsch-Gordan coefficients of SU(2), so we include a short discussion of the associated SU(2) spin-network diagrams in appendix \ref{sec:weighted_networks}. In section \ref{sec:computation_npt_thermal}, the general construction of the $n$-point thermal conformal block in the projection channel is presented. We derive the one-point block in detail and explore the low-temperature limit for the general case. We finally discuss our results in section \ref{sec:discussion} and give an outlook.  

Detailed computations are presented in the appendix. Appendix \ref{sec_app:angular_potentials} is dedicated to the discussion of non-vanishing angular potentials and their implementation in the calculations of thermal blocks. In appendix \ref{sec_app:properties_cgcs}, we list properties of Wigner-$\mathcal{D}$ matrices (for both SU(2) and its holomorphic extension to general $2\times 2$ matrices) and SU(2) intertwiners. A key result for the computation of thermal and vacuum conformal blocks in the main part of this paper is given in appendix \ref{sec_app:HigherIntegrals_Details}, where we evaluate general integrals over basis functions of $\mathcal{H}L^2(\mathbb{D}_4)$. For readers who are not familiar with the oscillator construction, appendix \ref{sec_app:2d_blocks} discusses the easier case of conformal blocks in the projection channel on the two-dimensional torus.

\section{Definition of Four-Dimensional Thermal Blocks in Projection Channel}
\label{sec:def_thermal_blocks}

In this work, we focus on Euclidean correlation functions, hence a field is parametrised by $x \in \mathbb{R}^4$. The fields are taken to be bosonic and scalar primary fields. The line element $\mathrm{d}s^2_{\mathbb{R}^4}$ transforms as $\mathrm{d}s^2_{\mathbb{R}^4} = \mathrm{e}^{2\tau} \mathrm{d}s^2_{\mathbb{R}^1 \times \mathbb{S}^3}$ under the transformation $r = \operatorname{e}^\tau$, where our coordinates on $\mathbb{R}^1 \times \mathbb{S}^3$ are given by $(\tau,\Omega)$ and $r^2 = x^\mu x_\mu$. Under the same transformation, the primary scalar field $\mathcal{O}_{\mathbb{R}^4}(x)$ with conformal weight $\Delta$ transforms as $\mathcal{O}_{\mathbb{R}^4}(x) = r^{-\Delta}\mathcal{O}_{\mathbb{R} \times \mathbb{S}^3}(\tau,\Omega)$; in the following we will drop the labels denoting the space unless they are needed for clarity. 

The thermal $n$-point correlation function for the cylinder $\mathbb{R} \times \mathbb{S}^3$ with coordinates $(\tau,\Omega)$ is given by
\begin{align} \label{ref:def_thermal_correlator}
   \expval{\mathcal{O}_{\Delta_1}(\tau_1,\Omega_1)\cdots \mathcal{O}_{\Delta_n}(\tau_n,\Omega_n)}_{\beta} =  \text{Tr}\left( \mathcal{O}_{\Delta_1}(\tau_1,\Omega_1) \cdots \mathcal{O}_{\Delta_n}(\tau_n,\Omega_n)\, q^D\right)
\end{align}
for $q = \mathrm{e}^{-\beta}$ and where the trace runs over the space of states of the CFT.  The thermal $n$-point correlator is naturally periodic in $\tau$, which can be easily seen by inserting $n$ copies of $\mathbb{1} = \operatorname{e}^{-\beta D} \operatorname{e}^{\beta D}$\,:
\begin{align} 
\expval{\mathcal{O}_{\Delta_1}(\tau_1,\Omega_1)\cdots \mathcal{O}_{\Delta_n}(\tau_n,\Omega_n)}_{\beta} & =\operatorname{Tr}\left[e^{\tau_1 D} \mathcal{O}_{\Delta_1}(0, \Omega_1) e^{-\tau_1 D} \cdots e^{\tau_n D} \mathcal{O}_{\Delta_n}(0, \Omega_n) e^{-\tau_n D} e^{-\beta D} \right] \notag\\
& =\operatorname{Tr}\left[\mathcal{O}_{\Delta_1}(\tau_1 + \beta, \Omega_1)  \cdots  \mathcal{O}_{\Delta_n}(\tau_n + \beta, \Omega_n) e^{-\beta D} \right] \notag\\ &=\expval{\mathcal{O}_{\Delta_1}(\tau_1 + \beta,\Omega_1)\cdots \mathcal{O}_{\Delta_n}(\tau_n + \beta,\Omega_n)}_{\beta}\,. \label{eq:KMS}
\end{align}
Hence, the thermal $n$-point correlation function with fields of coordinates on $\mathbb{R} \times \mathbb{S}^3$ is effectively defined on $\mathbb{S}^1_\beta \times \mathbb{S}^3$. Equation \eqref{eq:KMS} is known as the KMS condition; imposing it  for an explicit expression of the correlation function gives non-trivial so-called \textit{thermal crossing equations} which constrain the coefficient of the one-point correlation function in terms of the vacuum CFT data \cite{Iliesiu:2018fao}. 

One has to be careful when directly translating the KMS condition into the background geometry, as it obscures how the theory at hand is quantised, i.\,e. how the Hamiltonian is chosen. This translation leads to vastly different behaviour: according to above, we say that the thermal one-point function of a scalar operator defined over $\mathbb{R}^{4}$ through a flat slicing is actually defined over $\mathbb{S}^1 \times \mathbb{R}^{3}$ and has (inverse) monomial dependence in $\beta$, i.\,e. $\expval{ \mathcal{O}_\Delta}_{\mathbb{S}^1 \times \mathbb{R}^{3}} \sim \beta^{-\Delta}$. The thermal one-point function for the scalar primary $\mathcal{O}_{\mathbb{R}^4}(x)$ discussed before, however, is defined through radial quantization and has a hypergeometric dependence in $\beta$ \cite{Gobeil:2018fzy,Buric:2024kxo,Alkalaev:2024jxh}.

In general, there exists a plethora of channels for the thermal $n$-point block. We focus here on the so-called \textit{projection channel} and define it by inserting $n$ projectors\footnote{We are restricting ourselves to scalar exchange. In the more general case including internal spinning operators, the projector would also depend on the spins $j_1$ and $j_2$ in addition to the scaling dimension $\Deltabb$.} $\mathbb{P}_{\Deltabb_i}$ into the correlation function in the following way 
\begin{align}
    G^{(n)}_{\text{Proj.}}(q,(\tau_1,\Omega_1), \dots, (\tau_n,\Omega_n)) \equiv \text{Tr}\left( \mathbb{P}_{\Deltabb_1} q^D \mathcal{O}_1(\tau_1,\Omega_1) \mathbb{P}_{\Deltabb_2}\mathcal{O}_{\Delta_2}(\tau_2,\Omega_2) \cdots \mathbb{P}_{\Deltabb_{n}}\mathcal{O}_n(\tau_n,\Omega_n) \right). \label{def:block}
\end{align}
We will, however, compute this object for coordinates $x_i \in \mathbb{R}^4$ and use 
\begin{align}
    G^{(n)}_{\text{Proj.}} (q,(\tau_1,\Omega_1), \dots, (\tau_n,\Omega_n)) = r_1^{\Delta_1} \cdots r_n^{\Delta_n} G^{(n)}_{\text{Proj.}}(q,x_1,\dots,x_n)\, 
\end{align}
in the end to translate it back to $\mathbb{S}^1_\beta \times \mathbb{S}^3$.
We want to stress that also the block parametrised by coordinates $x_i$ is still defined on $\mathbb{S}^1_\beta \times \mathbb{S}^3$ because it is inherently defined through radial quantization. 

The projectors $\mathbb{P}_{\Deltabb_i}$ for $i = 1,\dots,n$ project onto a representation space of $\mathfrak{su}(2,2)$ by definition. We discuss that the natural choice for this space is the functional space $\mathcal{H}L^2(\mathbb{D}_4)$ consisting of square integrable holomorphic functions on the domain $\mathbb{D}_4$ in section \ref{sec:review_osc_constr}, where all further details are provided. The projector $\mathbb{P}_{\Deltabb_i}$ can then be expressed via an integral over $\mathbb{D}_4$ 
\begin{equation}
    \mathbb{P}_{\Deltabb_i} = \int_{\mathbb{D}_4} \left[\d U_i\right]_{\Deltabb_i} \dyad*{U_i^\dagger}{U_i}\,
\end{equation}
with coordinates $U_i$. Using this form of the projectors we can rewrite equation \eqref{def:block} as 
\begin{equation} \label{eq:projblock_as_gluing}
    G^{(n)}_{\text{Proj.}} = \int_{\mathbb{D}_4}\left[\d U_1\right]  \cdots \int_{\mathbb{D}_4}\left[\d U_n\right] \Omega(x_1,U_1,U_2^\dagger)\Omega(x_2,U_2,U_3^\dagger) \cdots \Omega(x_n, U_n, qU_1^\dagger)\,,
\end{equation}
for the wavefunctions
\begin{equation}
    \Omega(x_i,U_i,U_{i+1}^\dagger) \equiv \mel*{U_i}{\mathcal{O}_{\Delta_i}(x_i)}{U_{i+1}^\dagger}, \label{eq:Omega_mel_def}
\end{equation}
where cyclicity of the trace was used and the action of $q^D$ evaluated (see additional steps in section \ref{sec:computation_npt_thermal}). We introduce the corresponding diagram for equation \eqref{eq:projblock_as_gluing} as 
\begin{align} \label{dia:npt-block-glueing}
    &G^{(n)}_{\text{Proj.}}(q,x_1,\dots,x_n)\\ &= 
\def\tkzscl{0.5}
\begin{tikzpicture}[baseline={([yshift=-.5ex]current bounding box.center)},vertex/.style={anchor=base,
    circle,fill=black!25,minimum size=18pt,inner sep=2pt},scale=\tkzscl]
    \coordinate[label={[yshift=0.05cm]$U_1$}] (u1) at (-3,0);
    \coordinate[label={[yshift=0.04cm]$U_2^\dagger$}] (u12) at (-1,0);
        \coordinate[label=left:$x_1$\,] (z_1) at (-2,2);
        \coordinate[label={[yshift=0.05cm]$U_2$}] (u21) at (0,0);
        \coordinate[label=left:$x_2$\,] (z_2) at (1,2);
        \coordinate[label={[yshift=0.04cm]$U^\dagger_3$}] (u2) at (2,0);
        \draw[thick] (u1) -- (-2,0);
        \draw[thick] (z_1) -- (-2,0);
        \draw[thick] (-2,0) -- (u12);
        \draw[thick,densely dashed] (u12) -- (u21);
        \draw[thick] (z_2) -- (1,0);
        \draw[thick] (u2) -- (u21);

        \coordinate[label={[yshift=0.05cm]$U_{n-1}$}] (u3) at (4,0);
        \coordinate[label=left:$x_{n-1}$\,] (z_n-1) at (5,2);
        \coordinate[label={[yshift=0.1cm]$U_{n}^\dagger$}] (u32) at (6,0);
        \coordinate[label={[yshift=0.1cm]$U_n$}] (u41) at (7,0);
        \coordinate[label=left:$x_n$\,] (z_n) at (8,2);
        \coordinate[label={[yshift=0.04cm]$qU^\dagger_1$}] (u4) at (9,0);
        \draw[thick,loosely dotted] (u2) -- (u3);
        
        \draw[thick] (u3) -- (u32);
        \draw[thick] (z_n-1) -- (5,0);
        \draw[thick, densely dashed] (u32) -- (u41);
        \draw[thick] (z_n) -- (8,0);
        \draw[thick] (u4) -- (u41);

        \coordinate[] (u5) at (9,-2);
        \coordinate[] (u6) at (-3,-2);
        \draw[thick,dashed] (u5) -- (u6);
        \draw[thick,dashed] (9,-2) arc (-90:90:1);
        \draw[thick,dashed] (-3,0) arc (90:270:1);
        \fill (z_1) circle (6pt);
        \fill (z_2) circle (6pt);
        \fill (z_n-1) circle (6pt);
        \fill (z_n) circle (6pt);
        \filldraw [fill=white] (u1) circle (6pt);
        \filldraw [fill=white] (u2) circle (6pt);
        \filldraw (u2) circle (2pt);
        \draw (u2) circle (6pt);
        \filldraw [fill=white] (u3) circle (6pt);
        \filldraw [fill=white] (u4) circle (6pt);
        \filldraw (u4) circle (2pt);
        \draw (u4) circle (6pt);
        \filldraw [fill=white] (u21) circle (6pt);
        \filldraw [fill=white] (u12) circle (6pt);
        \filldraw (u12) circle (2pt);
        \draw (u12) circle (6pt);
        \filldraw [fill=white] (u41) circle (6pt);
        \filldraw [fill=white] (u32) circle (6pt);
        \filldraw (u32) circle (2pt);
        \draw (u32) circle (6pt);
\end{tikzpicture}
 = 
\def\tkzscl{0.5}
\begin{tikzpicture}[baseline={([yshift=-.5ex]current bounding box.center)},vertex/.style={anchor=base,
    circle,fill=black!25,minimum size=18pt,inner sep=2pt},scale=\tkzscl]
    \coordinate[] (u1) at (-3,0);
        \coordinate[label=left:$x_1$\,] (z_1) at (-2,2);
        \coordinate[label=left:$x_2$\,] (z_2) at (0,2);
        \coordinate[] (u2) at (1,0);
        \draw[thick] (u1) -- (-2,0);
        \draw[thick] (z_1) -- (-2,0);
        \draw[thick] (-2,0) -- (0,0);
        \draw[thick] (z_2) -- (0,0);
        \draw[thick] (u2) -- (0,0);

        \coordinate[] (u3) at (4,0);
        \coordinate[label=left:$x_{n-1}$\,] (z_n-1) at (5,2);
        \coordinate[label=left:$x_n$\,] (z_n) at (7,2);
        \coordinate[] (u4) at (8,0);
        \draw[thick,dotted] (u2) -- (u3);
        
        \draw[thick] (u3) -- (5,0);
        \draw[thick] (z_n-1) -- (5,0);
        \draw[thick] (5,0) -- (7,0);
        \draw[thick] (z_n) -- (7,0);
        \draw[thick] (u4) -- (7,0);

        \coordinate[] (u5) at (8,-2);
        \coordinate[] (u6) at (-3,-2);
        \draw[thick] (u5) -- (u6);
        \draw[thick] (8,-2) arc (-90:90:1);
        \draw[thick] (-3,0) arc (90:270:1);
        \fill (z_1) circle (6pt);
        \fill (z_2) circle (6pt);
        \fill (z_n-1) circle (6pt);
        \fill (z_n) circle (6pt);

\node at (-1,0.5) {$\Deltabb_2$};
\node at (6,0.5) {$\Deltabb_n$};
\node at (2.5,-1.5) {$\Deltabb_1$};
        
\end{tikzpicture}
 \notag
\end{align} 
where the integrations over $U_1, \dots, U_n$ are implicit. The external legs at $x_i$ carry each a weight of $\Delta_i$, respectively.

In \cite{Dolan:2000ut,Dolan:2003hv,Dolan:2011dv} it was shown that the vacuum four-point conformal block is determined by the Casimir equations derived by the Casimir elements of the conformal algebra. The vacuum higher-point case was recently discussed in \cite{Buric:2020dyz,Buric:2021ywo, Buric:2021ttm, Buric:2021kgy}. In the thermal case, the Casimir equation for the $d$-dimensional one-point block including angular potentials was derived in \cite{Gobeil:2018fzy}; the same strategy can be applied to the $n$-point block in the projection channel. In general, $\mathfrak{su}(2,2)$ has three Casimir operators $C_2$, $C_3$ and $C_4$. For the scalar representation, $C_3$ is identically zero. Furthermore, one can check that in this case $C_4 \mathbb{P}_\Deltabb = (\frac{1}{4}C_2^2 + C_2)\mathbb{P}_\Deltabb$ holds as an operator identity, meaning that the scalar representation of $\mathfrak{su}(2,2)$ is fully characterised by $C_2$.\footnote{Because we are focusing on scalar exchange, no additional vertex differential operators that were proposed in proposed in \cite{Buric:2020dyz,Buric:2021ywo, Buric:2021ttm, Buric:2021kgy} are necessary. The Casimir operator $C_4$ that commutes with the algebra in equation \eqref{eq:comm_relations_euclidPlus} is for the scalar representation given by \scriptsize\begin{align*}
    C_4&=\frac{1}{4}\left(K_{\mu}K^{\mu}P_{\nu}P^{\nu}+4K_\mu M^{\mu\nu}M_{\nu\rho}P^{\rho}+4K_\mu M^{\mu\nu}P_\nu(D-6)+\frac{3}{4}(M_{\mu\nu}M^{\mu\nu})^2\right.
    +M_{\mu\nu}M^{\mu\nu}(D^2-8D+C_2+22)\\ &\left.\quad\,-D^4+16D^3
    -80D^2+128D+36C_2-16C_2D+2C_2D^2\right)\,.
\end{align*}
}
The quadratic Casimir element for $\mathfrak{su}(2,2)$ is given by
\begin{align} \label{eq:c2}
    C_2 = D^2 - \frac{1}{2}M_{\mu \nu}M^{\mu \nu} + \frac{1}{2}\left(P_\mu K^\mu + K_\mu P^\mu\right)\,.
\end{align}
Inserting $C_2$ into the conformal block \eqref{def:block}, commuting through and using cyclicity, one can derive the respective Casimir differential equation in the same way as it was done for the one-point block in \cite{Gobeil:2018fzy}. As one can easily see from the derivation of the same Casimir equation in the two-dimensional case (see e.\,g. \cite{Kraus2017} and appendix \ref{sec_app:2d_blocks}) the difference between the one-point and the $n$-point Casimir equation in the projection channel, is the change of the differential operators $\mathcal{L}^{(x)}$ to multi-point differential operators 
\begin{align}
    \mathcal{L}^{(x_1,\dots,x_n)} = \sum_{i=1}^n \mathcal{L}^{(x_i)}\,, \qquad \mathcal{L}^{(x)}\mathcal{O}_\Delta(x) = [L,\mathcal{O}_\Delta(x)]\,,
\end{align}
in $x_1,\dots,x_n$.
Note that one needs to introduce angular potentials $\mathrm{e}^{\mu_2 M_{03}}$ and $\mathrm{e}^{\mu_3 M_{12}}$ for the derivation of any Casimir equation, even for the blocks with $\mu_2=\mu_3=0$, which poses an obstruction if one is interested only in the purely thermal case. We omit the explicit form of the Casimir equation as well as its derivation here because the method we present allows us to directly obtain the blocks at vanishing potential, without making this detour.

One can see from equation \eqref{eq:projblock_as_gluing} that the density matrix $q^D$ amounts to a finite transformation of the $U^\dagger_1$ coordinate, effectively changing boundary conditions before gluing. The presence of additional angular potentials $\mathrm{e}^{\mu_2 M_{03}}$ and $\mathrm{e}^{\mu_3 M_{12}}$, $M_{03}$ and $M_{12}$ being the other two elements of the Cartan subalgebra, gives rise to additional finite transformations of $U^\dagger_1$ in equation \eqref{eq:projblock_as_gluing}, leading to ``twists'' in the boundary conditions. These twists are then reflected in the oscillator diagram \eqref{dia:npt-block-glueing} by a label $\mathrm{e}^{\mu_2 M_{03}} \mathrm{e}^{\mu_3 M_{12}} q U_1^\dagger$ instead of just $q U_1^\dagger$.  Computing these actions and, thus, incorporating the angular potentials in our analysis is straight forward; we give more details on that in appendix \ref{sec_app:angular_potentials}.
We focus, however, on the limit of vanishing angular potentials $\mu_2 = \mu_3 =0$ in the following. Note that even though this limit is difficult to take for the respective Casimir equation, it does not pose any problems here, as the angular potentials act finitely.

\section{\texorpdfstring{$\bm{n}$}{n}-Point Blocks in the Comb Channel}
\label{sec:comb_channel_blocks}

The aim of this chapter is to compute the scalar $n$-point conformal block in the comb channel, which is the low-temperature limit of the $(n-2)$-point blocks at finite $T$ in the projection channel, as will be discussed in section \ref{sec:computation_npt_thermal}. This calculation simultaneously generalises the results in \cite{Ammon:2024axd} and sets the stage for the derivation of thermal higher-point blocks in the following chapter. We start by reviewing the oscillator construction that we use for the computation before moving on to the actual computation and conclude with some consistency checks concerning our result.

\subsection{Review of the Oscillator Construction}
\label{sec:review_osc_constr}
Our next objective is to introduce the oscillator method with which we aim to eventually compute higher-point thermal blocks. This approach was first used in \cite{Kraus2020} to rederive global blocks in two dimensions and the framework was generalised to four dimensions in \cite{Ammon:2024axd}. We begin by giving a quick review of the oscillator formalism in four dimensions.

Oscillator representations are unitary irreducible representations of the conformal algebra. The four-dimensional conformal group\footnote{Even though we here are interested in the Euclidean CFT, we study the corresponding unitary highest-weight representations of the Lorentzian conformal group. Furthermore, we really only care about the Lie algebra, so we use the term conformal group somewhat loosely. Since the Lie algebras of SO(4,2) and SU(2,2) are isomorphic, we refer to both simply as the conformal group. See for example \cite{Luscher:1974ez}.}  is given by $\text{SO}(4,2)$. Since the conformal algebra is isomorphic to $\mathfrak{su}(2,2)$, we start our construction from the quotient space 
\begin{equation}
\label{eq:su22_quotient}
    \mathbb{D}_4 \cong \text{SU}(2,2)/(\text{SU}(2) \times \text{SU}(2) \times \text{U}(1))\,
\end{equation}
of $\text{SU}(2,2)$ by its maximally compact subgroup.
The domain $\mathbb{D}_4$ consists of $(2\times 2)$-matrices $U$ such that $\mathbb{1} - U U^\dagger$ is positive-definite, and can thus be interpreted as a higher-dimensional generalisation of the complex unit disc $\mathbb{D}$ that plays an essential role in the two-dimensional oscillator formalism. 

As representation space we use the weighted Bergman space $\mathcal{H}L^2_{\Deltabb}(\Disc_4)$, the space of holomorphic functions on $\mathbb{D}_4$ with finite norm induced by the inner product
\begin{align}
    (f,g) &= c_\Deltabb \int_{\Disc_4} \frac{\d U}{\mathrm{det}^{4-\Deltabb}(\mathbb{1}-UU^\dagger)} \overline{f(U)}g(U) \equiv \int_{\mathbb{D}_4}[\d U]_\Deltabb\, \overline{f(U)}g(U)\,. \label{eq:D4_inner_product} 
\end{align}
Here, $\d U$ denotes the Lebesgue measure on $\mathbb{C}^{2\times 2}$ and the scaling dimension $\Deltabb$ labels our representations. The prefactor $c_{\mathbb{\Delta}} = \pi^{-4}(\mathbb{\Delta}-1)(\mathbb{\Delta}-2)^2(\mathbb{\Delta}-3)$ in \eqref{eq:D4_inner_product} is chosen in such a way that the constant function $1$ has unit norm. We usually drop the subscript of $[\d U]_\Deltabb$.

Based on this functional space, we represent the generators of conformal transformations as differential operators in terms of coordinates $w^\mu$ parametrizing the domain $\mathbb{D}_4$ through Pauli matrices $\sigma^j$ and unit matrix $\sigma^0$ as
\begin{equation}
    U=w_0\sigma^0 - iw_1\sigma^1 - iw_2\sigma^2- iw_3\sigma^3\,,  \label{eq:def_U_matrix}
\end{equation}
where the entries of $U$ are denoted by $u_{ij}$.
This allows us to represent the generators $D,P^\mu,K^\mu$  and $M^{\mu\nu}$ as
\begin{align}
\begin{split}\label{eq:4d_oscillator-gen}
    \mathfrak{P}^\mu = \partial^\mu\,, \quad \mathfrak{K}^\mu = w^2 \partial^\mu - 2w^\mu w_\rho \partial^\rho - 2w^\mu \mathbb{\Delta}\,,  \\
    \mathfrak{D} = w_\mu \partial^\mu + \mathbb{\Delta} \,, \qquad \mathfrak{M}^{\mu \nu} = w^\mu \partial^\nu - w^\nu \partial^\mu
    \,,
\end{split}
\end{align}
where $w^2 = w_\nu w^\nu$ is the Euclidean norm squared. One can check that these operators obey the correct algebra 
\begin{subequations}
\label{eq:comm_relations_euclidPlus}
\begin{alignat}{3}
    &[\mathfrak{P}^\mu,\,\mathfrak{P}^\nu]=0\,,\qquad &&[\mathfrak{D},\,\mathfrak{P}^\mu]=-\mathfrak{P}^\mu\,,\qquad &&[\mathfrak{M}^{\mu\nu},\,\mathfrak{P}^\rho]=-\delta^{\mu\rho}\mathfrak{P}^\nu+\delta^{\nu\rho}\mathfrak{P}^\mu\,,\\
    &[\mathfrak{K}^\mu,\,\mathfrak{K}^\nu]=0\,,\quad &&[\mathfrak{D},\,\mathfrak{K}^\mu]=\mathfrak{K}^\mu\,, \quad &&[\mathfrak{M}^{\mu\nu},\,\mathfrak{K}^\rho]=-\delta^{\mu\rho}\mathfrak{K}^\nu+\delta^{\nu\rho}\mathfrak{K}^\mu\,,\\
    &[\mathfrak{M}^{\mu\nu},\,\mathfrak{D}]=0\,,&&[\mathfrak{K}^\mu,\,\mathfrak{P}^\nu]=2(\mathfrak{M}^{\mu\nu}+\delta^{\mu\nu}\mathfrak{D})\,,&&\\
    &[\mathfrak{M}^{\mu\nu},\mathfrak{M}^{\rho\sigma}]=-\delta^{\mu\rho}&&\mathfrak{M}^{\nu\sigma}-\delta^{\nu\sigma}\mathfrak{M}^{\mu\rho}+\delta^{\nu\rho}\mathfrak{M}^{\mu\sigma}+&&\,\delta^{\mu\sigma}\mathfrak{M}^{\nu\rho}\,,
\end{alignat}
\end{subequations}
as well as the adjoint relations
\begin{align}
\label{eq:adjoint_relations_4d}
    \left(\mathfrak{P}^\mu\right)^\dagger&=-\mathfrak{K}^\mu\,, &
    \left(\mathfrak{M}^{\mu\nu}\right)^\dagger&=-\mathfrak{M}^{\mu\nu}\,, &
    \mathfrak{D}^\dagger&=\mathfrak{D}
\end{align}
with respect to the inner product of the Bergman space. 
The base polynomials
\begin{equation}
    \phi_{a,b}^{j,m} (U) = \text{det}^m(U) \mathcal{D}^j_{a,b}(U)
\end{equation}
are a modification of the (holomorphically extended) Wigner $\mathcal{D}$-matrices\footnote{In this expression $k$ runs from $\max(0,a + b)$ to $\min(j+a,j+b)$.}
\begin{equation} \label{eq:def_wignerD}
    \mathcal{D}^j_{a,b}(U) = \sqrt{\frac{(j + a)!(j-a)!}{(j+b)!(j - b)!}} \sum\limits_k \binom{j + b}{k} \binom{j-b}{k - a - b}
    u^k_{11} u^{j+a - k}_{12} u^{j+b - k}_{21} u^{k-a - b}_{22}\,,
\end{equation}
and depend on four parameters: $m\in\mathbb{N}_{0}$ and $j\in \mathbb{N}_{0}/2$, as well as $a,b=-j,-j+1,\dots,j$. In \cite{Calixto2010}, it was shown that these polynomials indeed form an orthogonal basis as their inner product is given by 
\begin{align}
    \braket{\phi_{a,b}^{j,m}}{\phi_{a',b'}^{j',m'}} = \int_{\mathbb{D}_4}[\d U]_\mathbb{\Delta} \overline{\phi_{a,b}^{j,m} (U)}\, \phi_{a',b'}^{j',m'} (U) = (\mathcal{N}^{j,m}_\mathbb{\Delta})^{-2}\delta_{j,j'}\delta_{m,m'}\delta_{a,a'}\delta_{b,b'}\,,\label{eq:4d_orthonormality_property} 
\end{align}
with normalisation factor
\begin{align} 
    \label{eq:normalisation_inner_prod}\mathcal{N}^{j,m}_\mathbb{\Delta} = \sqrt{\frac{d_j}{\mathbb{\Delta} -1} \binom{m+\mathbb{\Delta} -2}{\mathbb{\Delta} - 2} \binom{d_j + m+\mathbb{\Delta} - 2}{\mathbb{\Delta} - 2}}\,, \qquad d_j = 2j+1\,.
\end{align}
Note that we could also work in the orthonormal basis $\varphi_{a,b}^{j,m} (U) = \mathcal{N}^{j,m}_\mathbb{\Delta}\phi_{a,b}^{j,m} (U)$, however, this makes the base polynomials $\Deltabb$-dependent which we want to avoid. As was already pointed out in \cite{Ammon:2024axd}, it is necessary to analyse higher order products, such as $\braket{ \phi_{a_1,b_1}^{j_1,m_1}\phi_{a_2,b_2}^{j_2,m_2}}{\phi_{a_3,b_3}^{j_3,m_3}\phi_{a_4,b_4}^{j_4,m_4}}$, to compute higher-point blocks -- we dedicate appendix \ref{sec_app:HigherIntegrals_Details} to this. 

We need to introduce two more concepts of the oscillator method before turning to the higher order integrals. For weighted Bergman spaces like $\mathcal{H}L^2_{\Deltabb}(\Disc_4)$ there exist reproducing kernels, i.\,e. functions  $K_\mathbb{\Delta}(U,U')$ that are holomorphic (anti-holomorphic) in $U'$ ($U$) and obey a reproducing identity 
\begin{equation}
    f(U') = \int_{\mathbb{D}_4}[\d U]_\mathbb{\Delta} K_\Deltabb(U,U') f(U)\, \label{eq:D4_reproducing_identity}
\end{equation}
for all $f$ in $\mathcal{H}L^2_{\Deltabb}(\Disc_4)$.
In our case, this kernel takes a rather compact form and allows for a simple expansion in terms of the previously introduced base polynomials, namely 
\begin{equation}
\begin{split}
    K_\mathbb{\Delta}(U,U') &= \mathrm{det}^{-\mathbb{\Delta}}(\mathbb{1}-U^\dagger U')\\ 
    &= \sum\limits_{j \in \mathbb{N}/2}\sum\limits_{m=0}^\infty \sum\limits_{a,b=-j}^j (\mathcal{N}^{j,m}_\mathbb{\Delta})^2 \overline{\phi_{a,b}^{j,m} (U)} \phi_{a,b}^{j,m} (U') \label{eq:expansion_kernel}\\
    &\equiv\sum\limits_{a,b}^{j,m} (\mathcal{N}^{j,m}_\mathbb{\Delta})^2 \overline{\phi_{a,b}^{j,m} (U)} \phi_{a,b}^{j,m} (U')\, 
\end{split}
\end{equation}
as was shown in \cite{Calixto2010}. We will repeatedly make use of this kind of expansion throughout this paper. Note that the existence of the reproducing kernel implies the existence of a resolution of identity, which is given as the projection operator $\mathbb{P}_\Deltabb$.

Let us now turn towards the key ingredients for the computation of conformal blocks, the oscillator wavefunctions. Instead of working with Euclidean coordinates $x_\mu$ we can equivalently use a complex matrix 
\begin{equation} \label{eq:X_parametr}
    X = x_0 \sigma^0 + ix_1 \sigma^1 + ix_2 \sigma^2 + i x_3 \sigma^3\,.
\end{equation}
expressing the coordinate in the Pauli matrix basis.
The conformal generators then act in the usual way on the fields $[L,\mathcal{O}(x)] = \mathcal{L}^{(x)}\mathcal{O}(x)$, as operators $\mathcal{L}^{(x)}$ with respect to  $x_\mu$. We take $L$ as a placeholder for the conformal generators and the corresponding differential operators $\mathcal{L}^{(x)}$ explicitly read 
\begin{align} \label{eq:adjoint_generators}
\begin{split}
    \mathcal{P}_\mu &= \partial_\mu\,,\quad \mathcal{K}_\mu = x^2\partial_\mu-2x_\mu x^\rho\partial_\rho-2x_\mu\Delta\,,\\
    \mathcal{D} &= x^\mu \partial_\mu + \Delta\,,
    \quad
    \mathcal{M}_{\mu\nu} = x_\mu\p_\nu-x_\nu\p_\mu\,.
\end{split}
\end{align}
The remaining (second-level) oscillator wavefunctions, in addition to the one we already introduced  \eqref{eq:Omega_mel_def}, are defined as the matrix elements
\begin{align}     
    \chi(X_1,X_2;U^\dagger)&\equiv\chi_{\Delta_1,\Delta_2,\Deltabb}(X_1,X_2;U^\dagger) = \mel*{0}{\mathcal{O}_{\Delta_1}(X_1)\mathcal{O}_{\Delta_2}(X_2)}{U^\dagger}_{\Deltabb}\,,\label{eq:2ndlevel_chi_osc_eq}\\
    \psi\left(X_1,X_2;U\right)&\equiv\psi_{\Delta_1,\Delta_2,\Deltabb}(X_1,X_2;U) = {}_\Deltabb\hspace{-.08cm}\mel{U}{\mathcal{O}_{\Delta_1}(X_1)\mathcal{O}_{\Delta_2}(X_2)}{0}\,. \label{eq:2ndlevel_psi_osc_eq}
\end{align}
Using the conformal invariance of the vacuum state and commuting generator insertions with the primary operators, one arrives at the oscillator equations
\begin{align}
    0 &= \left(\mathfrak{L}^{(\bar{w})\phantom{\dagger}}\!\! + \mathcal{L}^{(x_1)} + \mathcal{L}^{(x_2)}\right)\chi(x_1,x_2;\bar{w})\,, \label{eq:4d_diffeq_chi_2nd}\\
    0 &= \left(\mathfrak{L}^{(w)\dagger} - \mathcal{L}^{(x_1)} - \mathcal{L}^{(x_2)} \right)\psi(x_1,x_2;w)\,, \label{eq:4d_diffeq_psi_2nd} \\
    0 &= \left(\mathfrak{L}^{(w_1)\dagger} - \mathfrak{L}^{(\bar{w}_2)}-\mathcal{L}^{(x)}\right) \Omega(x;w_1,\bar{w}_2)\,. \label{eq:4d_diffeq_omega_2nd}
\end{align}
The solutions of these systems of PDEs read
\begin{align}
    \chi(X_1,X_2;U^\dagger)&=  \mathrm{det}^{-\alpha}(X_1-U^\dagger)\, \mathrm{det}^{-\beta}(X_2-U^\dagger)\,\mathrm{det}^{-\gamma}(X_1-X_2)\,,\label{eq:sol_chi_2ndLvl_V1}\\
    \psi(X_1,X_2;U) &= \mathrm{det}^{-\alpha}\left(\mathbb{1} -U {X_1} \right) \mathrm{det}^{-\beta}\left(\mathbb{1} -U {X_2}\right)\mathrm{det}^{-\gamma}(X_1-X_2)\,, \label{eq:sol_psi_2ndLvl_V1} \\
    \Omega(X;U_1,U_2^{\dagger}) &=  \mathrm{det}^{-\Tilde{\alpha}}\left(\mathbb{1} -U_1 X\right) \mathrm{det}^{-\Tilde{\beta}}(\mathbb{1} -U_1 {U_2^{\dagger}})\, \mathrm{det}^{-\Tilde{\gamma}}(X-U_2^{\dagger})\,, \label{eq:sol_Omega_2ndLvl_V1}
\end{align}
with sets of exponents  
\begin{align}
    \alpha &= \frac{1}{2}(\Delta_1+\Deltabb-\Delta_2)\,, & \beta &= \frac{1}{2}(\Delta_2+\Deltabb-\Delta_1)\,, & \gamma &= \frac{1}{2}(\Delta_1+\Delta_2-\Deltabb)\,
\end{align}
and
\begin{align}
     \Tilde{\alpha} &= \frac{1}{2}(\Delta+\Deltabb_1-\Deltabb_2)\,,  & \Tilde{\beta} &= \frac{1}{2}(\Deltabb_2+\Deltabb_1-\Delta)\,,& \Tilde{\gamma} &= \frac{1}{2}(\Delta+\Deltabb_2-\Deltabb_1)\,. 
\end{align}
With this, essentially all the tools for the computation of the $n$-point comb blocks and thermal blocks are laid out.

\subsection{\texorpdfstring{Computing the $\bm{n}$-Point Comb Block}{Computation of the n-Point Comb block}}
\label{sec:computation_npt_comb}

Let us now turn to the explicit computations of scalar conformal blocks from the oscillator formalism. In \cite{Ammon:2024axd}, this method was already used to reproduce the results for the four-point block \cite{Dolan:2000ut,Dolan:2003hv} and a special case of the $n$-point block in the comb channel. Using the solution to what we call the $(2,2)$-integral, calculated in appendix \ref{sec_app:HigherIntegrals_Details}, we can now approach the general case of the $n$-point comb block with external and inner scalar weights. Note that this block was first computed in \cite{Fortin:2019zkm,Parikh:2019dvm}. 

Starting from the correlation function of $n$ scalar primary operators $\mathcal{O}_{\Delta_i}$ with scaling dimensions $\Delta_i$, the corresponding comb block 
\begin{equation}
    G^{(n)}_{\text{Comb}} = \expval{\mathcal{O}_{\Delta_1}(x_1)\mathcal{O}_{\Delta_2}(x_2)\mathbb{P}_{\Deltabb_1}\mathcal{O}_{\Delta_3}(x_3)\mathbb{P}_{\Deltabb_2}\dots \mathbb{P}_{\Deltabb_{n-3}}\mathcal{O}_{\Delta_{n-1}}(x_{n-1})\mathcal{O}_{\Delta_n}(x_n)}
\end{equation}
is obtained via the insertion of $(n-3)$ projectors $\mathbb{P}_{\Deltabb_j}$. Expressing the projectors through integrals over the domain $\mathbb{D}_4$, the block reads
\begin{align}
    G^{(n)}_{\text{Comb}} 
    =& \int [\d U_1]\dots [\d U_{n-3}]\chi_{\Delta_1,\Delta_2,\Deltabb_1}(X_1,X_2;U_1^\dagger) \prod_{i=2}^{n-3} \Omega_{\Delta_{i+1},\Deltabb_{i-1},\Deltabb_i}(X_{i+1};U_{i-1},U_i^\dagger) \notag \\
    &\quad\times \psi_{\Delta_{n-1},\Delta_{n},\Deltabb_{n-3}}(X_{n-1},X_n;U_{n-3}) 
    \label{eq:Comb_block_wavefunctions_ansatz}
\end{align}
in terms of the oscillator wavefunctions. The corresponding diagram looks like
\begin{align} \label{dia:comb_npt}
G^{(n)}_{\text{Comb}}(X_1,\dots,X_n) =
\def\tkzscl{0.5}
\begin{tikzpicture}[baseline={([yshift=-.5ex]current bounding box.center)},vertex/.style={anchor=base,
    circle,fill=black!25,minimum size=18pt,inner sep=2pt},scale=\tkzscl]
        \coordinate[label=left:$X_2$\,] (z_1) at (-2,2);
        \coordinate[label=left:$X_1$\,] (z_2) at (-2,-2);
        \coordinate[label=above:\,$U_1^\dagger\phantom{|}$] (u_1) at (3,0);
        \draw[thick] (z_1) -- (0,0);
        \draw[thick] (z_2) -- (0,0);
        \draw[thick, dashed] (0,0) -- node[above] {$\mathbb{\Delta}_1$} (u_1);
        \fill (z_1) circle (6pt);
        \fill (z_2) circle (6pt);
        \filldraw [fill=white] (u_1) circle (6pt);
        \filldraw (u_1) circle (2pt);
        \draw (u_1) circle (6pt);
\end{tikzpicture}
 \!\!\cdots \mkern-10mu \! \!
\def\tkzscl{0.5}
\begin{tikzpicture}[baseline={([yshift=-.5ex]current bounding box.center)},vertex/.style={anchor=base,
    circle,fill=black!25,minimum size=18pt,inner sep=2pt},scale=\tkzscl]
        \coordinate[label=left:$X_{n-2}$\,] (z_1) at (0,2.2);
        \coordinate[label=below:$U_{n-4}\phantom{|}$\,] (u_1) at (-3,0);
        \coordinate[label=below:\,$U^\dagger_{n-3}\phantom{|}$] (u_2) at (3,0);
        \draw[thick] (z_1) -- (0,0);
        \draw[thick, dashed] (u_2) --node[above] {$\mathbb{\Delta}_{n-3}$} (0,0);
        \draw[thick, dashed] (0,0) -- node[above] {$\mathbb{\Delta}_{n-4}$} (u_1);
        \fill (z_1) circle (6pt);
        \fill[white] (u_1) circle (6pt);
        \draw (u_1) circle (6pt);
        \fill[white] (u_2) circle (6pt);
        \filldraw (u_2) circle (2pt);
        \draw (u_2) circle (6pt);
        \draw[draw=none] (0,0) -- (0,-2.6);
\end{tikzpicture}
 \!\!
\def\tkzscl{0.5}
\begin{tikzpicture}[baseline={([yshift=-.5ex]current bounding box.center)},vertex/.style={anchor=base,
    circle,fill=black!25,minimum size=18pt,inner sep=2pt},scale=\tkzscl]
        \coordinate[label=left:$X_{n-1}$\,\,] (z_1) at (2,2);
        \coordinate[label=left:$X_n$\,] (z_2) at (2,-2);
        \coordinate[label=below:$U_{n-3}\phantom{|}$\,] (u_1) at (-3,0);
        \draw[thick] (z_1) -- (0,0);
        \draw[thick] (z_2) -- (0,0);
        \draw[thick, dashed] (0,0) -- node[above] {$\mathbb{\Delta}_{n-3}$} (u_1);
        \fill (z_1) circle (6pt);
        \fill (z_2) circle (6pt);
        \fill[white] (u_1) circle (6pt);
        \draw (u_1) circle (6pt);
\end{tikzpicture}
\end{align}
where we suppressed the integrations over $U_1$ to $U_{n-3}$. Using the oscillator equations \eqref{eq:4d_diffeq_chi_2nd}, \eqref{eq:4d_diffeq_psi_2nd} and \eqref{eq:4d_diffeq_omega_2nd}, we can show that the construction \eqref{eq:Comb_block_wavefunctions_ansatz} for the $n$-point block in the comb channel solves Casimir eigenvalue equations for $j = 1,\dots, n-3$ as well as the Ward identities 
\begin{align} \label{eq:c2_eq_comb}
    \mathcal{C}^{(X_1,\dots,X_{j+1})}_2 G_{\text{Comb}}^{(n)} &= \Deltabb_{j} (\Deltabb_{j}-4)\, G_{\text{Comb}}^{(n)}\,, & \mathcal{L}^{(X_1,\dots,X_n)} G_{\text{Comb}}^{(n)} = 0\,.
\end{align}
Here, $\mathcal{L}$ stands for the different multipoint generators in equation \eqref{eq:adjoint_generators}. All one has to do is use the ansatz \eqref{eq:Comb_block_wavefunctions_ansatz} and rewrite the differential operators acting on the wavefunctions through their respective oscillator equation. In this way, one ends up either with trivial actions $\mathfrak{C}_2 = \Deltabb(\Deltabb-4)$ or terms that drop out due to the adjoint relations \eqref{eq:adjoint_relations_4d}. 

The computation now runs as follows: We use the explicit expressions for the wavefunctions \eqref{eq:sol_chi_2ndLvl_V1}, \eqref{eq:sol_psi_2ndLvl_V1} as well as \eqref{eq:sol_Omega_2ndLvl_V1} and expand each determinant in base polynomials similarly to the kernel \eqref{eq:expansion_kernel}. This reduces the integrals to higher-order products that we discuss in the appendix \ref{sec_app:HigherIntegrals_Details}. To simplify the calculation a bit, we use conformal transformations to set $x_1\to\infty$ and $x_{n-1}=0$.\footnote{We compensate the divergence from the $x_1\to\infty$ limit with an additional factor of ${\det}^{\Delta_1}{(X_1)}$. The choice of setting $x_{n-1}=0$ instead of $x_{n}=0$ is somewhat arbitrary; the only thing that would be different in the result is the change $\Delta_{n-1}\leftrightarrow\Delta_n$ or equivalently $\alpha_{n-2}\leftrightarrow\beta_{n-2}$.} Due to this choice there only remains one factor to expand in both $\chi$ and $\psi$, so we find
\begin{equation}
\begin{split}
    \chi_{\Delta_1,\Delta_2,\Deltabb_1}(X_1,X_2;U_1^\dagger) &= {\det}^{-\alpha_1}(X_1-U_1^\dagger)\, {\det}^{-\beta_1}(X_2-U_1^\dagger) \, {\det}^{-\gamma_1}(X_1-X_2) \\
    &\!\overset{X_1\to\infty}{\longrightarrow} {\det}^{-\beta_1}(X_2)\,{\det}^{-\beta_1}(\mathbb{1}-X_2^{-1}U_1^\dagger) \\
    &=  {\det}^{-\beta_1}(X_2)\, \sum_{a_1,b_1}^{j_1,m_1} \left(\mathcal{N}^{j_1,m_1}_{\beta_1}\right)^2 \overline{\phi^{j_1,m_1}_{a_1,b_1}(U_1)} \phi^{j_1,m_1}_{a_1,b_1}(X_2^{-1})\,,
\end{split} \label{eq:comb_chi_expansion}
\end{equation}
and
\begin{equation}
\begin{split}
    &\psi_{\Delta_{n-1},\Delta_{n},\Deltabb_{n-3}}(X_{n-1},X_n;U_{n-3}) \\
    &= {\det}^{-\alpha_{n-2}}(\mathbb{1}-U_{n-3}X_{n-1})\, {\det}^{-\beta_{n-2}}(\mathbb{1}-U_{n-3}X_{n}) \, {\det}^{-\gamma_{n-2}}(X_{n-1}-X_n) \\
    &\!\overset{X_{n-1}\to 0}{\longrightarrow} {\det}^{-\gamma_{n-2}}(X_n)\,{\det}^{-\beta_{n-2}}(\mathbb{1}-U_{n-3}X_{n}) \\
    &=  {\det}^{-\gamma_{n-2}}(X_n)\, \sum_{a_{n-2},b_{n-2}}^{j_{n-2},m_{n-2}} \left(\mathcal{N}^{j_{n-2},m_{n-2}}_{\beta_{n-2}}\right)^2 \phi_{a_{n-2},b_{n-2}}^{j_{n-2},m_{n-2}}(U_{n-3}) \phi_{a_{n-2},b_{n-2}}^{j_{n-2},m_{n-2}}(X_n^T)\,.
\end{split}\label{eq:comb_psi_expansion}
\end{equation}
Slightly more involved is the expansion of the $\Omega$-wavefunctions, since each has three factors that need to be expanded. To make our notation more compact, let us introduce some short-hand notation. First, we group the four indices of each base polynomial in one block-index such as $(i) \equiv (j_i,m_i,a_i,b_i)$. This includes writing the sum over all included indices simply as $\sum_{(i)}$ and the polynomials as $\phi^{(i)}$. Moreover, if we have primed indices such as $j_i'$, we indicate that in the block-index as $(i')$.
If $j_i$ and $m_i$ appear as a sum, we abbreviate it with $k_i = j_i+m_i$. 
Now, we can express the expansion of each $\Omega$-wavefunction as  
\begin{equation}
\begin{split}
    &\Omega_{\Delta_{i+1},\Deltabb_{i-1},\Deltabb_i}(X_{i+1};U_{i-1},U_i^\dagger) \\
    &= {\det}^{-\hat{\alpha}_i}(\mathbb{1}-U_{i-1}X_{i+1})\,{\det}^{-\hat{\beta}_i}(\mathbb{1}-U_{i-1}U_{i}^\dagger)\,{\det}^{-\hat{\gamma}_i}(X_{i+1}-U_i^\dagger) \\
    &= {\det}^{-\hat{\gamma}_i}(X_{i+1}) \sum_{(i)} \sum_{(i')} \sum_{(i'')} \left(\mathcal{N}^{j_i,m_i}_{\hat{\alpha}_i}\mathcal{N}^{j_i',m_i'}_{\hat{\beta}_i}\mathcal{N}^{j_i'',m_i''}_{\hat{\gamma}_i}\right)^2  \\
    &\qquad\times \phi^{(i)}(X_{i+1}^T)\phi^{(i)}(U_{i-1}) \phi^{(i')}(U_{i-1})
    \overline{\phi^{(i')}(U_{i})} \phi^{(i'')}(X_{i+1}^{-1})
    \overline{\phi^{(i'')}(U_{i})} \,, \label{eq:comb_Omega_expansion}
\end{split}
\end{equation}
with $i=2,3,\dots,n-3$. The exponents in equations \eqref{eq:comb_chi_expansion}, \eqref{eq:comb_psi_expansion} and \eqref{eq:comb_Omega_expansion} are given by
\begin{align}
    \alpha_1 &= \frac{\Delta_1+\Deltabb_1-\Delta_2}{2}\,, & \beta_1 &=\frac{\Delta_2+\Deltabb_1-\Delta_1}{2}\,, & \gamma_1 &= \frac{\Delta_1+\Delta_2-\Deltabb_1}{2}\,, \notag\\
    \hat{\alpha}_i &= \frac{\Delta_{i+1}+\Deltabb_{i-1}-\Deltabb_i}{2}\,, & \hat{\beta}_i &=\frac{\Deltabb_i+\Deltabb_{i-1}-\Delta_{i+1}}{2}\,, & \hat{\gamma}_i &= \frac{\Delta_{i+1}+\Deltabb_{i}-\Deltabb_{i-1}}{2}\,, \notag\\
    \alpha_{n-2} &= \frac{\Delta_{n-1}+\Deltabb_{n-3}-\Delta_{n}}{2}\,, & \beta_{n-2} &=\frac{\Delta_n+\Deltabb_{n-3}-\Delta_{n-1}}{2}\,, & \gamma_{n-2} &= \frac{\Delta_{n-1}+\Delta_n-\Deltabb_{n-3}}{2}\,. \label{eq:exponents_conf_weights}
\end{align}
Plugging in the expansions and expressing the $\mathbb{D}_4$-integrals as higher-order inner products of base functions, we obtain 
\begin{align}
    &G^{(n)}_{\text{Comb}}(X_1,\dots,X_n) \label{eq:intermed_result_comb} \\
    &= {\det}^{-\beta_1}(X_2)\,\left(\prod_{i=2}^{n-3} {\det}^{-\hat{\gamma}_i}(X_{i+1}) \right)\,{\det}^{-\gamma_{n-2}}(X_n)\, \sum_{(1)} \left(\prod_{i=2}^{n-3} \sum_{(i)} \sum_{(i')} \sum_{(i'')}\right) \sum_{(n-2)} \notag\\
    &\times \hspace{-.1cm} \left( \mathcal{N}^{j_1,m_1}_{\beta_1}\left( \prod_{i=2}^{n-3} \mathcal{N}^{j_i,m_i}_{\hat{\alpha}_i}\mathcal{N}^{j_i',m_i'}_{\hat{\beta}_i}\mathcal{N}^{j_i'',m_i''}_{\hat{\gamma}_i}\right)  \mathcal{N}^{j_{n-2},m_{n-2}}_{\beta_{n-2}}  \right)^2  \hspace{-.2cm}  \phi^{(1)}(X_2^{-1}) \left( \prod_{i=2}^{n-3} \phi^{(i)}(X_{i+1}^T) \phi^{(i'')}(X_{i+1}^{-1}) \right) \notag\\
    &\times \phi^{(n-2)}(X_n^T)\braket{\phi^{(1)}}{\phi^{(2)}\cdot\phi^{(2')}} \left(\prod_{i=2}^{n-4} \hspace{-.1cm} \braket{\phi^{(i')}\cdot\phi^{(i'')}}{\phi^{(i+1)}\cdot\phi^{(i+1')}} \right) \hspace{-.1cm} \braket{\phi^{(n-3')}\cdot\phi^{(n-3'')}}{\phi^{(n-2)}}\hspace{-.08cm}.\notag
\end{align}
We calculated the above integrals, which we call the $(1,2)$-, $(2,1)$- and the $(2,2)$-integral, in appendix \ref{sec_app:HigherIntegrals_Details} as
\begin{align}
&\braket{\phi^{j_1,m_1}_{a_1,b_1} \phi^{j_2,m_2}_{a_2,b_2} }{\phi^{j_3,m_3}_{a_3,b_3} \phi^{j_4,m_4}_{a_4,b_4} } \\ 
&\quad= \sum_{p_a,p_b}^{l,n} \left(\mathcal{N}^{l,n}_{\Deltabb}\right)^{-2} 
C^{j_1j_2l}_{a_1,a_2,p_a} 
C^{j_3j_4l}_{a_3,a_4,p_a} 
C^{j_1j_2l}_{b_1,b_2,p_b} C^{j_3j_4l}_{b_3,b_4,p_b}   \delta^{l+n}_{k_1+k_2} \delta^{l+n}_{k_3+k_4}\,, \notag
\end{align}
with $\mathcal{N}$-coefficients given in equation \eqref{eq:normalisation_inner_prod} and the SU(2) Clebsch-Gordan coefficients. It is convenient to group the $\mathcal{N}$-coefficients in equation \eqref{eq:intermed_result_comb} together with the Clebsch-Gordan coefficients, defining the weighted intertwiners $\widetilde{\mathcal{I}}$ and base polynomials as $\varphi^{(1)}_{\alpha}(U) \equiv \mathcal{N}^{j_1,m_1}_\alpha\phi^{(1)}(U)$. Taken together, we define the $\mathcal{T}$-functions 
\begin{align}
    \mathcal{T}_{(1),(4)}\left(\begin{matrix}\alpha,\beta,\gamma,\delta \\ \Deltabb \end{matrix}\,\Bigg| X_1,X_2\right) \equiv \sum_{(2),(3)} \widetilde{\mathcal{I}}_{(1),(2),(3),(4)}^{\alpha\beta\gamma\delta} \varphi^{(2)}_\beta(X_1) \varphi^{(3)}_\gamma(X_2)\,. 
\end{align}
We elaborate on this in appendix \eqref{sec:adjusted_coeffs}. The $n$-point conformal block in the comb channel can then be expressed as 
\begin{align}
    &G^{(n)}_{\text{Comb}}(X_1,\dots,X_n) \notag\\
    &= {\det}^{-\beta_1}(X_2)\,\left(\prod_{i=2}^{n-3} {\det}^{-\hat{\gamma}_i}(X_{i+1}) \right)\,{\det}^{-\gamma_{n-2}}(X_n)\,  \sum_{(2)}\cdots \sum_{(n-3)}  \notag\\
    &\quad\times \mathcal{T}_{0,(2)}\left(\begin{matrix}\,-\,,\beta_1,\hat{\alpha}_2,\hat{\beta}_2 \\ \Deltabb_1 \end{matrix}\,\Bigg| X_2^{-1},X_3^T\right) \left(\prod_{i=2}^{n-4} \mathcal{T}_{(i),(i+1)}\left(\begin{matrix} \hat{\beta}_i,\hat{\gamma}_i,\hat{\alpha}_{i+1},\hat{\beta}_{i+1} \\ \Deltabb_i \end{matrix}\,\Bigg| X_{i+1}^{-1},X_{i+2}^T\right) \right) \notag\\
    &\quad\times \mathcal{T}_{(n-3),0}\left(\begin{matrix}\hat{\beta}_{n-3},\hat{\gamma}_{n-3},\beta_{n-2},\,-\, \\ \Deltabb_{n-3} \end{matrix}\,\Bigg| X_{n-2}^{-1},X_n^T\right)\,. \label{eq:npt_comb_tau}
\end{align}
Since the $n$-point block is a series of weighted SU(2) Clebsch-Gordan coefficients, we can use SU(2) graphical notation to visualise equation \eqref{eq:npt_comb_tau} as\footnote{We always use equation \eqref{eq:rewrite_intertwiners} to rewrite $\widetilde{I}$-tensors into $I$-tensors, because these are more standard, giving additional $d_j$-factors for every $\mathcal{T}$-function. This is why the $\mathcal{T}$-functions depend on matrices $X_i^T$, but the diagrams on $X^\dagger_i$.}
\begin{align} \label{eq:su2_vac_npt}
\begin{split} G^{(n)}_{\text{Comb}}(X_1,\dots,X_n) \quad = \quad \begin{tikzpicture}[scale=1.1,baseline={([yshift=-.5ex]current bounding box.center)},vertex/.style={anchor=base,
    circle,fill=black!25,minimum size=18pt,inner sep=2pt}]

\def\dx{1.2}
\def\dy{1.2}
\def\boxW{0.86}
\def\boxH{0.8}
\def\lineW{1pt}
\def\gap{0.6}

\pgfmathsetmacro{\xLone}{0*\dx}
\draw[line width=\lineW] (\xLone - \boxW/2, -\boxH/2) rectangle (\xLone + \boxW/2, \boxH/2);
\node at (\xLone, 0) {$X_2^{-1}$};

\pgfmathsetmacro{\xLtwo}{1*\dx}
\draw[line width=\lineW] (\xLtwo - \boxW/2, -\boxH/2) rectangle (\xLtwo + \boxW/2, \boxH/2);
\node at (\xLtwo, 0) {$X_3^{\dagger}$};

\pgfmathsetmacro{\xLthree}{2*\dx}
\draw[line width=\lineW] (\xLthree - \boxW/2, -\boxH/2) rectangle (\xLthree + \boxW/2, \boxH/2);
\node at (\xLthree, 0) {$X_3^{-1}$};

\foreach \i in {1,2} {
  \pgfmathsetmacro{\x}{\i*\dx}
  \draw[line width=\lineW] (\x, \boxH/2) -- (\x, \boxH/2 + \dy);
  \draw[line width=\lineW] (\x, -\boxH/2) -- (\x, -\boxH/2 - \dy);
}

\pgfmathsetmacro{\xtopL}{1*\dx}
\pgfmathsetmacro{\xtopR}{2*\dx}
\pgfmathsetmacro{\ext}{\gap}
\draw[line width=\lineW] (\xtopL, \boxH/2 + \dy) -- (\xtopR + \ext, \boxH/2 + \dy);
\draw[line width=\lineW] (\xtopL, -\boxH/2 - \dy) -- (\xtopR + \ext, -\boxH/2 - \dy);

\node at (0.5*\dx+0.3, \boxH/2 + 0.5) {$\hat{\alpha}_2$};
\node at (1.5*\dx+0.3, \boxH/2 + 0.5) {$\hat{\gamma}_2$};
\node at (3*\dx+0.2, \boxH/2 + 0.5) {$\hat{\alpha}_{n-3}$\,};
\node at (4*\dx+0.2, \boxH/2 + 0.5) {$\hat{\gamma}_{n-3}$};
\node at (0.5*\dx+0.3, -\boxH/2 - 0.5) {$\hat{\alpha}_2$};
\node at (1.5*\dx+0.3, -\boxH/2 - 0.5) {$\hat{\gamma}_2$};
\node at (3*\dx+0.2, -\boxH/2 - 0.5) {$\hat{\alpha}_{n-3}$\,};
\node at (4*\dx+0.2, -\boxH/2 - 0.5) {$\hat{\gamma}_{n-3}$};

\pgfmathsetmacro{\xcdots}{\xtopR + \gap + 0.35}
\node at (\xcdots, 0) {\Large $\cdots$};

\pgfmathsetmacro{\xRone}{3*\dx + \gap}
\draw[line width=\lineW] (\xRone - \boxW/2, -\boxH/2) rectangle (\xRone + \boxW/2, \boxH/2);
\node at (\xRone, 0) {$X_{n-2}^{\dagger}$};

\pgfmathsetmacro{\xRtwo}{3*\dx + \gap + 1*\dx}
\draw[line width=\lineW] (\xRtwo - \boxW/2, -\boxH/2) rectangle (\xRtwo + \boxW/2, \boxH/2);
\node at (\xRtwo, 0) {$X_{n-2}^{-1}$};

\pgfmathsetmacro{\xRthree}{3*\dx + \gap + 2*\dx}
\draw[line width=\lineW] (\xRthree - \boxW/2, -\boxH/2) rectangle (\xRthree + \boxW/2, \boxH/2);
\node at (\xRthree, 0) {$X_n^{\dagger}$};

\foreach \j in {0,1} {
  \pgfmathsetmacro{\x}{3*\dx + \gap + \j*\dx}
  \draw[line width=\lineW] (\x, \boxH/2) -- (\x, \boxH/2 + \dy);
  \draw[line width=\lineW] (\x, -\boxH/2) -- (\x, -\boxH/2 - \dy);
}

\pgfmathsetmacro{\xrightL}{3*\dx + \gap}
\pgfmathsetmacro{\xrightR}{\xrightL + 1*\dx}
\draw[line width=\lineW] (\xrightL - \ext, \boxH/2 + \dy) -- (\xrightR, \boxH/2 + \dy);
\draw[line width=\lineW] (\xrightL - \ext, -\boxH/2 - \dy) -- (\xrightR, -\boxH/2 - \dy);
\draw[line width=\lineW, dashed] (3, -\boxH/2 - \dy) -- (3.6, -\boxH/2 - \dy);
\draw[line width=\lineW, dashed] (3, \boxH/2 + \dy) -- (3.6, \boxH/2 + \dy);

\node at (1.5*\dx, \boxH/2 + \dy + 0.3) {$\hat{\beta}_2$};
\node at (4*\dx, \boxH/2 + \dy + 0.3) {$\hat{\beta}_{n-3}$};
\node at (1.5*\dx, -\boxH/2 - \dy - 0.3) {$\hat{\beta}_2$};
\node at (4*\dx, -\boxH/2 - \dy - 0.3) {$\hat{\beta}_{n-3}$};

\pgfmathsetmacro{\xstart}{0}
\pgfmathsetmacro{\xtopLx}{\xtopL}
\pgfmathsetmacro{\ctrlXleft}{0.5*\xtopLx + 0.5*\xstart}
\draw[line width=\lineW] 
  (\xtopLx, \boxH/2 + \dy) .. controls (\xtopLx -1.1, \boxH/2 + \dy) .. 
  (\xstart, \boxH/2);

\draw[line width=\lineW] 
  (\xtopLx, -\boxH/2 - \dy) .. controls 
  (\xtopLx-1.1, -\boxH/2 - \dy) .. 
  (\xstart, -\boxH/2);

\node at (\xstart - 0.3, \boxH/2 + 0.5) {$\beta_1$};
\node at (\xstart - 0.3, -\boxH/2 - 0.5) {$\beta_1$};

\pgfmathsetmacro{\xend}{3*\dx + \gap + 2*\dx}
\pgfmathsetmacro{\xprev}{3*\dx + \gap + 1*\dx}
\pgfmathsetmacro{\ctrlXright}{0.5*\xend + 0.5*\xprev}
\draw[line width=\lineW] 
  (\xend, \boxH/2) .. controls 
  (\xprev+1.1, \boxH/2 + \dy) .. 
  (\xprev, \boxH/2 + \dy);

\draw[line width=\lineW] 
  (\xend, -\boxH/2) .. controls 
  (\xprev+1.1, -\boxH/2 - \dy) .. 
  (\xprev, -\boxH/2 - \dy);

\node at (\xend + 0.4, \boxH/2 + 0.5) {$\beta_{n-2}$};
\node at (\xend + 0.4, -\boxH/2 - 0.5) {$\beta_{n-2}$};

\end{tikzpicture}
\end{split}
\end{align}
Whereas usual SU(2) spin-networks have spin labels at every line, the labels in the diagram \eqref{eq:su2_vac_npt} are combinations of conformal weights and stand each for a whole block-index $(j,m,a,b)$, which are contracted. The vertices are labelled by the Kronecker deltas in the definition of the $\mathcal{T}$-function. For additional details see appendix \ref{sec:weighted_networks}.

We can verify that the $n$-point block correctly reduces to an $(n-1)$-point block if one sets one of the external weights $\Delta_{i+1}$ to zero and adjusts the neighbouring projectors accordingly, i.\,e. sets $\Deltabb_{i-1} = \Deltabb_{i}$, which manifests as $\hat{\alpha}_i=\hat{\gamma}_i = 0$ and $\hat{\beta}_i = \Deltabb_i$ in terms of the wavefunction exponents. Since $\mathcal{N}^{j,m}_0 = \delta_{m,0}\delta_{j,0}$, the restriction of the scaling dimensions enforces
\begin{multline}
    \sum_{(i)} \mathcal{T}_{(i-1),(i)}\left(\begin{matrix} \hat{\beta}_{i-1},\hat{\gamma}_{i-1},0,\Deltabb_{i} \\ \Deltabb_i \end{matrix}\,\Bigg| X_{i}^{-1},X_{i+1}^T\right) 
    \mathcal{T}_{(i),(i+1)}\left(\begin{matrix} \Deltabb_i,0,\hat{\alpha}_{i+1},\hat{\beta}_{i+1} \\ \Deltabb_i \end{matrix}\,\Bigg| X_{i+1}^{-1},X_{i+2}^T\right) \\ 
    = \mathcal{T}_{(i-1),(i+1)}\left(\begin{matrix} \hat{\beta}_{i-1},\hat{\gamma}_{i-1},\hat{\alpha}_{i+1},\hat{\beta}_{i+1} \\ \Deltabb_i \end{matrix}\,\Bigg| X_{i}^{-1},X_{i+2}^T\right) \,.
\end{multline}
In this way, the $n$-point comb block clearly reduces to the $(n-1)$-point block 
\begin{equation}
    \eval{G_{\Deltabb_1,\dots,\Deltabb_{n-3}}^{\Delta_1,\dots,\Delta_n}}_{\Delta_{i+1} = 0,\  \Deltabb_{i-1} = \Deltabb_{i}} = G_{\Deltabb_1,\dots,\Deltabb_{i-2},\Deltabb_{i},\dots,\Deltabb_{n-3}}^{\Delta_1,\dots,\Delta_i,\Delta_{i+2},\dots,\Delta_n}\,,
\end{equation}
as it should.

\section{\texorpdfstring{Thermal $\bm{n}$-Point Blocks in the Projection Channel}{n-Point Thermal Blocks in Projection Channel}}

\label{sec:computation_npt_thermal}

We have already defined the thermal $n$-point projection channel block on $\mathbb{S}^1_\beta \times \mathbb{S}^3$ in section \ref{sec:def_thermal_blocks}. In this section, we give explicit expressions for said conformal blocks, using the $(2,2)$-integral computed in appendix \ref{sec_app:HigherIntegrals_Details}, and discuss the low-temperature limit that results in the comb channel blocks from section \ref{sec:comb_channel_blocks}.

\subsection{General Construction}

The $n$-point thermal block in the projection channel is defined as
\begin{equation}
\label{eq:def_npt_proj_block}
    G^{(n)}_{\text{Proj.}}(q,X_1,\dots,X_n) = \text{Tr}\left(\mathbb{P}_{\Deltabb_1} q^D \mathcal{O}_{\Delta_1}(X_1) \mathbb{P}_{\Deltabb_2}\mathcal{O}_{\Delta_2}(X_2) \cdots \mathbb{P}_{\Deltabb_n}\mathcal{O}_{\Delta_n}(X_n)\right)\,.
\end{equation}
Evaluating this object is easiest demonstrated in the one-point case 
\begin{align} 
\begin{split}
    \text{Tr}\left[\mathbb{P}_\Deltabb \mathcal{O}_\Delta(x) q^D\right] &= \sum_n \int_{\mathbb{D}_4}[\mathrm{d}U]_\Deltabb \braket*{ n}{ U^\dagger} \mel*{U}{\mathcal{O}_\Delta q^D}{n} \\
    &= \sum_n \int_{\mathbb{D}_4}[\mathrm{d}U]_\Deltabb  \mel*{U}{\mathcal{O}_\Delta q^D}{n}\,\braket*{n}{ U^\dagger} \\
    &= \int_{\mathbb{D}_4}[\mathrm{d}U]_\Deltabb  \mel*{U}{\mathcal{O}_\Delta q^D}{U^\dagger}\,.
\end{split}
\end{align}
The remaining matrix element can then be evaluated as 
\begin{align} \label{eq:draw_D_out}
\begin{split}
    \mel*{U}{\mathcal{O}_\Delta \operatorname{e}^{-\beta D}}{U^\dagger}
    &= \operatorname{e}^{-\beta \mathfrak{D}(\overline{w})} \Omega(x,U,U^\dagger)\\
    &= q^{ \Deltabb}\, \Omega(x,U,qU^\dagger)\,,
\end{split}
\end{align}
where we used in the last step that $\mathfrak{D}(\bar{w}') = \bar{w}_\mu \partial_{\bar{w}} + \Deltabb$ and that hence each coordinate $\bar{w}_\mu$ is scaled separately.\footnote{That one scales here only $U^\dagger$ and not $U$ can be also seen from consistency with the general $n$-point block case.} Thus, we arrive at the expression  
\begin{equation} \label{eq:def_1pt_block}
     G^{(1)} (q) = q^{ \Deltabb} \int_{\mathbb{D}_4}[\d U]_\Deltabb \Omega(x,U,qU^\dagger)
\end{equation}
for the one-point block.
Coming back to equation \eqref{eq:def_npt_proj_block}, the more general computation shows that the $n$-point projection channel block on $\mathbb{S}^1_\beta \times \mathbb{R}^3$ can be computed by 
\begin{equation} \label{def:npt_block_proj_channel_2}
    G^{(n)}_{\text{Proj.}} = q^\Deltabb \int_{\mathbb{D}_4}\left[\d U_1\right]  \cdots \int_{\mathbb{D}_4}\left[\d U_n\right] \Omega(X_1,U_1,U_2^\dagger)\Omega(X_2,U_2,U_3^\dagger) \cdots \Omega(X_n, U_n, qU_1^\dagger)\,.
\end{equation}
We will compute the block on $\mathbb{S}^1_\beta \times \mathbb{S}^3$ in $X_i$-coordinates and add the corresponding Weyl factors $r^{\Delta_i}$ in the end.

\subsection{One-Point Block}
\label{sec:1pt_thermal}

We have derived in the last section that the one-point block on $\mathbb{S}^1_\beta \times \mathbb{S}^3$ can be computed through equation \eqref{eq:def_1pt_block}. Note that the single projector insertion in the one-point correlator means we equate the oscillator variables $U_1 = U_2$ as well as the inner weights $\Deltabb_1=\Deltabb_2$ giving us the exponents $\tilde{\alpha} = \Delta/2$, $\tilde{\beta} = \Deltabb - \Delta/2$ and $\tilde{\gamma} = \Delta/2$ in the $\Omega$-wavefunction. The computation is then analogous to the vacuum case before, with the addition of using homogeneity of the base functions
\begin{align}
\begin{split}
       \phi^{j,m}_{q_a,q_b}(qX) &= \text{det}^m(qX) \mathcal{D}^j_{q_a,q_b}(qX) \\
       &= q^{2(j+m)} \phi^{j,m}_{q_a,q_b}(X)
\end{split}
\end{align}
to pull out $q$ from the basis. Expanding then every determinant in $\Omega$ from equation \eqref{eq:sol_Omega_2ndLvl_V1} gives us 
\begin{align}  
    G^{(1)} (q)
    &= q^{ \Deltabb} \text{det}^{-\Delta/2}(X_1)  \sum^{j_1,m_1}_{a_1,b_1} \sum^{j_2,m_2}_{a_2,b_2} \sum^{j_3,m_3}_{a_3,b_3} \left(\mathcal{N}^{j_1,m_1}_{\Delta/2} \mathcal{N}^{j_2,m_2}_{\Delta/2} \mathcal{N}^{j_3,m_3}_{\Deltabb - \Delta/2}\right)^2  \phi^{j_1,m_1}_{a_1,b_1}(X_1^T) \phi^{j_2,m_2}_{a_2,b_2}(X_1^{-1}) \notag\\
    & \quad\ q^{2(k_2 + k_3)} \int_{\mathbb{D}_4} [\mathrm{d}U]_\Deltabb \overline{\phi^{j_3,m_3}_{a_3,b_3}( U) \phi^{j_2,m_2}_{a_2,b_2}(U)} \phi^{j_1,m_1}_{a_1,b_1}(U) \phi^{j_3,m_3}_{a_3,b_3}(U)\,. \label{eq:1pt_block_thermal_2}
\end{align}
We evaluated the general $(2,2)$-integral in the appendix according to \eqref{eq:22_integral} and use that $\mathrm{det}^{-\Delta/2}(X_1) = r_1^{-\Delta}$. One gets in terms of the $\mathcal{T}$-function introduced in section \eqref{sec:adjusted_coeffs}
\begin{align} 
    G^{(1)} (q) = \frac{q^\Deltabb}{r_1^\Delta} \sum_{(3)}\mathcal{T}_{(3),(3)}\left(\begin{matrix}\Deltabb - \frac{\Delta}{2},\frac{\Delta}{2},\frac{\Delta}{2}, \Deltabb - \frac{\Delta}{2} \\ \Deltabb \end{matrix}\,\Bigg| qX^{-1}_1,X^T_1\right) q^{2(j_3+m_3)}\,,
\end{align}
for $(3) = (j_3,m_3,a_3,b_3)$. In this special case of the one-point block, we can simplify the $\mathcal{T}$-function drastically. Following equation \eqref{eq:22_integral_2}, we can write this particular $(2,2)$-integral as 
\begin{align} 
\begin{split}
    &\braket{\phi^{j_3,m_3}_{a_3,b_3} \phi^{j_2,m_2}_{a_2,b_2}}{\phi^{j_1,m_1}_{a_1,b_1} \phi^{j_3,m_3}_{a_3,b_3}}\\
    &= \sum_l \sum_{p_a,p_b} \left(\mathcal{N}^{l,k_2 + k_3 - l}_{\Deltabb}\right)^{-2} C^{j_3j_1l}_{a_3,a_1,p_a} C^{j_3j_2l}_{a_3,a_2,p_a} C^{j_3j_1l}_{b_3,b_1,p_b} C^{j_3j_2l}_{b_3,b_2,p_b} \delta_{k_1,k_2}\,. 
\end{split}
\end{align}
Note that this includes the implicit Kronecker-deltas $\delta_{a_1,a_2}$ and $\delta_{b_1,b_2}$. We can use orthogonality \eqref{eq:cgc_orthog_2} to collapse the Clebsch-Gordan coefficients
\begin{align}
    \sum_{a_3,p_a} C^{j_3j_1l}_{a_3,a_1,p_a} C^{j_3j_2l}_{a_3,a_2,p_a} = d_l d_{j_2}^{-1} \delta_{j_1,j_2} \delta_{a_1,a_2}\,, \qquad \sum_{b_3,p_b} C^{j_3j_1l}_{b_3,b_1,p_b} C^{j_3j_2l}_{b_3,b_2,p_b} = d_l d_{j_2}^{-1} \delta_{j_1,j_2} \delta_{b_1,b_2}\,,
\end{align}
which allows us to to split $\delta_{k_1,k_2}$ into $\delta_{j_1,j_2}$ and $\delta_{m_1,m_2}$. We can subsequently evaluate the coordinate dependent part
\begin{equation} \label{eq:coord_indep}
    \sum_{a_2,b_2}\phi^{j_2, m_2}_{a_2,b_2}(X_1^T) \phi^{j_2, m_2}_{a_2,b_2}(X_1^{-1}) = d_{j_2}\,.
\end{equation}
Finally, we have to add a factor of $r_1^\Delta$ to obtain the result on $\mathbb{S}^1_\beta \times \mathbb{S}^3$. Altogether, the result is
\begin{align}
    \label{eq:1pt_block_result} G^{(1)} (q) = q^\Deltabb \sum_{m_2,m_3}^{j_2,j_3} a^{j_2,j_3}_{m_2,m_3} q^{2(j_2+j_3+m_2+m_3)} 
\end{align}
with 
\begin{equation}
\label{eq:1pt_block_result_coeffs}
    a^{j_2,j_3}_{m_2,m_3} = d_{j_2}^{-1} \left(\mathcal{N}^{j_2,m_2}_{\Delta/2}\right)^4 \left(\mathcal{N}^{j_3,m_3}_{\Deltabb - \Delta/2}\right)^2  \sum_l (d_l)^2\left(\mathcal{N}^{l,j_2+m_2+j_3+m_3-l}_{\Deltabb}\right)^{-2}\,,
\end{equation}
where $l$ runs from $|j_2 - j_3|$ to $j_2 + j_3$. We have checked up to $\mathcal{O}(q^{25})$ that this matches the result in \cite{Gobeil:2018fzy,Buric:2024kxo}. The calculation above, which shows that the one-point block is coordinate independent, can be conveniently expressed through diagrammatic rules close\footnote{One uses here that cutting a line in a SU(2) diagram that separates it into two disconnected parts inserts $\sim \delta_{j_2,j_3}$.} to the known rules in the SU(2) case: 
\begin{align} \label{eq:1pt_block_network}
\begin{split} G^{(1)} (q) \quad  = \quad \begin{tikzpicture}[scale=1.5,baseline={([yshift=-.5ex]current bounding box.center)}]

\def\extra{0.3}
\def\boxWidth{0.6}
\def\boxTopY{2.4}
\def\boxBottomY{1.8}
\def\topEllipseY{2.9 + \extra} 
\def\bottomEllipseY{1.3 - \extra} 
\def\ellipseXRadius{1.5}
\def\ellipseYRadius{0.15}
\def\touchA{0.08}
\def\touchB{0.03}

\pgfmathsetmacro{\xA}{-0.5 * (\boxWidth + 0.2)} 
\pgfmathsetmacro{\xB}{0.5 * (\boxWidth + 0.2)} 
\pgfmathsetmacro{\xC}{0.5 * (\boxWidth + 0.2)} 
\pgfmathsetmacro{\xD}{1.5 * (\boxWidth + 0.2)} 
\pgfmathsetmacro{\xE}{1.5 * (\boxWidth + 0.2)} 
\pgfmathsetmacro{\xF}{2.5 * (\boxWidth + 0.2)} 

\draw[line width=1pt]
  (-\ellipseXRadius+1.5, \topEllipseY - \ellipseYRadius)
  arc[start angle=270, end angle=90, radius=\ellipseYRadius]
  -- (\ellipseXRadius, \topEllipseY + \ellipseYRadius)
  arc[start angle=90, end angle=-90, radius=\ellipseYRadius]
  -- cycle;

\draw[line width=1pt]
  (-\ellipseXRadius+1.5, \bottomEllipseY - \ellipseYRadius)
  arc[start angle=270, end angle=90, radius=\ellipseYRadius]
  -- (\ellipseXRadius, \bottomEllipseY + \ellipseYRadius)
  arc[start angle=90, end angle=-90, radius=\ellipseYRadius]
  -- cycle;

\draw[line width=1pt] (\xC - \boxWidth/2, \boxBottomY) rectangle (\xC + \boxWidth/2, \boxTopY);
\node at (\xC, 2.1) { $X^{-1}$};
\draw[line width=1pt] (\xC, \boxTopY) -- (\xC, \topEllipseY - \ellipseYRadius);
\draw[line width=1pt] (\xC, \boxBottomY) -- (\xC, \bottomEllipseY + \ellipseYRadius);

\draw[line width=1pt] (\xD - \boxWidth/2, \boxBottomY) rectangle (\xD + \boxWidth/2, \boxTopY);
\node at (\xD, 2.1) { $X^{\dagger}$};
\draw[line width=1pt] (\xD, \boxTopY) -- (\xD, \topEllipseY - \ellipseYRadius);
\draw[line width=1pt] (\xD, \boxBottomY) -- (\xD, \bottomEllipseY + \ellipseYRadius);

\node at (\xC - 0.15, 2.6 + 0.5*\extra) { \small $\frac{\Delta}{2}$}; 
\node at (\xC - 0.15, 1.6 - 0.5*\extra) {\small $\frac{\Delta}{2}$}; 
\node at (\xD + 0.15, 2.6 + 0.5*\extra) { \small $\frac{\Delta}{2}$}; 
\node at (\xD + 0.15, 1.6 - 0.5*\extra) { \small $\frac{\Delta}{2}$}; 

\node at (0.75, \topEllipseY - 0.3) {\small $\Deltabb$};  

\node at (0.75, \topEllipseY + 0.3+0.5*\extra) {\small $\Deltabb - \frac{\Delta}{2}$};  

\node at (0.75, \bottomEllipseY - 0.3-0.5*\extra) {\small $\Deltabb-\frac{\Delta}{2}$};  

\node at (0.75, \bottomEllipseY + 0.3) {\small $\Deltabb$};

\end{tikzpicture} \qquad = \qquad \begin{tikzpicture}[scale=1.5,baseline={([yshift=-.5ex]current bounding box.center)},
    vertex/.style={anchor=base,circle,fill=black!25,minimum size=18pt,inner sep=2pt}]

\def\postop{3.1}
\def\posbot{0.5}

\draw[line width=1pt] (0,\postop) circle (0.5);
\draw[line width=1pt] (-0.5,\postop) -- (0.5,\postop);

\draw[line width=1pt] (0,\posbot) circle (0.5);
\draw[line width=1pt] (-0.5,\posbot) -- (0.5,\posbot);

\def\xshift{0.5}         
\def\boxWidth{0.6}
\def\boxTop{2.1}
\def\boxBottom{1.5}
\def\arcHeight{0.35}      

\draw[line width=1pt] (-\xshift - 0.3, \boxBottom) rectangle (-\xshift + 0.3, \boxTop);
\draw[line width=1pt] (\xshift - 0.3, \boxBottom) rectangle (\xshift + 0.3, \boxTop);

\node at (-\xshift, 1.8) { $X^{-1}$};
\node at (\xshift, 1.8) { $X^{\dagger}$};

\draw[line width=1pt]
  (-\xshift, \boxTop)
  .. controls (-\xshift, \boxTop + \arcHeight) and (\xshift, \boxTop + \arcHeight)
  .. (\xshift, \boxTop);

\draw[line width=1pt]
  (-\xshift, \boxBottom)
  .. controls (-\xshift, \boxBottom - \arcHeight) and (\xshift, \boxBottom - \arcHeight)
  .. (\xshift, \boxBottom);

\node at (0, \postop+0.75) {\small $\Deltabb - \frac{\Delta}{2}$};
\node at (0, \posbot-0.75) {\small $\Deltabb - \frac{\Delta}{2}$};

\node at (0, \postop-0.2) {\small $\Deltabb$};
\node at (0, \posbot + 0.2) {\small $\Deltabb$};

\node at (0+0.65, \postop-0.2) {\small $\frac{\Delta}{2}$};
\node at (0+0.65, \posbot+0.2) {\small $\frac{\Delta}{2}$};
\node at (0-0.5, \postop-0.65) {\small $\frac{\Delta}{2}$};
\node at (0-0.5, \posbot+0.65) {\small $\frac{\Delta}{2}$};

\end{tikzpicture} \qquad = \qquad 
\begin{tikzpicture}[scale=1.5,baseline={([yshift=-.5ex]current bounding box.center)},vertex/.style={anchor=base,
    circle,fill=black!25,minimum size=18pt,inner sep=2pt}]

\draw[line width=1pt] (0,2.8) circle (0.5);
\draw[line width=1pt] (-0.5,2.8) -- (0.5,2.8); 

\draw[line width=1pt] (0,1.5) circle (0.5);

\draw[line width=1pt] (0,0.2) circle (0.5);
\draw[line width=1pt] (-0.5,0.2) -- (0.5,0.2); 

\def\postop{3}
\def\posbot{0.7}

\node at (0, \postop+0.5) {\small $\Deltabb - \frac{\Delta}{2}$};
\node at (0, \posbot-1.2) {\small $\Deltabb - \frac{\Delta}{2}$};

\node at (0, \postop-0.4) {\small $\Deltabb$};
\node at (0, \posbot - 0.3) {\small $\Deltabb$};

\node at (0+0.65, \postop-0.4) {\small $\frac{\Delta}{2}$};
\node at (0+0.65, \posbot-0.3) {\small $\frac{\Delta}{2}$};
\node at (0.65, 1.5) {\small $\frac{\Delta}{2}$};

\end{tikzpicture}
\end{split}
\end{align}
The assignment of weights works as in the vacuum case, with the additional information that the thermal loops and the vertical lines of $X^{-1}$ get an extra factor of $q^{k_i}$, respectively. Note that the assignment rules (see appendix \ref{sec:weighted_networks}) are valid also for the other diagrams in equation \eqref{eq:1pt_block_network} if we assign the bubble diagram a weight of $d_j \left(\mathcal{N}^{j,m}_{\Delta/2}\right)^2$. 

The limit $\Delta = 0$ restricts the one-point block to the ``zero-point block'' $\text{Tr}(\mathbb{P}_\Deltabb q^D)$. It is supposed to compute the same trace as the one in the character formula of the four-dimensional conformal algebra (see e.\,g. \cite{Dolan:2005wy})
\begin{equation}
    G^{\Delta = 0}_\Deltabb(q) = \frac{q^\Deltabb}{(1-q)^4}
\end{equation}
and we can check order-by-order that \eqref{eq:1pt_block_result} satisfies this. That it also has the right low-temperature behaviour for $q \to 0$ can be easily seen by setting $k_2 = k_3 = 0$ in equations \eqref{eq:1pt_block_result} and \eqref{eq:1pt_block_result_coeffs}.

\subsection{Computing the Projection Channel \texorpdfstring{$\bm{n}$}{n}-Point Block} 
\label{sec:npt_thermal}

The $n$-point block is then a straight-forward generalisation of the one-point calculation. Using the definition \eqref{def:npt_block_proj_channel_2} (or equivalently the diagram \eqref{dia:npt-block-glueing}), we expand wavefunctions $\Omega$ of the form 
\begin{equation}
    \Omega(X_i;U_i,U_{i+1}^{\dagger}) =  \operatorname{det}^{-\Tilde{\alpha}_i}(\mathbb{1} -U_i X_i)\, \text{det}^{-\Tilde{\beta}_i}(\mathbb{1} -U_i {U_{i+1}^{\dagger}})\, \text{det}^{-\Tilde{\gamma}_i}(X_i-U_{i+1}^{\dagger})
\end{equation}
in the basis. The weights $\alpha_i$, $\beta_i$ and $\gamma_i$ are given as
\begin{equation} \label{eq:weights_thermal}
    \tilde{\alpha}_i = \frac{1}{2}(\Delta_i + \Deltabb_i - \Deltabb_{i+1})\,, \quad \tilde{\beta}_i = \frac{1}{2}(-\Delta_i + \Deltabb_i + \Deltabb_{i+1})\,, \quad \tilde{\gamma}_i = \frac{1}{2}(\Delta_i - \Deltabb_i + \Deltabb_{i+1})\,.
\end{equation}
After integrating out the $U$-variables using the inner product \eqref{eq:4d_orthonormality_property}, adding the Weyl factors $r_i^{\Delta_i}$ and using $\operatorname{det}^{-\tilde{\gamma}_i}(X_i) = r_i^{-2\tilde{\gamma}_i}$, the result is given by 
\begin{align} \label{eq:result_thermal_npt}
\begin{split}
    &G^{(n)}_{\text{Proj.}}(q,X_1,\dots,X_n) \\&= q^{\Deltabb_1} \left(\prod_{i=1}^n r_i^{\Delta_i-2\tilde{\gamma}_i}\right) \sum_{(1),\dots,(n)}q^{2(j_n + m_n)} \prod_{i=1}^n \mathcal{T}_{(i-1),(i)}\left(\begin{matrix}\tilde{\beta}_{i-1},\tilde{\gamma}_{i-1},\tilde{\alpha}_{i},\tilde{\beta}_{i} \\ \Deltabb_{i} \end{matrix}\,\Bigg| X^{-1}_{i-1},X_{i}^T\right)
\end{split}
\end{align}
which in the case of $i=1$, $i-1$ evaluates as $n$ and $X^{-1}_{i-1}$ as $q X^{-1}_n$. If we again consider the block as a series of weighted SU(2) spin network diagrams, one gets
\begin{align} \label{eq:su2_therm_npt}
\begin{split} G^{(n)}_{\text{Proj.}}(q,X_1,\dots,X_n) = \begin{tikzpicture}[scale=1.8,baseline={([yshift=-.5ex]current bounding box.center)}]

\def\extra{0.1}
\def\boxWidth{0.6}
\def\boxTopY{2.4}
\def\extend{\boxWidth+0.1}
\def\boxBottomY{1.8}
\def\topEllipseY{2.9+\extra}
\def\bottomEllipseY{1.3-\extra}
\def\ellipseXRadius{2.2 + \extend}
\def\ellipseYRadius{0.175} 

\draw[line width=1pt]
  (-\ellipseXRadius, \topEllipseY - \ellipseYRadius)
  arc[start angle=270, end angle=90, radius=\ellipseYRadius]
  -- (\ellipseXRadius, \topEllipseY + \ellipseYRadius )
  arc[start angle=90, end angle=-90, radius=\ellipseYRadius]
  -- cycle;

\draw[line width=1pt]
  (-\ellipseXRadius, \bottomEllipseY - \ellipseYRadius)
  arc[start angle=270, end angle=90, radius=\ellipseYRadius]
  -- (\ellipseXRadius, \bottomEllipseY + \ellipseYRadius)
  arc[start angle=90, end angle=-90, radius=\ellipseYRadius]
  -- cycle;

\pgfmathsetmacro{\xA}{(-2.5) * (\boxWidth + 0.1)+\extend}
\draw[line width=1pt] (\xA - \boxWidth/2, \boxBottomY) rectangle (\xA + \boxWidth/2, \boxTopY);
\node at (\xA, 2.1) { $X_n^{-1}$};
\draw[line width=1pt] (\xA, \boxTopY) -- (\xA, \topEllipseY - \ellipseYRadius);
\draw[line width=1pt] (\xA, \boxBottomY) -- (\xA, \bottomEllipseY + \ellipseYRadius);
\node at (\xA - 0.15, 2.6) { \small $\Tilde{\gamma}_n$};
\node at (\xA - 0.15, 1.6) { \small $\Tilde{\gamma}_n$};

\pgfmathsetmacro{\xB}{(-1.5) * (\boxWidth + 0.1)+\extend}
\draw[line width=1pt] (\xB - \boxWidth/2, \boxBottomY) rectangle (\xB + \boxWidth/2, \boxTopY);
\node at (\xB, 2.1) { $X_1^{\dagger}$};
\draw[line width=1pt] (\xB, \boxTopY) -- (\xB, \topEllipseY - \ellipseYRadius);
\draw[line width=1pt] (\xB, \boxBottomY) -- (\xB, \bottomEllipseY + \ellipseYRadius);
\node at (\xB + 0.15, 2.6) { \small $\Tilde{\alpha}_1$};
\node at (\xB + 0.15, 1.6) { \small $\Tilde{\alpha}_1$};

\pgfmathsetmacro{\xE}{(-0.5) * (\boxWidth + 0.1)+\extend}
\draw[line width=1pt] (\xE - \boxWidth/2, \boxBottomY) rectangle (\xE + \boxWidth/2, \boxTopY);
\node at (\xE, 2.1) { $X_1^{-1}$};
\draw[line width=1pt] (\xE, \boxTopY) -- (\xE, \topEllipseY - \ellipseYRadius);
\draw[line width=1pt] (\xE, \boxBottomY) -- (\xE, \bottomEllipseY + \ellipseYRadius);
\node at (\xE - 0.15, 2.6) { \small $\Tilde{\gamma}_1$};
\node at (\xE - 0.15, 1.6) { \small $\Tilde{\gamma}_1$};

\node at (\xB - 0.3, 2.6 + 0.5*\extra) { \small $\Deltabb_1$};
\node at (\xB - 0.3, 1.6-0.5*\extra) { \small $\Deltabb_1$};

\node at (\xE+0.725, 2.1) {\large$\cdots$};

\pgfmathsetmacro{\xbeta}{(\xB + 0.15) + 0.5*((\xE - 0.15)-(\xB + 0.15))}
\node at (\extend, 3.2+0.025+\extra) {\small $\Tilde{\beta}_n$};
\node at (\extend, 1-0.025-\extra) {\small $\Tilde{\beta}_n$};
\node at (\xbeta, 2.9+0.75*\extra) {\small $\Tilde{\beta}_1$};
\node at (\xbeta, 1.3-0.75*\extra) {\small $\Tilde{\beta}_1$};

\pgfmathsetmacro{\xC}{(1.5) * (\boxWidth + 0.1)+\extend}
\draw[line width=1pt] (\xC - \boxWidth/2, \boxBottomY) rectangle (\xC + \boxWidth/2, \boxTopY);
\node at (\xC, 2.1) { $X_{n-1}^{-1}$};
\draw[line width=1pt] (\xC, \boxTopY) -- (\xC, \topEllipseY - \ellipseYRadius);
\draw[line width=1pt] (\xC, \boxBottomY) -- (\xC, \bottomEllipseY + \ellipseYRadius);
\node at (\xC - 0.25, 2.6) { \small $\tilde{\gamma}_{n-1}$};
\node at (\xC - 0.25, 1.6) { \small $\tilde{\gamma}_{n-1}$};

\pgfmathsetmacro{\xD}{(2.5) * (\boxWidth + 0.1)+\extend}
\draw[line width=1pt] (\xD - \boxWidth/2, \boxBottomY) rectangle (\xD + \boxWidth/2, \boxTopY);
\node at (\xD, 2.1) { $X_n^{\dagger}$};
\draw[line width=1pt] (\xD, \boxTopY) -- (\xD, \topEllipseY - \ellipseYRadius);
\draw[line width=1pt] (\xD, \boxBottomY) -- (\xD, \bottomEllipseY + \ellipseYRadius);
\node at (\xD + 0.15, 2.6) { \small $\tilde{\alpha}_n$};
\node at (\xD + 0.15, 1.6) { \small $\tilde{\alpha}_n$};

\node at (\xD - 0.3, 2.6+0.5*\extra) { \small $\Deltabb_n$}; 
\node at (\xD - 0.3, 1.6-0.5*\extra) { \small $\Deltabb_n$};

\end{tikzpicture}
\end{split}
\end{align}
where again summation is implicit. 

The low-temperature limit $q \to 0$ is supposed to give the vacuum blocks we have discussed in section \ref{sec:computation_npt_comb}. We see from equation \eqref{eq:result_thermal_npt} that this means setting the corresponding $j_i$ and $m_i$ in $\mathcal{T}_{(n),(1)}$ to zero. Using equation \eqref{eq:special_cases_su22_intertwiner}, this sets the $\mathcal{T}_{(n),(1)}$-function to $1$, while also producing a zero in the adjacent $\mathcal{T}$-functions $\mathcal{T}_{(0),(2)}$ and $\mathcal{T}_{(n-1),(0)}$. This is easiest understood with an example. The two-point block is according to \eqref{eq:result_thermal_npt} and the definition of $\mathcal{T}$-functions in equation \eqref{eq:def_tau_fcts}, disregarding $q^\Deltabb$ and Weyl factors $r_1^{\Delta_1}r_2^{\Delta_2}$, given by
\begin{align} \label{eq:2pt_block_intertwiner}
    &G^{(2)}_{\text{Proj.}}(q,X_1,X_2) \\ &\sim r_1^{-2\tilde{\gamma}_1} r_2^{-2\tilde{\gamma}_2} \hspace{-.1cm} \sum_{(1),\dots,(6)} q^{2(k_2 + k_3)} \hspace{-.1cm} \widetilde{\mathcal{I}}_{(2),(3),(4),(1)}^{\tilde{\beta_2}\tilde{\gamma}_2\tilde{\alpha}_1\tilde{\beta}_1} \varphi^{(3)}_{\tilde{\gamma}_2}(X_2^{-1})\varphi^{(4)}_{\tilde{\alpha}_1}(X_1^T) \widetilde{\mathcal{I}}_{(1),(5),(6),(2)}^{\tilde{\beta_1}\tilde{\gamma}_1\tilde{\alpha}_2\tilde{\beta}_2} \varphi^{(5)}_{\tilde{\gamma}_1}(X_1^{-1})\varphi^{(6)}_{\tilde{\alpha}_2}(X_2^T)\,.\notag
\end{align}
Using equation \eqref{eq:special_cases_su22_intertwiner}, the $q \to 0$ limit cuts the corresponding diagram \eqref{dia:npt-block-glueing} at the $\Deltabb_1$-line and reduces it to the corresponding four-point diagram in \eqref{dia:comb_npt}. Equation \eqref{eq:2pt_block_intertwiner} reduces to 
\begin{align} \label{eq:limit_2pt_to_4pt}
\begin{split}
    \eval{G^{(2)}_{\text{Proj.}}(q,X_1,X_2)}_{q \to 0} &\sim r_1^{-2\tilde{\gamma}_1} r_2^{-2\tilde{\gamma}_2}\sum_{(5),(6)} \widetilde{\mathcal{I}}_{0,(5),(6),0}^{\,-\,\tilde{\gamma}_1\tilde{\alpha}_2\,-} \varphi^{(5)}_{\tilde{\gamma}_1}(X_1^{-1})\varphi^{(6)}_{\tilde{\alpha}_2}(X_2^T) 
    \\&= r_1^{-2\tilde{\gamma}_1} r_2^{-2\tilde{\gamma}_2}\sum_{(5)} \frac{\mathcal{N}^{(5)}_{\tilde{\gamma}_1} \mathcal{N}^{(5)}_{\tilde{\alpha}_2}}{\left(\mathcal{N}^{(5)}_\Deltabb\right)^2} \varphi^{(5)}_{\tilde{\gamma}_1}(X_1^{-1})\varphi^{(5)}_{\tilde{\alpha}_2}(X_2^T)\,. 
\end{split}
\end{align}
With weights defined in \eqref{eq:weights_thermal} and after the appropriate renaming of points and weights, equation \eqref{eq:limit_2pt_to_4pt} is exactly the vacuum four-point block with $X_1 \to \infty$ and $X_4 \to 0$, see \cite{Ammon:2024axd}.  In terms of tensor contractions, the $q\to 0$ limit is easily visualised. The reduction from the two-point block in the projection channel to the vacuum four-point block from above is given by 
\begin{align}
\begin{tikzpicture}[scale=1.5,baseline={([yshift=-.5ex]current bounding box.center)},
    vertex/.style={anchor=base,circle,fill=black!25,minimum size=18pt,inner sep=2pt}]

\def\boxWidth{0.6}
\def\boxTopY{2.4}
\def\boxBottomY{1.8}
\def\topEllipseY{2.9}
\def\bottomEllipseY{1.3}
\def\ellipseXRadius{2.6}
\def\ellipseYRadius{0.15}
\def\touchA{0.08}
\def\touchB{0.03}

\pgfmathsetmacro{\xA}{-0.5 * (\boxWidth + 0.2)} 
\pgfmathsetmacro{\xB}{0.5 * (\boxWidth + 0.2)} 
\pgfmathsetmacro{\xC}{1.5 * (\boxWidth + 0.2)} 
\pgfmathsetmacro{\xD}{2.5 * (\boxWidth + 0.2} 
\pgfmathsetmacro{\xE}{ 1.5 * (\boxWidth + 0.2)} 
\pgfmathsetmacro{\xF}{ 2.5 * (\boxWidth + 0.2)}

\draw[line width=1pt, dashed]
  (-\ellipseXRadius+1.5, \topEllipseY - \ellipseYRadius)
  arc[start angle=270, end angle=90, radius=\ellipseYRadius]
  -- (\ellipseXRadius, \topEllipseY + \ellipseYRadius)
  arc[start angle=90, end angle=-90, radius=\ellipseYRadius]
  -- cycle;

\draw[line width=1pt, dashed]
  (-\ellipseXRadius+1.5, \bottomEllipseY - \ellipseYRadius)
  arc[start angle=270, end angle=90, radius=\ellipseYRadius]
  -- (\ellipseXRadius, \bottomEllipseY + \ellipseYRadius)
  arc[start angle=90, end angle=-90, radius=\ellipseYRadius]
  -- cycle;


\draw[line width=1pt, dashed] (\xA - \boxWidth/2, \boxBottomY) rectangle (\xA + \boxWidth/2, \boxTopY);
\node at (\xA, 2.1) {$X_2^{-1}$};
\draw[line width=1pt, dashed] (\xA, \boxTopY) -- (\xA, \topEllipseY - \ellipseYRadius);
\draw[line width=1pt, dashed] (\xA, \boxBottomY) -- (\xA, \bottomEllipseY + \ellipseYRadius);

\draw[line width=1pt, dashed] (\xB - \boxWidth/2, \boxBottomY) rectangle (\xB + \boxWidth/2, \boxTopY);
\node at (\xB, 2.1) { $X_1^{\dagger}$};
\draw[line width=1pt, dashed] (\xB, \boxTopY) -- (\xB, \topEllipseY - \ellipseYRadius);
\draw[line width=1pt, dashed] (\xB, \boxBottomY) -- (\xB, \bottomEllipseY + \ellipseYRadius);

\draw[line width=1pt] (\xC - \boxWidth/2, \boxBottomY) rectangle (\xC + \boxWidth/2, \boxTopY);
\node at (\xC, 2.1) {$X_1^{-1}$};
\draw[line width=1pt] (\xC, \boxTopY) -- (\xC, \topEllipseY - \ellipseYRadius);
\draw[line width=1pt] (\xC, \boxBottomY) -- (\xC, \bottomEllipseY + \ellipseYRadius);

\draw[line width=1pt] (\xD - \boxWidth/2, \boxBottomY) rectangle (\xD + \boxWidth/2, \boxTopY);
\node at (\xD, 2.1) { $X_2^{\dagger}$};
\draw[line width=1pt] (\xD, \boxTopY) -- (\xD, \topEllipseY - \ellipseYRadius);
\draw[line width=1pt] (\xD, \boxBottomY) -- (\xD, \bottomEllipseY + \ellipseYRadius);



\draw[line width=1pt] (\xC-0.0125, \topEllipseY - \ellipseYRadius) -- (\xF+0.0125, \topEllipseY - \ellipseYRadius);
 \draw[line width=1pt] (\xC-0.0125, \bottomEllipseY +\ellipseYRadius) -- (\xF+0.0125, \bottomEllipseY + \ellipseYRadius);

\end{tikzpicture} \quad \xlongrightarrow{\quad q \to 0 \quad } \quad  \begin{tikzpicture}[scale=1.5,baseline={([yshift=-.5ex]current bounding box.center)}]

\def\boxW{0.6}
\def\boxH{0.6}
\def\dx{1.5}       
\def\arcHeight{0.5} 
\def\lineW{1pt}

\def\xA{0}
\def\xB{\dx}

\draw[line width=\lineW] (\xA - \boxW/2, -\boxH/2) rectangle (\xA + \boxW/2, \boxH/2);
\node at (\xA, 0) {$X_1^{-1}$};

\draw[line width=\lineW] (\xB - \boxW/2, -\boxH/2) rectangle (\xB + \boxW/2, \boxH/2);
\node at (\xB, 0) {$X_2^{\dagger}$};

\draw[line width=\lineW]
  (\xA, \boxH/2) .. controls (\xA, \boxH/2 + \arcHeight) and (\xB, \boxH/2 + \arcHeight) .. (\xB, \boxH/2);

\draw[line width=\lineW]
  (\xA, -\boxH/2) .. controls (\xA, -\boxH/2 - \arcHeight) and (\xB, -\boxH/2 - \arcHeight) .. (\xB, -\boxH/2);

\end{tikzpicture}
\end{align}
where we dashed the lines that are deleted by the low-temperature limit and suppressed the labels. The corners of the sub-diagram on the left-hand side are naturally smoothed out because of the associated Kronecker deltas we discussed in section \ref{sec:weighted_networks}. The general $n$-point block works analogously, the left-most boxes are deleted because $q \to 0$ removes (i.\,a.) $\mathcal{T}_{(n),(1)}$ in equation \eqref{eq:result_thermal_npt}.

\section{Discussion and Outlook}
\label{sec:discussion}

We defined and computed the thermal $n$-point conformal blocks in the projection channel in four Euclidean dimensions. To our knowledge, this is a new result for $n \geq 2$. We checked that their low-temperature limit is given by the corresponding vacuum blocks. This involved the computation of general integrals of the basis functions $\phi^{j,m}_{a,b}$ of $\mathcal{H}L^2(\mathbb{D}_4)$ in appendix \ref{sec_app:HigherIntegrals_Details}. The generalisation to non-vanishing angular potentials is presented in appendix \ref{sec_app:angular_potentials}.
    
Admittedly, our results for both vacuum and thermal $n$-point conformal blocks in \eqref{eq:npt_comb_tau} and \eqref{eq:result_thermal_npt} appear intricate, but they have the advantage that their basic constituent is the SU(2) Clebsch-Gordan coefficient $C^{j_1 j_2 j}_{a_1,a_2,c}$. There is a vast literature discussing the properties of these coefficients; we provided a small subset in appendix \ref{sec_app:properties_cgcs} and emphasised throughout the text how the appearance of said coefficients made some calculations almost trivial. For example, the computation of the thermal one-point block on $\mathbb{S}^1_\beta \times \mathbb{S}^3$ in section \ref{sec:1pt_thermal} boils down to column switches \eqref{eq:switching} and the orthogonality relation \eqref{eq:cgc_orthog_2}; the low-temperature limit discussed in section \ref{sec:npt_thermal} is just evaluating trivial cases of $C^{j_1 j_2 j}_{a_1,a_2,c}$ according to equation \eqref{eq:zero_cgc_delta}. 

We have presented the vacuum and thermal $n$-point conformal blocks as weighted series of SU(2) spin network contractions. The SU(2) Clebsch-Gordan coefficients that define this representation can be expressed in terms of terminating ${}_3F_2$-hypergeometric functions, i.\,e. as finite sums (see equation \eqref{eq:cgc_as_3F2}). This structure suggests that established numerical methods for spin network contractions may be adaptable to our setting. In particular, recent advances in the numerical evaluation of contractions between four-valent SU(2) intertwiners and Wigner-$\mathcal{D}$ functions with SU(2) and $\text{SL}(2,\mathbb{C})$ (originally developed in the spin foam literature) could provide efficient tools for evaluating expressions of the type found in this work \cite{Gozzini:2021kbt, Dona:2018nev}. These implementations, written in \texttt{C} with more recent interfaces in \texttt{Julia}, leverage high-performance computing techniques; see \cite{Dona:2022dxs} for a pedagogical introduction. Although these packages are tailored to spin-foam amplitudes, it may be possible to adapt them to the contractions involving Wigner-$\mathcal{D}$ matrices of a general $\text{GL}(2,\mathbb{C})$ argument as defined in equation \eqref{eq:def_wignerD}.  More generally, \texttt{Julia} packages such as \texttt{TensorOperations.jl} \cite{TensorOperationsjl} offer powerful frameworks for contracting general tensor networks. Using such tools, one could define the Wigner-$\mathcal{D}$ matrices from equation \eqref{eq:def_wignerD} and evaluate the finite amplitudes, i.\,e. the contraction over magnetic indices, in equations \eqref{eq:npt_comb_tau} and \eqref{eq:result_thermal_npt}. The full conformal blocks then follow as weighted sums over these amplitudes. From this perspective, the vacuum and thermal conformal blocks constructed here can be viewed as generalisations of spin foams.

From the aforementioned SU(2) perspective, the conformal blocks can be expanded in an infinite sum of spin-network contractions that themselves are finite sums. Alternatively, one can seek functions that are intrinsically associated with the full SU(2,2) symmetry. We define the $\mathcal{T}$-functions in equation \eqref{eq:def_tau_fcts}, which are by definition infinite sums\footnote{The coefficients $\mathcal{C}_{(3),(2),(1)}$ in equation \eqref{eq:coeff_321_weighted} are finite because the SU(2) CGC are, according to equation \eqref{eq:cgc_as_3F2}. Every further block index contraction in the definition of $\widetilde{\mathcal{I}}$ and $\mathcal{T}$ then adds two infinite and two finite sums.} but which give us the compact expression \eqref{eq:result_thermal_npt} that resembles the two dimensional result. There, the $n$-point projection channel block can be written as
    \begin{align} \label{eq:npt_block_2d_2}
    \mathcal{G}^{(n)}_{\text{Proj.}}(q,w_1,\dots,w_n) = q^{\hbb_1 - \frac{c}{24}} \sum_{s_1,\dots s_n=0}^\infty \prod_{i=1}^n x_{i,i+1}^{-\hbb_{i+1}}  x_{i,i+1}^{-s_{i+1}} \frac{\tau_{s_{i},s_{i+1}}(\hbb_i,h_i,\hbb_{i+1})}{s_{i}!(2\hbb_i)_{s_{i}}} q^{s_1}\,, 
\end{align}
as we showed in appendix \ref{sec_app:2d_blocks} and the resulting comb channel block in the $ q \to 0$ limit has the same form as in four dimensions as well. Because one can directly relate the defining coefficients $\tau_{m,n}$ to the SU(1,1) Clebsch-Gordan coefficients and the $\Omega$-wavefunction to its generating function, we expect that the same holds for $\mathcal{T}_{(M),(N)}$ and the four-dimensional $\Omega$-wavefunction in equation \eqref{eq:sol_Omega_2ndLvl_V1} with respect to the four-dimensional conformal group SU(2,2).

In \cite{Alkalaev:2020yvq}, the computation of toroidal geodesic Witten diagrams in thermal $\text{AdS}_3$ yielded expressions for torus conformal blocks on the boundary, expressed through Clebsch-Gordan coefficients for finite dimensional SU(1,1) representations. It would be interesting to first generalise this to unitary representations of SU(1,1) and second, to do the same for $\text{AdS}_5$, maybe similarly to the $d$-dimensional computation of \cite{Gobeil:2018fzy} for the one-point block. 

We have not yet discussed the high-temperature limit of the $n$-point block, i.\,e. its reduction to $\mathbb{S}^1 \times \mathbb{R}^3$. In two dimensions, the rich theory of the ${}_2F_1$-hypergeometric function enables one to take the $q \to 1$ limit for the one-point block. However, already the higher-point blocks in the projection channel are only known in the form of equation \eqref{eq:npt_block_2d_2} for which this is not possible anymore. In four dimensions, we are in a similar situation in the case of the one-point block, where the closed form in terms of a generalised ${}_3F_2$-hypergeometric function allows to take the $q \to 1$ limit \cite{Gobeil:2018fzy,Buric:2024kxo} (similarly in \cite{Alkalaev:2024jxh} in a different form). Because our result \eqref{eq:result_thermal_npt} resembles for higher points the two-dimensional one in equation \eqref{eq:npt_block_2d_2}, it turns out to be harder to do anything similar. The connection formula used in \cite{Perlmutter:2015iya} for the one-point block in two dimensions has the effect of transforming $q \to 1-q$ in the argument of the hypergeometric function. A possible future direction from here is to attempt a similar transformation for \eqref{eq:result_thermal_npt}  by using one of the various generating functions for the SU(2) Clebsch-Gordan coefficients; see, for example, \cite{varsh1988}. 

In the two-dimensional case, there is by now ample discussion on different channels of global conformal blocks, see for example \cite{Cho:2017oxl,Fortin:2020bfq,Fortin:2020yjz}. We tried to show that in four dimensions, the computation of thermal conformal blocks in the projection channel is straightforward once the corresponding integrals over $\mathbb{D}_4$ are computed, and hope that the same applies to blocks in different channels, as well as different topologies of spaces, see e.\,g. \cite{Benjamin:2023qsc,Shaghoulian:2015kta,Belin:2016yll}. Finally, we have so far restricted ourselves so far to the scalar exchange between external scalar operators. However, general conformal blocks with external scalar operators principally involve also more general spinning representations in the exchange. We plan to address this in future work.

\section*{Acknowledgements}
The authors thank Alexander Jercher, Christoph Sieling, José Diogo Simão, Sebastian Steinhaus, Julio Virrueta and Andreas Wipf for related discussions. 

The work of JH, TH and KW is funded by the \emph{Deutsche Forschungsgemeinschaft (DFG)} under Grant No.\,406116891 within the Research Training Group RTG\,2522/1. JH is funded by a \emph{Landesgraduiertenstipendium} of the federal state of Thuringia. 

\appendix

\section{Generalisation to Non-Vanishing Angular Potentials}\label{sec_app:angular_potentials}

For the more general case of non-vanishing angular potentials, additional factors of $\mathrm{e}^{\mu_2 M_{03}}$ and $\mathrm{e}^{\mu_3 M_{12}}$ are present in the definition of the $n$-point projection channel block 
\begin{align} \label{eq:def_npt_potentials}
    G^{(n)}_{\text{Proj.}}(q,\mu_2,\mu_3) \equiv \text{Tr}\left( \mathbb{P}_{\Deltabb_1} q^D \mathrm{e}^{\mu_2 M_{03}} \mathrm{e}^{\mu_3 M_{12}} \mathcal{O}_1 \mathbb{P}_{\Deltabb_2}\mathcal{O}_{\Delta_2} \cdots \mathbb{P}_{\Deltabb_{n}}\mathcal{O}_n\right). 
\end{align}
We can pull the exponentials out in the same way as in the the purely thermal case in equation \eqref{eq:draw_D_out}, giving in the case of the one-point block
\begin{align} \label{eq:1pt_action_potentials}
    G^{(1)}_{\text{Proj.}}(q,\mu_2,\mu_3) = q^{ \Deltabb}\int_{\mathbb{D}_4}[\mathrm{d}U]_\Deltabb    \mathrm{e}^{\mu_2 \mathfrak{M}_{03}} \mathrm{e}^{\mu_3 \mathfrak{M}_{12}} \, \Omega(x,U,qU^\dagger)\,.
\end{align}
Because one can show that
\begin{align}
    \mathfrak{M}_{03} \phi_{a,b}^{j,m}(U) &= -i(a+b)\phi_{a,b}^{j,m}(U)\,, & \mathfrak{M}_{12} \phi_{a,b}^{j,m}(U) &= -i(a-b)\phi_{a,b}^{j,m}(U)\,,
\end{align}
the action of these transformations can be evaluated explicitly. In the case of the one-point block in equation \eqref{eq:1pt_block_thermal_2}, the expansion of $\Omega$ gives base functions $\phi^{j_2,m_2}_{a_2,b_2}(U)$ and $\phi^{j_3,m_3}_{a_3,b_3}(U)$ and the action of $\mathrm{e}^{\mu_2 M_{03}}\mathrm{e}^{\mu_3 M_{12}}$  on those evaluates as
\begin{equation} \label{eq:phase_potentials}
    \mathrm{e}^{\mu_2 M_{03}}\mathrm{e}^{\mu_3 M_{12}} \phi^{j_2,m_2}_{a_2,b_2}\phi^{j_3,m_3}_{a_3,b_3} = \mathrm{e}^{-i\mu_2(a_2 + a_3 + b_2 + b_3)} \mathrm{e}^{-i\mu_3(a_2 + a_3 - b_2 - b_3)} \phi^{j_2,m_2}_{a_2,b_2}\phi^{j_3,m_3}_{a_3,b_3}
\end{equation}
which can be shown by using equation \eqref{eq:expansion_d4}. The coordinate independence of the one-point block at vanishing angular potential followed from equation \eqref{eq:coord_indep}. The presence of the phase from equation \eqref{eq:phase_potentials} makes it difficult to evaluate \eqref{eq:coord_indep} as before, consistent with the fact that the presence of angular potentials should make the one-point block not translationally invariant. Note that, as in the purely thermal case, the transformations in equation \eqref{eq:1pt_action_potentials} act on only one $\Omega$-wavefunction, even in the $n$-point block. As a result, the general case \eqref{eq:def_npt_potentials} works exactly in the same way.

\section{Auxiliary Results on Clebsch-Gordan Coefficients and Representations}
\label{sec_app:properties_cgcs}

We give identities for both the SU(2) Wigner-$\mathcal{D}$ matrices and their analytic extensions. For SU(2), we roughly follow the classic reference \cite{varsh1988} where not stated otherwise.

\subsection{Identities for Wigner-\texorpdfstring{$\mathcal{D}$}{D} Matrices and  Clebsch-Gordan Coefficients}
We introduced in equation \eqref{eq:def_wignerD} the Wigner-$\mathcal{D}$ matrix for an arbitrary $2\times 2$ complex matrix $X$. Two important properties are 
    \begin{equation}
        \sum_{c = -j}^{j} \mathcal{D}^{j}_{a,c}(X) \mathcal{D}^{j}_{c,b}(Y) = \mathcal{D}^{j}_{a,b}(XY)\,, \qquad \mathcal{D}^j_{a,b}(X) = \mathcal{D}^j_{b,a}(X^T)\,, \label{eq:homom_prop_D}
    \end{equation}
where $j \in \mathbb{N}_0/2$ and $a,b = -j,-j+1,\dots,j-1,j$. For $X \in \text{GL}(2,\mathbb{C})$, one can show\footnote{For example by generalising the proof in \cite{ruehl_1970}[Ch.2.4] for $a \in \text{SL}(2,\mathbb{C})$ to $X = \sqrt{\det(X)} a \in \text{GL}(2,\mathbb{C})$.} that 
\begin{align} \label{eq:inverse_wigner}
    \mathcal{D}^j_{a,b}((X^{-1})^T) = (-1)^{a-b}\operatorname{det}^{-2j}(X) \mathcal{D}^j_{-a,-b}(X)\,.
\end{align}
In the case where $X \in \text{SU}(2)$, equation \eqref{eq:inverse_wigner} leads to the standard statement about complex conjugation 
    \begin{equation}
    \label{eq:wigner_cc}
        \mathcal{D}^j_{a,b}(X) = (-1)^{a-b} \overline{\mathcal{D}^j_{-a,-b}(X)}\,.
    \end{equation}
Note that one can check using \eqref{eq:inverse_wigner} that \eqref{eq:wigner_cc} also holds for $X$ defined in equation \eqref{eq:X_parametr}, i.\,e. $X = x_0 \sigma^0 + ix_j \sigma^j$ with $x_\mu$ real. A possibility to introduce the SU(2) Clebsch-Gordan coefficients $C^{j_1j_2j}_{a_1a_2c}$ is then through the expansion of the product of two Wigner-$\mathcal{D}$ matrices as 
    \begin{equation}
        \label{eq:exp_wigner_mat}\mathcal{D}^{j_1}_{a_1,b_1}(X) \mathcal{D}^{j_2}_{a_2,b_2}(X) = \sum_{j} \sum_{c_1 c_2} C^{j_1j_2j}_{a_1,a_2,c_1} C^{j_1j_2j}_{b_1,b_2,c_2} \mathcal{D}^j_{c_1,c_2}(X)\,,
    \end{equation}
    where the coefficients vanish unless the conditions 
    \begin{equation}
        |j_1 - j_2| \leq j \leq j_1 + j_2\,, \qquad 
         c_1 = a_1 + a_2\quad (c_2 = b_1 + b_2)
    \end{equation}
are fulfilled. This means that both the $c_1$ and the $c_2$ sum in equation \eqref{eq:exp_wigner_mat} are formal in the sense that only one summand is non-zero. The Clebsch-Gordan coefficients satisfy two orthogonality conditions
    \begin{align}
        \sum_{j = |j_1 - j_2|}^{j_1 + j_2}\sum_{c = -j}^j C^{j_1j_2j}_{a_1,a_2,c} C^{j_1j_2j}_{b_1,b_2,c} &= \delta_{a_1,b_1} \delta_{a_2,b_2}\,, \label{eq:cgc_orthog_1}\\
        \sum_{a_1=-j_1}^{j_1} \sum_{a_2 = -j_2}^{j_2} C^{j_1j_2j}_{a_1,a_2,c} C^{j_1j_2j'}_{a_1,a_2,c'} &= \delta_{j,j'} \delta_{c,c'}\,, \label{eq:cgc_orthog_2}
    \end{align}    
meaning that we choose the Clebsch-Gordan coefficients to be real. Switching columns has the following effect
\begin{subequations} \label{eq:switching}
\begin{align}
\label{eq:switch_column_1}
    C^{j_1 j_2 j}_{a_1, a_2, c} &= (-1)^{j_2 + a_2} \sqrt{d_jd_{j_1}^{-1}} C^{j j_2 j_1}_{-c,a_2,-a_1} = (-1)^{j_2 + a_2} \sqrt{d_j d_{j_1}^{-1}} C^{j_2 j j_1}_{-a_2,c,a_1}\\ &= (-1)^{j_1 - a_1} \sqrt{d_j d_{j_2}^{-1}} C^{j_1 j j_2}_{a_1,-c,-a_2} = (-1)^{j_1 - a_1}\sqrt{d_j d_{j_2}^{-1}} C^{j j_1 j_2}_{c,-a_1,a_2} \label{eq:switch_column_2} \\
    &= (-1)^{j_1 + j_2 - j} C^{j_2 j_1 j}_{a_2, a_1, c} = (-1)^{j_1 + j_2 - j} C^{j_1 j_2 j}_{-a_1, -a_2, -c} \,, \label{eq:cgc_symmetry}
\end{align}
\end{subequations}
where the last equation is a symmetry and we used the notation $d_j = 2j+1$. Finally, an important special case of $C^{jj_1j_2}_{c,a_1,a_2}$ is given by
\begin{equation} \label{eq:zero_cgc_delta}
    C^{0j_1j_2}_{0,a_1,a_2} = \delta_{j_1,j_2}\delta_{a_1,a_2}\,.
\end{equation}
Note that a Clebsch-Gordan coefficient can be represented as a terminating ${}_3F_2$-hypergeometric function of unit argument , and as such, a finite sum. We give as an example one of the possible expressions as
\begin{align} \label{eq:cgc_as_3F2}
    C^{j_1 j_2 j}_{a_1, a_2, c} &= \delta_{c,a_1 + a_2} \frac{\Delta(j_1 j_2 j) }{(j_1 + j_2  - j)! (-j_2 + j + a_1)!(-j_1 +j - a_2)!}   \\
    &\sqrt{\frac{d_j!(j_1 + a_1)!(j_2 - a_2)! (j+ c)!(j - c)!}{(j_1 - a_1)!(j_2 + a_2)!}} {}_3F_2\left[\begin{matrix}
        -j_1 - j_2+j ,\,-j_1 + a_1,\,-j_2-a_2 \\ 1-j_1+j-a_2\,, 1-j_2+j+a_1
    \end{matrix} \Bigg| 1\right] \notag
\end{align}
for 
\begin{equation}
    \Delta(j_1j_2j) = \sqrt{\frac{(j_1 + j_2 - j)!(j_1 - j_2 + j)!(-j_1 + j_2 + j)!}{(1+j_1 + j_2 + j)!}}\,.
\end{equation}

\subsection{Intertwiners and Integration over SU(2)}
In the case where the Wigner $D$-matrices depend solely on an element of SU(2), they define an irreducible representation of the group. They fulfil the following orthogonality relation when integrated over them 
\begin{equation}
    \int_{\text{SU}(2)} \d g\, \overline{\mathcal{D}^{j_1}_{a_1,b_1}(g)}\mathcal{D}^{j_2}_{a_2,b_2}(g) = d_{j_1}^{-1} \delta_{j_1,j_2} \delta_{a_1,a_2}\delta_{b_1,b_2}
\end{equation}
where $\d g$ is the Haar measure\footnote{There is some nuance in choosing the correct numerical prefactor in $\d g$ when integrating over SU(2), see \cite{varsh1988}. However, since both the conformal blocks and the oscillator wavefunctions used are only determined up to an overall constant, we ignore this factor in the following.} of SU(2) \cite{Makinen:2019rou}. This can be generalised to 
\begin{equation}
\label{eq:su2_12_integral}
    \int_{\text{SU}(2)} \d g\, \overline{\mathcal{D}^{j_1}_{a_1,b_1}(g)}\mathcal{D}^{j_2}_{a_2,b_2}(g) \mathcal{D}^{j_3}_{a_3,b_3}(g) = d_{j_1}^{-1} C^{j_3j_2j_1}_{a_3,a_2,a_1} C^{j_3j_2j_1}_{b_3,b_2,b_1}\,. 
\end{equation}
Every further addition of a Wigner-$\mathcal{D}$ matrix to the integrand of equation \eqref{eq:su2_12_integral} can be absorbed by expanding the product of two matrices in terms of a third, using equation \eqref{eq:exp_wigner_mat}. This leads to the general formula
\begin{equation}\label{eq:su22_(1,N-1)_integral}
    \int_{\text{SU}(2)}\d g\, \overline{\mathcal{D}^{j_1}_{a_1,b_1}(g)} \mathcal{D}^{j_2}_{a_2,b_2}(g) \cdots \mathcal{D}^{j_{N-1}}_{a_{N-1},b_{N-1}}(g) = \sum_{l_1\dots l_{N-3}} I^{l_1\dots l_{N-3}}_{a_N,\cdots,a_1} I^{l_1\dots l_{N-3}}_{b_N,\cdots,b_1}\,,
\end{equation}
where the $N$-valent intertwiner $I^{l_1,\dots,l_{N-3}}_{a_N,\cdots,a_1}$ is given by
\begin{equation}
\label{eq:def_intertwiner}
    I^{l_1\dots l_{N-3}}_{a_N,\cdots,a_1} = \sum_{p_1,\dots,p_{N-3}} C^{j_N j_{N-1}l_1}_{a_N,a_{N-1},p_1} \left(\prod_{i=2}^{N-3} C^{l_{i-1}j_{N-i}l_i}_{p_{i-1}a_{N-i}p_i}\right)C^{l_{N-3}j_2 j_1}_{p_{N-3},a_2,a_1}\,.
\end{equation}
The $p_i$ run between $-l_i$ and $l_i$ and the $l_i$ as
\begin{gather}
    |j_N-j_{N-1}| \leq l_1 \leq j_N + j_{N-1}\,, \notag\\ |l_1-j_{N-2}| \leq l_2 \leq l_1 + j_{N-2}\,, \dots \,,|l_{N-4}-j_{3}| \leq l_{N-3} \leq l_{N-4} + j_{3}\,.
\end{gather}
The only example we really use in our calculation of blocks is the four-valent one, given by
\begin{equation} \label{eq:1_valent_intertw}
    I^l_{a_4,a_3,a_2,a_1} = \sum_{p_a} C^{j_4j_3l}_{a_4a_3p_a}C^{l j_2 j_1}_{p_a,a_2,a_1}\,.
\end{equation}
For convenience, we can introduce the following tensors that are defined through an exchange of columns in one of the two defining Clebsch-Gordan coefficients as in equations \eqref{eq:switch_column_1} and \eqref{eq:switch_column_2}
\begin{align}\label{eq:def_I_tilde}
    \widetilde{I}^{l}_{a_4,a_3,a_2,a_1} 
    := \sum_{p_a}  C^{j_4j_3l}_{a_4,a_3,p_a}C^{j_2 j_1 l}_{a_2,a_1,p_a} 
    = (-1)^{j_2 - a_2} \sqrt{d_l d_{j_1}^{-1}}  I^{l}_{a_4,a_3,-a_2,a_1}\,.
\end{align}
We will use this rewriting only in the context where, due to its contraction with the basis, the switch from $\widetilde{I}$ to $I$ amounts simply to a complex conjugation together with a dimensional factor. This is the case for the matrices $X = x_0\sigma^0 + ix_j\sigma^j$ in equation \eqref{eq:X_parametr}. One then has
\begin{equation} \label{eq:rewrite_intertwiners}
\begin{split}
    &\sum_{a_2,b_2} \sum_{a_3,b_3} \widetilde{I}^l_{a_1,\dots,a_4} \widetilde{I}^l_{b_1,\dots,b_4} \mathcal{D}^{j_2}_{a_2,b_2}(X) \mathcal{D}^{j_3}_{a_3,b_3}(Y)  \\
    &= d_l d_{j_4}^{-1}  \sum_{a_2,b_2} \sum_{a_3,b_3} I^l_{a_1,\dots,a_4} I^l_{b_1,\dots,b_4} \mathcal{D}^{j_2}_{a_2,b_2}(X) \overline{\mathcal{D}^{j_3}_{a_3,b_3}(Y)}\,,
\end{split}
\end{equation}
where one uses equation \eqref{eq:wigner_cc}. Because in this case also $\operatorname{det}(Y)$ is real, the same holds for the analogous contraction involving $\phi^{j_3,m_3}_{a_3,b_3}(Y)$.

\section{Details of the Computation of Higher Integrals}\label{sec_app:HigherIntegrals_Details}

The computation of the scalar four-point block requires the inner product of two basis elements of $\mathcal{H}L^2_{\Deltabb}(\mathbb{D}_4)$ as
\begin{equation} \label{eq:inner_product}
 \braket{\phi^{j_1,m_1}_{a_1,b_1}}{\phi^{j_2,m_2}_{a_2,b_2}} = \int_{\mathbb{D}_4} [\d U]_{\Deltabb} \overline{\phi^{j_1,m_1}_{a_1,b_1}(U)} \phi^{j_2,m_2}_{a_2,b_2}(U) = \left(\mathcal{N}^{j_1,m_1}_\Deltabb\right)^{-2} \delta_{j_1,j_2}\delta_{m_1,m_2}\delta_{a_1,a_2} \delta_{b_1,b_2}
\end{equation}
We will in the following, hopefully intuitively, call this case the $(1,1)$-integral. Building on that, we will first compute the $(1,2)$-integral and generalise the computation to the case $(1,N-1)$. We then proceed to calculate the $(2,2)$-integral, which is the fundamental integral for the main part of this publication. We do this by using the $(1,2)$-integral and an expansion of the basis elements similar to the expansion of Wigner-$\mathcal{D}$ matrices in \eqref{eq:exp_wigner_mat}. The general integral can be straightforwardly derived from there, but is not important for this publication. Note that the computations of these integrals generalise the computation of the inner product in \cite{Calixto_2011}.  

\subsection{\texorpdfstring{Decomposition of Elements of $\mathbb{D}_4$}{Decomposition of Elements of Bergman Domain}}
Any element $M \in \text{SU}(2,2)$ has to satisfy 
\begin{equation}
\label{eq:u22_conditions}
    M^\dagger H M = H\,, \qquad H = \text{diag}\left(-\mathbb{1},\mathbb{1}\right)\,,
\end{equation}
by definition together with $\text{det}(M) = 1$. We can take the following ansatz
\begin{align}
    \begin{pmatrix}
        A&B\\C &D
     \end{pmatrix} 
    = \begin{pmatrix}
        N_1 & U N_2 \\ U^\dagger N_1 & N_2
    \end{pmatrix} \begin{pmatrix}
        U_a & 0 \\ 0 & U_b
    \end{pmatrix}
\end{align}
for $U_{a,b} \in \text{U}(2)$, $N_{1,2}$ two positive-definite Hermitian matrices and at this point $U$ an arbitrary complex $2 \times 2$ matrix. We justify this\footnote{Furthermore, this is a generalisation of decompositions for $\text{SU}(1,1)$ (adapted to $\mathbb{D}$) and $\text{SU}(2)$ (adapted to $\mathbb{S}^2)$, see for example equation \eqref{eq:u2_decomp}.} by already knowing at this point that we want to consider a Hilbert space over quotient of $\text{SU}(2,2)$ by its maximally compact subgroup $\text{S}(\text{U}(2) \times \text{U}(2))$.\footnote{The Lie algebras of $\text{S}(\text{U}(2) \times \text{U}(2))$ and $\text{SU}(2) \times \text{SU}(2) \times \text{U}(1)$ are isomorphic.} In other words, this is the decomposition adjusted to the specific structure of this symmetric space. It was checked in \cite{Ruehl:1972jy} that this ansatz satisfies the defining $\text{U}(2,2)$ conditions \eqref{eq:u22_conditions} in the case where $U \in \mathbb{D}_4$ (see definition \eqref{eq:su22_quotient}) and they furthermore determine for example 
\begin{equation}
  N_1 = \left(\mathbb{1}-U U^\dagger \right)^{-1/2}\,, \quad  N_2 =  \left(\mathbb{1}-U^\dagger U \right)^{-1/2}\,, \quad U_a = N_1^{-1}A\,, \quad U_b = N_2^{-1}D\,,
\end{equation}
amounting to a polar decomposition of $A$ and $D$.

Then, it is convenient to drop the determinant condition $\text{det}(U_a U_b) = 1$ in the following, i.\,e. to consider $\text{U}(2,2)/\text{U}(2)^2$ instead. This allows us to apply the analogue decomposition to the elements in both of the $\text{U}(2)$ factors
\begin{equation} \label{eq:u2_decomp}
    \begin{pmatrix}
        a & b \\
        c & d
    \end{pmatrix}
    = \begin{pmatrix}
        \delta & z\delta \\
        -\bar{z}\delta & \delta 
    \end{pmatrix}
     \begin{pmatrix}
            \mathrm{e}^{i\beta_1} & 0 \\
            0 & \mathrm{e}^{i\beta_2} 
    \end{pmatrix}\,,
\end{equation} 
for $\delta = (1-z \overline{z})^{-1/2}$. Note that both of these decompositions lead in turn to decompositions of the integral measure $\d \mu(g)$
\begin{gather}
    \d\mu(g) = \d \mu(g)\big |_{\mathbb{D}_4} \d \nu (U_a) \d \nu (U_b)\,, \qquad \d \mu(g)\big |_{\mathbb{D}_4} = \mathrm{det}^{-4}\left(\mathbb{1}-U U^\dagger \right) \d U\\
    \d \nu (U_{a,b}) = (1+z \overline{z})^{-2} \d z\, \beta_1 \d \beta_1\, \beta_2 \d \beta_2 \label{eq:measure_u2}
\end{gather}
where $\d U$ and $\d z$ are the Lebesgue measures on $\mathbb{C}^4$ and $\mathbb{C}$, respectively. When considering the weighted Bergman space $\mathcal{H}L^2_{\Deltabb}(\mathbb{D}_4)$, the measure on $\mathbb{D}_4$ gets adjusted to $\left[\d U\right]_\Deltabb$ in equation \eqref{eq:D4_inner_product}, after which $\int_{\mathbb{D}_4}\left[\d U\right]_\Deltabb 1 = 1$ (see e.\,g. \cite{Calixto_2011}).

Because $\mathbb{D}_4$ is a symmetric space, we can by definition use transformations of the subgroup to simplify its elements, in this case unitary transformations $U_{a,b}$ to diagonalise $U \in \mathbb{D}_4$. This means in turn that we can parametrie any $U \in \mathbb{D}_4$ as\footnote{This can be seen as a less restrictive singular value decomposition, allowing for complex $\xi_1$ and $\xi_2$.}
\begin{equation}
\label{eq:d4_decomp}
    U = U_a \Xi U^\dagger_b
\end{equation}
where $U_{a,b}$ are unitary and $\Xi = \text{diag}(\xi_1,\xi_2)$ is a complex diagonal matrix with $\xi_{1,2} \in \mathbb{D}$. Because this decomposition is only unique up to a phase, $U_{a,b}$ should be considered elements of $\text{U}(2)/\text{U}(1)^2 \cong \mathbb{S}^2$. Note that this means that later integrations over $U_{a,b}$ have $\beta_1$ and $\beta_2$ in equation \eqref{eq:u2_decomp} set to zero, such that the measure in equation \eqref{eq:measure_u2} reduces to the measure of $\mathbb{S}^2$
\begin{equation}
    \d s(U_{a,b}) := (1+z \overline{z})^{-2} \d z\,.
\end{equation}

We proceed to parametrise $\xi_{1,2} \in \mathbb{D}$ as $\xi_{1,2} = \rho_{1,2} \mathrm{e}^{i\theta_{1,2}}$ in standard polar coordinates. It is now straightforward to check that the weight-function under the decomposition in equation \eqref{eq:d4_decomp} reduces to 
\begin{equation}
\label{eq:def_omega}
    \text{det}^{\Deltabb - 4}\left(\mathbb{1}-UU^\dagger\right) = \left((1-\rho_1^2) (1-\rho_2^2)\right)^{\Deltabb - 4} =: \Omega(\rho_1,\rho_2)\,.
\end{equation}
Furthermore, because $\mathbb{D}_4 \subset \mathbb{C}^4$, we can rewrite the Lebesgue measure on the latter as
\begin{equation}
\label{eq:def_jacobian}
    \d U = J(\rho_1,\rho_2) \d \xi_1 \d \xi_2 \d s(U_a) \d s(U_b)\,, \qquad J(\rho_1,\rho_2) = \frac{1}{2}\left(\rho_1^2 - \rho_2^2\right)^2\,,
\end{equation}
where $J(\rho_1,\rho_2)$ is the Jacobian determinant. Altogether, this leads to the parametrization of $\left[\d U\right]_\Deltabb$ as
\begin{equation}
    \left[\d U\right]_\Deltabb = J(\rho_1,\rho_2) \Omega(\rho_1,\rho_2) \d \xi_1 \d \xi_2 \d s(U_a) \d s (U_b)\,.
\end{equation}

\subsection{\texorpdfstring{Computation of $(1,2)$-Integral}{Computation of (1,2)-Integral}}
According to the discussion in the previous section, choosing $U = U_a \Xi U_b^\dagger$ gives us for the $(1,2)$-integral 
\begin{align}
    & \braket{\phi_{a_1,b_1}^{j_1,m_1}}{\phi_{a_2,b_2}^{j_2,m_2}\phi_{a_3,b_3}^{j_3,m_3}} \notag\\&= \int_{\mathbb{D}_4} [\d U]_\Deltabb\overline{\mathrm{det}^{m_1}(U) \mathcal{D}_{a_1,b_1}^{j_1}(U)}\mathrm{det}^{m_2}(U) \mathcal{D}_{a_2,b_2}^{j_2}(U) \mathrm{det}^{m_3}(U) \mathcal{D}_{a_3,b_3}^{j_3}(U) \notag\\
    &=  \sum_{c_1,c_2,c_3} c_{\mathbb{\Delta}} \int_{\mathbb{D}^2} \d\xi_1\d\xi_2\, J(\xi_1,\xi_2) \Omega(\xi_1,\xi_2) \mathrm{det}^{m_1}(\overline{\Xi}) \mathrm{det}^{m_2+m_3}(\Xi) 
     \mathcal{D}^{j_1}_{c_1,c_1}(\overline{\Xi}) \mathcal{D}^{j_2}_{c_2,c_2}(\Xi) \mathcal{D}^{j_3}_{c_3,c_3}(\Xi) \notag\\
    &\,\,\,\, \qquad \quad\, \times \int_{\mathrm{U}(2)/\mathrm{U}(1)^2} \d s(U_a) \mathcal{D}^{j_1}_{a_1,c_1}(\overline{U_a}) \mathcal{D}^{j_2}_{a_2,c_2}(U_a)  \mathcal{D}^{j_3}_{a_3,c_3}(U_a) \notag\\
    &\,\,\,\, \qquad \quad\, \times \int_{\mathrm{U}(2)/\mathrm{U}(1)^2} \d s(U_b) \mathcal{D}^{j_1}_{b_1,c_1}(U_b) \mathcal{D}^{j_2}_{b_2,c_2}(\overline{U}_b)  \mathcal{D}^{j_3}_{b_3,c_3}(\overline{U}_b)\,, \label{eq:12_integral}
\end{align}
where we have used that $\mathrm{det}(U_{a,b}) = 1$ (following from equation \eqref{eq:u2_decomp}) and equation \eqref{eq:homom_prop_D}. We consider first the disk integrations. Because 
\begin{equation}
    \mathcal{D}_{a,b}^j(\Xi) = \delta_{a,b} \xi_1^{j+a} \xi_2^{j-a}\,, \qquad \mathrm{det}(\Xi) = \xi_1 \xi_2\,, \qquad \xi_{1,2} = \rho_{1,2}\mathrm{e}^{i\theta_{1,2}}\,,
\end{equation}
we get for the integrand of the disk integrations
\begin{align}
\begin{split}
    &\rho_1^{j_1 + j_2 + j_3 + c_1 + c_2 + c_3 + m_1 + m_2 + m_3+1} \rho_2^{j_1 + j_2 + j_3 - c_1 - c_2 - c_3 + m_1 + m_2 + m_3+1} 
    \\
    &\times \mathrm{e}^{i\theta_1 (-j_1+j_2+j_3 - c_1+c_2+c_3-m_1+m_2+m_3)} \mathrm{e}^{i\theta_2 (-j_1+j_2+j_3 + c_1-c_2-c_3-m_1+m_2+m_3)}\,.
    \end{split}
\end{align}
Integration over the angular part gives then
\begin{equation}
    4\pi^2 \delta_{c_1,c_2 + c_3} \delta_{k_1,k_2 + k_3} \label{eq:12_angular_result}
\end{equation}
where we abbreviated $k_i = j_i + m_i$. When taking into account these Kronecker deltas, the radial part (together with the factor of $4\pi^2$ from equation \eqref{eq:12_angular_result}) is given by 
\begin{align}
\begin{split}
    R_{c_1}^{k} &:= 4\pi^2 c_\Deltabb \int_0^1 \int_0^1 \rho_1 \d \rho_1 \rho_2 \d \rho_2J(\rho_1,\rho_2) \Omega (\rho_1,\rho_2) \rho_1^{2(k+c_1)} \rho_2^{2(k-c_1)}\\
    &= 2\pi^2 c_\Deltabb \int_0^1 \d\rho_1\,\int_0^1 \d\rho_2\, \Big\{ \rho_1^{2(k+c_1+2)+1}(1-\rho_1^2)^{\Deltabb-4}\rho_2^{2(k-c_1)+1}(1-\rho_2^2)^{\Deltabb-4} \\
    &\quad \,+ \rho_1^{2(k+c_1)+1}(1-\rho_1^2)^{\Deltabb-4}\rho_2^{2(k-c_1+2)+1}(1-\rho_2^2)^{\Deltabb-4} \\
    &\quad \, - 2 \rho_1^{2(k+c_1+1)+1}(1-\rho_1^2)^{\Deltabb-4}\rho_2^{2(k-c_1+1)+1}(1-\rho_2^2)^{\Deltabb-4} \Big\}\,,
\end{split}
\end{align}
where we used equations \eqref{eq:def_omega} and \eqref{eq:def_jacobian}. We can now use in every summand the elementary integral
\begin{equation}
    \int_0^1 \d x\, x^{2a+1}(1-x^2)^b = \frac{\Gamma(a+1)\Gamma(b+1)}{\Gamma(a+b+2)}\,,
\end{equation}
and together with $c_\Deltabb=\pi^{-4}(\Deltabb-1)(\Deltabb-2)^2(\Deltabb-3)$ we arrive at 
\begin{align}
\label{eq:result_radial_integral}
    R^k_{c_1} &= \frac{c_1^2(2\Deltabb-5) + k^2 + \Deltabb(k+1)-1}{\pi^2(\Deltabb-1)} \binom{k+c_1+\Deltabb-1}{ \Deltabb-1}^{-1} \binom{k-c_1+\Deltabb-1}{\Deltabb-1}^{-1}\,.
\end{align}

We continue with the spherical integrals in equation \eqref{eq:12_integral}. Both integrals are a special case of the SU(2) integrals given in equation \eqref{eq:su2_12_integral}. Because of the invariance of $U_{a}$, we can add an extra integration and compensate by the appropriate volume factor, ending up with\footnote{This is again up to constant prefactor.} 
\begin{equation}
    \int_{\mathrm{U}(2)/\mathrm{U}(1)^2} \d s(U_a) \overline{\mathcal{D}^{j_1}_{a_1,c_1}(U_a)} \mathcal{D}^{j_2}_{a_2,c_2}(U_a)  \mathcal{D}^{j_3}_{a_3,c_3}(U_a) = \frac{\pi}{d_{j_1}} C^{j_3j_2j_1}_{a_3,a_2,a_1} C^{j_3j_2j_1}_{c_3,c_2,c_1}\,.
\end{equation}
The $U_b$ integral gives the analogue result, if one carefully accounts for the phases after using equation \eqref{eq:wigner_cc} and \eqref{eq:cgc_symmetry}. We can then evaluate the $c_2$ and $c_3$ sums in equation \eqref{eq:12_integral}, using the orthogonality relation \eqref{eq:cgc_orthog_2}. The remaining $c_1$ sum can then be evaluated over the result of the radial integral in equation \eqref{eq:result_radial_integral}. Grouping together with the numerical prefactors from the $U_{a,b}$-integrals, one obtains the same coefficient as in the inner product \eqref{eq:inner_product}
\begin{align}
\begin{split}
    \frac{\pi^2}{d_{j_1}^2} \sum\limits_{c_1=-j_1}^{j_1} R^{j_1+m_1}_{c_1} &= \frac{\Deltabb - 1}{d_{j_1}} \binom{m_1+\Deltabb-2}{\Deltabb-2}^{-1} \binom{d_{j_1}+m_1+\Deltabb-2}{\Deltabb-2}^{-1} \\
    &= (\mathcal{N}^{j,m}_\Deltabb)^{-2}\,. 
\end{split}
\end{align}
Putting everything together again, the $(1,2)$-integral evaluates hence to 
\begin{equation}
    \braket{\phi_{a_1,b_1}^{j_1,m_1}}{\phi_{a_2,b_2}^{j_2,m_2}\phi_{a_3,b_3}^{j_3,m_3}} = \left(\mathcal{N}^{j_1,m_1}_\Deltabb\right)^{-2} C^{j_3j_2j_1}_{a_3,a_2,a_1}
 C^{j_3 j_2 j_1}_{b_3, b_2, b_1} \delta^{k_1}_{k_2 + k_3}\,. \label{eq:12_integral_evaluated}
\end{equation}
This calculation is easily generalised to the $(1,N-1)$-integral by consequent expansion of products of Wigner-$\mathcal{D}$ matrices using \eqref{eq:exp_wigner_mat} and then collapsing a chain of orthogonality relations \eqref{eq:cgc_orthog_2}, yielding
\begin{equation} \label{eq:(1,N-1)_integral}
     \braket{\phi_{a_1,b_1}^{j_1,m_1}}{\phi_{a_2,b_2}^{j_2,m_2} \cdots \phi_{a_{N-1},b_{N-1}}^{j_{N-1},m_{N-1}}} = \left(\mathcal{N}^{j_1,m_1}_\Deltabb\right)^{-2} \delta^{k_1}_{k_2+\dots + k_{N-1}} \sum_{l_1,\dots,l_{N-3}} I^{l_1\dots l_{N-3}}_{a_N,\cdots,a_1} I^{l_1\dots l_{N-3}}_{b_N,\cdots,b_1}\,,
\end{equation}
where the SU(2) intertwiners are defined in equation \eqref{eq:def_intertwiner}.

\subsection{\texorpdfstring{Computation of the $(2,2)$- and Higher Integrals}{Computation of the (2,2)- and Higher Integrals}}
Having computed the $(1,2)$-integral, we can now compute in theory every other integral, because we can expand the product of two basis functions in terms of a single basis function, similarly to the SU(2) case in \eqref{eq:exp_wigner_mat}. We present in the following only the $(2,2)$-integral since that suffices for the purpose of this work. From orthogonality we get  
\begin{align} \label{eq:expansion_d4}
    \phi^{j_1,m_1}_{a_1,b_1}(U) \phi^{j_2,m_2}_{a_2,b_2}(U) = \sum_{p_a,p_b}^{l,n} \left(\mathcal{N}^{l,n}_{\Deltabb}\right)^2 \braket{\phi^{l,n}_{p_a,p_b}}{\phi^{j_1,m_1}_{a_1,b_1} \phi^{j_2,m_2}_{a_2,b_2}} \phi^{l,n}_{p_a,p_b}(U)\,
\end{align}
for $U \in \mathbb{D}_4$. 
Note that this expansion reduces to the SU(2) expansion in equation \eqref{eq:exp_wigner_mat} for $\det(U) = 1$.
We can use now equation \eqref{eq:expansion_d4} to compute the $(2,2)$-integral 
\begin{align} \label{eq:22_integral} 
\begin{split}
    &\braket{\phi^{j_1,m_1}_{a_1,b_1} \phi^{j_2,m_2}_{a_2,b_2} }{\phi^{j_3,m_3}_{a_3,b_3} \phi^{j_4,m_4}_{a_4,b_4} } \\ &= \sum_{p_a,p_b}^{l,n} \left(\mathcal{N}^{l,n}_{\Deltabb}\right)^{-2} C^{j_4j_3l}_{a_4,a_3,p_a} C^{j_2j_1l}_{a_2,a_1,p_a} C^{j_4j_3l}_{b_4,b_3,p_b} C^{j_2j_1l}_{b_2,b_1,p_b}   \delta^{l+n}_{k_1+k_2} \delta^{l+n}_{k_3+k_4} \\
    &=: \sum_{l,n} \left(\mathcal{N}^{l,n}_{\Deltabb}\right)^{-2} \widetilde{I}^{l}_{a_4,a_3,a_2,a_1} \widetilde{I}^{l}_{b_4,b_3,b_2,b_1} \delta^{l+n}_{k_1+k_2} \delta^{l+n}_{k_3+k_4}\\
    &= \sum_{l,n} \left(\mathcal{N}^{l,n}_{\Deltabb}\right)^{-2} \widetilde{I}^{l}_{a_1,a_2,a_3,a_4} \widetilde{I}^{l}_{b_1,b_2,b_3,b_4} \delta^{l+n}_{k_1+k_2} \delta^{l+n}_{k_3+k_4}
\end{split}
\end{align}
where we used the definition of $\widetilde{I}$-tensors in equation \eqref{eq:def_I_tilde} and switched columns according to \eqref{eq:cgc_symmetry}. We can also evaluate the $n$-sum at this point, leading to 
\begin{align} \label{eq:22_integral_2}
\begin{split}
    \braket{\phi^{j_1,m_1}_{a_1,b_1} \phi^{j_2,m_2}_{a_2,b_2}}{\phi^{j_3,m_3}_{a_3,b_3} \phi^{j_4,m_4}_{a_4,b_4}} &= \sum_{l} \left(\mathcal{N}^{l,k_1+k_2 - l}_{\Deltabb}\right)^{-2} \widetilde{I}^{l}_{a_1,a_2,a_3,a_4} \widetilde{I}^{l}_{b_1,b_2,b_3,b_4} \delta^{k_1+k_2}_{k_3+k_4}\,.
\end{split}
\end{align}
We can check that this is still normalised in the right way for e.\,g. $\phi^{j_1,m_1}_{a_1,b_1} = \phi^{j_2,m_2}_{a_2,b_2} = 1$. Note that the range of summation for $l$ is implicitly determined by the Clebsch-Gordan coefficients.

Finally, note that any of the expressions like e.\,g. the $(1,N-1)$-integral in equation \eqref{eq:(1,N-1)_integral} are conceptually similar to SU(2) expressions like in equation \eqref{eq:su22_(1,N-1)_integral}, which are in the literature also known as Haar projectors (see for example \cite{Makinen:2019rou,Martin-Dussaud:2019ypf}). To prove that these integrals are actual projectors and as such idempotent, one uses compactness of SU(2) and runs into problems for non-compact examples like e.\,g. $\text{SL}(2,\mathbb{C})$. In our case, however, we are working on $\mathbb{D}_4$ and because $\int_{\mathbb{D}_4}[\d U]_\Deltabb 1 = 1$ with $[\d U]_\Deltabb$ defined in equation \eqref{eq:D4_inner_product}, we can proceed in the same way.

\subsection{SU(2,2) Adapted Coefficients} \label{sec:adjusted_coeffs}
The coefficients we encountered in the computation of vacuum and thermal conformal blocks in sections \ref{sec:computation_npt_comb} and \ref{sec:computation_npt_thermal} are SU(2) Clebsch-Gordan coefficients and 4-valent intertwiners weighted with general linear combinations of conformal weights. We define
\begin{equation}
\label{eq:coeff_321_weighted}
    \mathcal{C}_{(3),(2),(1)}^{\alpha\beta\gamma} = \left(\sqrt{\frac{\mathcal{N}^{j_3,m_3}_{\alpha} \mathcal{N}^{j_2,m_2}_\beta}{{N}^{j_1,m_1}_\gamma}}C^{j_3j_2j_1}_{a_3,a_2,a_1}\right) \left(\sqrt{\frac{\mathcal{N}^{j_3,m_3}_\alpha \mathcal{N}^{j_2,m_2}_\beta}{{N}^{j_1,m_1}_\gamma}}C^{j_3 j_2 j_1}_{b_3, b_2, b_1}\right) \delta^{k_1}_{k_2 + k_3}
\end{equation}
where we again used the block index notation, e.\,g. $(1) = (j_1,m_1,a_1,b_1)$. Similarly 
\begin{align}
    &\widetilde{\mathcal{I}}_{(1),(2),(3),(4)}^{\alpha\beta\gamma\delta} \\ 
    &=  \sum_{(E)} \mathcal{C}_{(1),(2),(E)}^{\alpha\beta\Deltabb} \mathcal{C}_{(3),(4),(E)}^{\Deltabb\gamma\delta} \notag\\
    &= \sum_{p_a,p_b}^{l,n} \left(\frac{\mathcal{N}^{j_1,m_1}_\alpha\mathcal{N}^{j_2,m_2}_\beta}{\mathcal{N}^{l,n}_\Deltabb}C^{j_1j_2l}_{a_1,a_2,p_a}  C^{j_1j_2l}_{b_1,b_2,p_b}\right) \left(\frac{\mathcal{N}^{j_3,m_3}_\gamma \mathcal{N}^{j_4,m_4}_\delta}{\mathcal{N}^{l,n}_\Deltabb}C^{j_3j_4l}_{a_3,a_4,p_a} C^{j_3j_4l}_{b_3,b_4,p_b}\right)    \delta^{l+n}_{k_1+k_2} \delta^{l+n}_{k_3+k_4}\,. \notag
\end{align}
where $(E) = (l,n,p_a,p_b)$ is just the tupel we sum over. We define the weighted basis polynomials
\begin{equation} \label{eq:weighted_basis}
    \varphi^{(1)}_{\alpha}(U) = \mathcal{N}^{j_1,m_1}_\alpha\phi^{j_1,m_1}_{a_1,b_1}(U)
\end{equation}
for $\alpha$ a general linear combination of conformal weights and which are for $\alpha = \Deltabb$ normalised with respect to the inner product \eqref{eq:inner_product}.\footnote{For such normalised basis elements, the coefficient \eqref{eq:coeff_321_weighted} is just the coefficient appearing in the expansion \eqref{eq:expansion_d4}, i.\,e. $\varphi^{(1)}_\Deltabb \varphi^{(2)}_\Deltabb = \sum_{(E)} \mathcal{C}_{(1),(2),(E)}^{\Deltabb\Deltabb\Deltabb} \varphi^{(E)}_\Deltabb$.} The $(2,2)$- and the $(1,2)$-integrals of the weighted basis elements \eqref{eq:weighted_basis} are given by
\begin{align} 
    \braket{ \varphi^{(1)}_{\alpha} \varphi^{(2)}_{\beta}}{ \varphi^{(3)}_{\gamma} \varphi^{(4)}_{\delta}}
    = \widetilde{\mathcal{I}}^{\alpha\beta\gamma\delta}_{(1),(2),(3),(4)}\,, \qquad 
    \braket{ \varphi^{(1)}_{\alpha} }{\varphi^{(2)}_{\beta} \varphi^{(3)}_{\gamma}}
    = \widetilde{\mathcal{I}}^{\,-\,\alpha\beta\gamma}_{0,(1),(2),(3)}\,.
\end{align}
where the integrals are always with respect to the weight $\Deltabb$. The defining functions in our computations are now defined as a contraction of $\widetilde{\mathcal{I}}_{(1),(2),(3),(4)}^{\alpha\beta\gamma\delta}$ and two weighted basis elements 
\begin{align} \label{eq:def_tau_fcts}
    \mathcal{T}_{(1),(4)}\left(\begin{matrix}\alpha,\beta,\gamma,\delta \\ \Deltabb \end{matrix}\,\Bigg| X_1,X_2\right) \equiv \sum_{(2),(3)} \widetilde{\mathcal{I}}_{(1),(2),(3),(4)}^{\alpha\beta\gamma\delta} \varphi^{(2)}_\beta(X_1) \varphi^{(3)}_\gamma(X_2)\,,
\end{align}
where $\Deltabb$ indicates the weight of the defining contraction in $\widetilde{\mathcal{I}}$. We often leave out the dependence on the conformal weights for readability. 

Properties of these weighted coefficients are generally different from the original SU(2) Clebsch-Gordan coefficients, for example $\mathcal{C}^{\alpha \beta \gamma}_{(1),(2),(3)} = \mathcal{C}^{\beta\alpha\gamma}_{(2),(1),(3)}$. However, one still has $ \mathcal{C}_{(1),0,(1)}^{\alpha\beta\gamma} = 1$. Further special cases that we use in the context of the low-temperature limit of conformal blocks are
\begin{align} \label{eq:special_cases_su22_intertwiner}
     \widetilde{\mathcal{I}}_{0,0,(3),(4)}^{\alpha\beta\gamma\delta} &= \delta_{(3),0} \delta_{(4),0}\,, & \widetilde{\mathcal{I}}_{0,(2),(3),0}^{\alpha\beta\gamma\delta} &= \frac{\mathcal{N}^{(2)}_\beta \mathcal{N}^{(3)}_\gamma}{\mathcal{N}^{(2)}_\Deltabb \mathcal{N}^{(3)}_\Deltabb} \delta_{(2),(3)}\,, \nonumber\\
     \widetilde{\mathcal{I}}_{0,(2),(3),(4)}^{\alpha\beta\gamma\delta} &= \left(\mathcal{N}^{(2)}_\Deltabb\right)^{-1} \mathcal{C}^{\gamma\delta\beta}_{(3),(4),(2)} \,, & \widetilde{\mathcal{I}}_{(1),(2),(3),0}^{\alpha\beta\gamma\delta} &= \left(\mathcal{N}^{(3)}_\Deltabb\right)^{-1} \mathcal{C}^{\alpha\beta\gamma}_{(1),(2),(3)} \,,
\end{align}
where we used equation \eqref{eq:zero_cgc_delta}. Clearly, $\widetilde{\mathcal{I}}_{0,0,0,0}^{\alpha\beta\gamma\delta} = 1$.

\subsection{Weighted SU(2) Spin-Networks}
\label{sec:weighted_networks}
We found in the main text that the conformal blocks we calculated are naturally series of weighted SU(2) Clebsch-Gordan coefficients. Consequently, we can introduce weighted versions of SU(2) spin-network diagrams. We denote $C^{j_1j_2l}_{a_1,a_2,c}$ and $I_{a_1,a_2,a_3,a_4}^{l}$ as\footnote{We use a really rudimentary version of the actual standard SU(2) diagrammatic rules given in for example \cite{Makinen:2019rou,Martin-Dussaud:2019ypf}. For example, we do not care about orientations of the diagrams and rely more on context. Note that according to equation \eqref{eq:rewrite_intertwiners}, for the conformal blocks the left side of the $I^l$ equation should actually say $\sqrt{d_l d_{j_4}^{-1}}I_{a_1,a_2,a_3,a_4}^{l}$, we leave these additional factors out for simplicity. } 
\begin{align}
    C^{j_1j_2l}_{a_1,a_2,c} = \begin{tikzpicture}[scale=0.8, baseline={(current bounding box.center)}]

\def\lineW{1pt}

\coordinate (O) at (0,0);
\coordinate (L1) at (-0.8,-1.0);
\coordinate (L2) at (-0.8,1.0);
\coordinate (R) at (1.3,0);

\draw[line width=\lineW] (L1) -- (O);
\draw[line width=\lineW] (L2) -- (O);
\draw[line width=\lineW] (O) -- (R);

\node at (-1.05,-0.5) {$j_1$};
\node at (-1.05,0.5) {$j_2$};
\node at (1.55,0) {$l$};

\end{tikzpicture}\,, \qquad I_{a_1,a_2,a_3,a_4}^{l} = \begin{tikzpicture}[scale=0.8, baseline={(current bounding box.center)}]

\def\lineW{1pt}

\coordinate (A) at (-2,0);
\coordinate (B) at (-1,1.2);
\coordinate (B2) at (-1.0,1.20);
\coordinate (C) at (1,1.2);
\coordinate (D) at (2,0);

\coordinate (Bbot) at (-1,0);
\coordinate (Cbot) at (1,0);

\draw[line width=\lineW,shorten >=-0.3pt] (A) -- (B);
\draw[line width=\lineW,shorten >=-0.3pt] (D) -- (C);

\draw[line width=\lineW] (B) -- (Bbot);
\draw[line width=\lineW] (C) -- (Cbot);

\draw[line width=\lineW,shorten <=-0.1pt,shorten >=-0.1pt] (B) -- (C);
\node at (0,1.45) {$l$};

\node at (-2,0.5) {$j_1$};
\node at (-0.5, 0.5) {$j_2$};
\node at (0.5, 0.5) {$j_3$};
\node at (2,0.5) {$j_4$};

\end{tikzpicture} \label{dia:rules_su2}
\end{align}
The lines are only labelled by the parameters $j_i$ that label the SU(2) representation, not the magnetic numbers $a_i,b_i$. Usually, the lines are contracted and hence the magnetic numbers summed over anyways. One can now adjust the diagrammatics \eqref{dia:rules_su2} by attaching a factor of $\mathcal{N}^{j,m}_{\delta}$ to a line with label $j$, (where $\delta$ is a stand-in for a general combination of conformal weights). Furthermore, lines are labelled with $\delta$ instead, where $\delta$ denotes the whole tupel $(\delta,j,m,a,n)$ and the block-index $(j,m,a,b)$ is summed over when the lines are contracted. From the expansion in equation \eqref{eq:expansion_d4} we find that the spin-coupled $l$-lines in \eqref{dia:rules_su2} are weighted with $(\mathcal{N}^{l,n}_{\Deltabb})^{-1}$. Moreover, from the same expansion we find that each vertex is labelled with the appropriate Kronecker delta: for example, the vertex in the left diagram in \eqref{dia:rules_su2} gets weighted with $\delta^{l+n}_{k_1 + k_2}$ for $k_i = j_i + m_i$, the vertices in the right diagram with $\delta^{l+n}_{k_1 + k_2}$ and $\delta^{l+n}_{k_3 + k_4}$. In the thermal diagrams \eqref{eq:1pt_block_network} and \eqref{eq:su2_therm_npt}, the thermal loops together with the vertical lines of the left-most $X$-matrix are weighted with $q^{k_i}$. 

We have decided to label the diagrams for comb and thermal $n$-point blocks, see for example \eqref{eq:su2_vac_npt} and \eqref{eq:su2_therm_npt} with the respective conformal weights, which in theory allows to read off the explicit expressions from just the diagrams. Keeping the SU(2) spin notation as in the diagrams \eqref{dia:rules_su2} would emphasise the contractions in the magnetic numbers and view the block as a weighted series of such a SU(2) amplitude (with the caveat that involved Wigner-$\mathcal{D}$ are more general than SU(2)). This has the advantage that there are very efficient numerical packages for the evaluation SU(2) spin-networks, we commented on this in section \ref{sec:discussion}.

\section{Computation of Projection Channel Blocks in Two Dimensions}
\label{sec_app:2d_blocks}
The aim of this section is to pedagogically illustrate how the oscillator method is used for thermal blocks in the simpler case of two dimensions to get a better understanding of the method. The construction was reviewed for comb blocks on the two-dimensional plane in \cite{Ammon:2024axd}, the idea being to introduce the weighted Bergman space $\mathcal{H}L^2_\hbb(\mathbb{D})$. This is a natural irreducible representation of one sector of the global conformal algebra $\mathfrak{su}(1,1)$ because $\mathbb{D} \cong \text{SU}(1,1)/\text{U}(1)$. Details on the basics of the formalism in two dimensions (like the explicit form of the $\mathfrak{su}(1,1)$ generators and the definition of the inner product) can be found in \cite{Ammon:2024axd}.

To obtain thermal blocks, we map the (punctured) plane to a cylinder and then to a torus using conformal transformations, which does not change the parametrization of the $\mathfrak{L}_n$ generators, but of the adjoint action on primaries. Let us demonstrate this procedure by considering the matrix element $\Omega(z;u_1,\bar{u}_2) = \mel{u_1}{\mathcal{O}_h(z)}{\bar{u}_2}$ on the plane, see \cite{Ammon:2024axd}. Mapping the plane to the cylinder, we introduce a new coordinate via $z = \mathrm{e}^w$ and taking the transformation behaviour of the primary into account, the matrix element on the cylinder reads 
\begin{equation}
    \Omega(w;u_1,\bar{u}_2) = \mathrm{e}^{hw} (\mathrm{e}^w-\bar{u}_2)^{\hbb_1-\hbb_2-h} (1-\mathrm{e}^w u_1)^{\hbb_2-\hbb_1-h} (1-u_1\bar{u}_2)^{h-\hbb_1-\hbb_2}\,. \label{eq:omega_cylinder}
\end{equation} 
We could derive the same result by transforming $\mathfrak{L}^{(z)}_n$ to $\mathfrak{L}^{(w)}_n$ and solving the corresponding oscillator equation
\begin{align}
    0 = \left(\mathfrak{L}_n^{(u_1)} - \bar{\mathfrak{L}}_{-n}^{(\bar{u}_2)}+\mathcal{L}_n^{(w)}\right)\; \Omega(w;u_1,\bar{u}_2)\,.
\end{align}
The one-point correlation function on the torus for $q = \mathrm{e}^{2\pi i \tau}$ can be defined as
\begin{equation}
    \expval{\mathcal{O}} = (q \bar{q})^{-\frac{c}{24}} \text{Tr}\left(q^{L_0}q^{\bar{L}_0} \mathcal{O}\right)\,,
\end{equation}
where we will in the following consider only the holomorphic sector. Inserting then a projection operator into the one-point function and expressing it through an oscillator integral, we arrive at an expression for the one-point block of the holomorphic sector involving the matrix element in equation \eqref{eq:omega_cylinder}
\begin{equation}
    \begin{split}
    G^{(1)}(q) =  q^{-\frac{c}{24}}\mathrm{Tr}_{\hbb}\left(\mathbb{P}_{\mathbb{h}} \mathcal{O}_h(w) q^{L_0} \right)
    = q^{\hbb-\frac{c}{24}}\int_{\mathbb{D}} [\d u]\,  \Omega(w;u,q\bar{u})\,,
\end{split}
\end{equation}
in terms of the cylinder matrix element. We have used that the operator $q^{\bar{\mathfrak{L}}_0(\bar{u})}$ acting on the anti-holomorphic oscillator variable generates a dilation (this is a special case of the four-dimensional calculation in equation \eqref{eq:def_npt_proj_block}). This result in terms of oscillator diagrams corresponds to the following intuitive gluing rule for the torus one-point block:
\begin{align*}
    \label{dia:1pblock}
    G^{(1)}(q) \quad = \quad
    \begin{tikzpicture}[baseline={([yshift=-0.65ex]current bounding box.center)},vertex/.style={anchor=base,
    circle,fill=black!25,minimum size=18pt,inner sep=2pt}]
        \coordinate[] (u) at (5,1);
        \coordinate[] (belowu) at (5.25,1);
        \coordinate[] (ubar) at (5,-1);
        \coordinate[] (aboveubar) at (5.25,-1);
        \coordinate[label=above:$w$] (w_1) at (2.5,0);
        \draw[thick,dashed] (u) -- node[above left] {$\hbb$} (4,0);
        \draw[thick,dashed] (ubar) -- node[below left] {$\hbb$} (4,0);
        \draw[thick] (4,0) --node[above] {$h$} (w_1);
        \fill (w_1) circle (3pt);
        \draw (u) circle (3pt);
        \draw[{Latex[round]}-{Latex[round]},solid] (aboveubar) to[out=-10,in=10, looseness=1.5] (belowu);
        \fill[white] (u) circle (3pt);
        \fill[white] (ubar) circle (3pt);
        \filldraw (ubar) circle (1pt);
        \draw (ubar) circle (3pt);
    \end{tikzpicture}
    \quad = \quad 
    \begin{tikzpicture}[baseline={([yshift=-.5ex]current bounding box.center)},vertex/.style={anchor=base,
    circle,fill=black!25,minimum size=18pt,inner sep=2pt}]
    \coordinate[label=above:$w$] (w_1) at (2.5,0);
    \fill (w_1) circle (3pt);
    \draw[thick] (3.95,0) --node[above] {$h$} (w_1);
     \draw[thick,dashed] (5,0) circle (30pt);
     \coordinate[label=above:$\mathbb{h}$] (h) at (6.25,-0.25);
    \end{tikzpicture}
     \numberthis
\end{align*}
The actual computations proceeds in the ordinary fashion, that is, one expands $\Omega$ in monomials and uses the orthogonality relation $(u^n,u^m) = \frac{n!}{(2\hbb)_n}\delta_{n,m}$ to arrive at the power series in $q$
\begin{align} \label{eq:series_exp_1pt_block_2d}
&q^{\hbb-\frac{c}{24}}\int_{\mathbb{D}} [\d u]  \Omega(w; u, q\bar{u})   \notag \\ &=  q^{\mathbb{h}-\frac{c}{24}} \sum_{k_1, k_2, k_3 = 0}^\infty (-1)^{k_1 + k_2 + k_3} q^{k_1+k_3} \mathrm{e}^{-k_1w} \mathrm{e}^{k_2 w} \binom{-h}{k_1} \binom{-h}{k_2} \binom{h-2\mathbb{h}}{k_3}   \int_{\mathbb{D}} [d^2u] \bar{u}^{k_1 + k_3} u^{k_2 + k_3} \notag\\  
&= q^{\mathbb{h}-\frac{c}{24}} \sum_{k_2 = 0}^\infty \sum_{k_3 = 0}^\infty  \frac{(h)_{k_2}}{k_2!} \frac{ (h)_{k_2} (2\mathbb{h}-h)_{k_3} (k_2 + k_3)!}{(2\mathbb{h})_{k_2 + k_3}} \frac{q^{k_2}}{k_2!} \frac{q^{k_3}}{ k_3!} \notag\\
&= \frac{q^{\hbb - \frac{c}{24}}}{(1-q)^h} {}_2F_1(1-h,2\hbb-h,2\hbb,q)\,,
\end{align}
where in the last step several Pochhammer identities and manipulations of the sums were applied. One can verify that this result -- as is implied by the construction -- fulfils the two-dimensional Casimir equation 
\begin{align} \label{eq:c2_eq_2d}
    \left(Q_\hbb + \mathcal{C}_2^{(w)} \right)\mathcal{G}^{(1)}(q) = \left(Q_\hbb - h(h-1)\right)\mathcal{G}^{(1)}(q) = 0
\end{align}
with the differential operator $Q_\hbb$ given by
\begin{align}
    Q_\hbb = q(1-q)^2 \partial_q^2 - 2q(1-q)\partial_q - \hbb(\hbb-1)q^{-1}(1-q)^2
\end{align}
and using that $C_2^{(w_1)} \mathcal{O}_{h_1}(w_1) = -h(h-1) \mathcal{O}_{h_1}(w_1)$. This can be achieved (starting for example from the second to last line in \eqref{eq:series_exp_1pt_block_2d}) by evaluating the derivatives in $q$ and performing careful index shifts in $k_2$ and $k_3$ to account for the different orders in $q$ due to the various prefactors. Equation \eqref{eq:c2_eq_2d} can be generalised to the $n$-point case using the multi-point Casimir operator $\mathcal{C}_2^{(w_1,\dots,w_n)}$
\begin{gather}
        \left(Q_\hbb + \mathcal{C}_2^{(w_1,\dots,w_n)} \right)\mathcal{G}^{(n)}(q,w_1,\dots,w_n) = 0\,, \\
    \mathcal{C}_2^{(w_1,\dots,w_n)} = \frac{1}{2}\bigg\{ \mathfrak{L}^{(w_1,\dots,w_n)}_1\,,\,\mathfrak{L}^{(w_1,\dots,w_n)}_{-1}\bigg\}-\left(\mathfrak{L}^{(w_1,\dots,w_n)}_0\right)^2\,.
\end{gather}

Note how the one-point block turns out to be independent of $w$ as to be expected from its $\mathfrak{u}(1)$-symmetry. One recovers in the $h=0$ limit the zero-point block 
\begin{equation}
    \textnormal{Tr}\left(\mathbb{P}_{\hbb}q^{L_0-\frac{c}{24}}\right) = q^{\hbb-\frac{c}{24}} \sum_{n=0}^\infty q^n = \frac{q^{\hbb-\frac{c}{24}}}{1-q}\,,
\end{equation}
i.\,e. the character of the $\mathfrak{sl}(2,\mathbb{R})$ highest-weight representation. The series expansion of the one-point block in equation \eqref{eq:series_exp_1pt_block_2d} can be expressed as
\begin{equation}
    G^{(1)}(q) = q^{\hbb-\frac{c}{24}} \sum_{n=0}^\infty q^n \frac{\tau_{n,n}(\hbb,h,\hbb)}{n! (2\hbb)_{n}}
\end{equation}
where we introduced the $\tau$-coefficients 
\begin{equation}
    \tau_{k,l}(\hbb_1,h,\hbb_2) =  \sum\limits_{m=0}^{\min(k,l)} \frac{(\mathbb{h}_1+\mathbb{h}_2-h)_{m} (\mathbb{h}_1-\mathbb{h}_2+h)_{k-m}(\mathbb{h}_2-\mathbb{h}_1+h)_{l-m} }{m!(k-m)!(l-m)!}\,, \label{eq:tau_coeffs}
\end{equation}
already encountered in the five-point block on the plane \cite{Ammon:2024axd}. 

In essentially the same way, one obtains the $n$-point projection channel block 
\begin{align}
    G^{(n)}_{\text{Proj.}}(q,w_1,\dots,w_n) &= q^{\hbb-\frac{c}{24}} \int_{\mathbb{
    D}^n} [\d u]\, \Omega(w_1,u_1,\bar{u}_2) \Omega(w_2,u_2,\bar{u}_3) \cdots \Omega(w_n,u_n,q\bar{u}_1)
\end{align}
by gluing multiple $\Omega$-wavefunctions. Diagrammatically, this corresponds to the same diagram we used for the four-dimensional case in equation \eqref{dia:npt-block-glueing}. Note that only the last matrix element carries the dependence on the modular parameter $q$ while the $(n-1)$ other wavefunctions are the standard matrix elements on the cylinder. Expanding each wavefunction and using the inner product in the usual way, one obtains 
\begin{align} \label{eq:npt_block_2d}
    G^{(n)}_{\text{Proj.}}(q,w_1,\dots,w_n) = q^{\hbb_1 - \frac{c}{24}} \sum_{s_1,\dots s_n=0}^\infty \prod_{i=1}^n x_{i,i+1}^{-\hbb_{i+1}}  x_{i,i+1}^{-s_{i+1}} \frac{\tau_{s_{i},s_{i+1}}(\hbb_i,h_i,\hbb_{i+1})}{s_{i}!(2\hbb_i)_{s_{i}}} q^{s_1}\,, 
\end{align}
with $x_{ij} = \mathrm{e}^{w_i-w_j}$. Note that this can be brought into the form given in \cite{Alkalaev:2022kal} by inserting\footnote{That their $\tau$-coefficients match ours in equation \eqref{eq:tau_coeffs} can be checked by calculating their respective (exponential) generating functions.} 
\begin{equation}
    \frac{x_{n,1}^{\hbb_1}}{x_{n,1}^{\hbb_1}} = \frac{\prod^{n-1}_{k=1} x^{\hbb_1}_{k+1,k}}{\prod^{n-1}_{k=1} x^{\hbb_1}_{k+1,k}}\,, \qquad \frac{x_{n,1}^{s_1}}{x_{n,1}^{s_1}} = \frac{\prod^{n-1}_{k=1} x^{s_1}_{k+1,k}}{\prod^{n-1}_{k=1} x^{s_1}_{k+1,k}}\,, 
\end{equation}
using the denominator terms to rewrite the first $n-1$ terms of $\prod^n_{i=1}x^{-\hbb_{i+1}}_{i,i+1} x^{-s_{i+1}}_{i,i+1}$ and cancel the numerator terms with the $n$-th term through $x_{n,1}^{\hbb_1} x_{n,1}^{\hbb_1} = 1$ and $x_{n,1}^{s_1} x_{n,1}^{s_1} = 1$, finally defining $x_1 = q/(x_2 \cdots x_n)$. In the low-temperature limit $q\to 0$, this thermal block reduces to the comb-channel $(n+2)$-block on the plane with $z_1 \to \infty$ and $z_{n+2} = 0$. Because the leading order in $q \to 0$ sets $s_1 = 0$ in equation \eqref{eq:npt_block_2d}, we find that the resulting comb channel block is very similar, with the difference of being dependent on the coefficients $\tau_{0,s_2}$ and $\tau_{s_n,0}$. By using that the $\tau$-coefficients in equation \eqref{eq:tau_coeffs} reduce as $\tau_{0,l} = (\hbb_2 + h - \hbb_1)_l$ and $\tau_{k,0} = (\hbb_1+h-\hbb_2)_k$, we see that the $q \to 0$ limit for $n=2$ is just the usual expression for the four-point block.

\begin{appendices}
\addtocontents{toc}{\protect\setcounter{tocdepth}{2}}

\end{appendices}

\bibliography{main}

\providecommand{\href}[2]{#2}\begingroup\raggedright\begin{thebibliography}{10}

\bibitem{Ferrara:1973yt}
S.~Ferrara, A.F.~Grillo and R.~Gatto, \emph{{Tensor representations of conformal algebra and conformally covariant operator product expansion}}, \href{https://doi.org/10.1016/0003-4916(73)90446-6}{\emph{Annals Phys.} {\bfseries 76} (1973) 161}.

\bibitem{Polyakov:1974gs}
A.M.~Polyakov, \emph{{Nonhamiltonian approach to conformal quantum field theory}}, {\emph{Zh. Eksp. Teor. Fiz.} {\bfseries 66} (1974) 23}.

\bibitem{Rattazzi:2008pe}
R.~Rattazzi, V.S.~Rychkov, E.~Tonni and A.~Vichi, \emph{{Bounding scalar operator dimensions in 4D CFT}}, \href{https://doi.org/10.1088/1126-6708/2008/12/031}{\emph{JHEP} {\bfseries 12} (2008) 031} [\href{https://arxiv.org/abs/0807.0004}{{\ttfamily 0807.0004}}].

\bibitem{El-Showk:2012cjh}
S.~El-Showk, M.F.~Paulos, D.~Poland, S.~Rychkov, D.~Simmons-Duffin and A.~Vichi, \emph{{Solving the 3D Ising Model with the Conformal Bootstrap}}, \href{https://doi.org/10.1103/PhysRevD.86.025022}{\emph{Phys. Rev. D} {\bfseries 86} (2012) 025022} [\href{https://arxiv.org/abs/1203.6064}{{\ttfamily 1203.6064}}].

\bibitem{Hartman:2022zik}
T.~Hartman, D.~Mazac, D.~Simmons-Duffin and A.~Zhiboedov, \emph{{Snowmass White Paper: The Analytic Conformal Bootstrap}},  in \emph{{Snowmass 2021}}, 2, 2022 [\href{https://arxiv.org/abs/2202.11012}{{\ttfamily 2202.11012}}].

\bibitem{Antunes:2020yzv}
A.~Antunes, S.~Harris, A.~Kaviraj and V.~Schomerus, \emph{{Lining up a positive semi-definite six-point bootstrap}}, \href{https://doi.org/10.1007/JHEP06(2024)058}{\emph{JHEP} {\bfseries 24} (2020) 058} [\href{https://arxiv.org/abs/2312.11660}{{\ttfamily 2312.11660}}].

\bibitem{Buric:2020zlb}
I.~Buric, \emph{{Harmonic Analysis in d-dimensional Superconformal Field Theory}}, \href{https://doi.org/10.3842/SIGMA.2021.007}{\emph{SIGMA} {\bfseries 17} (2021) 007} [\href{https://arxiv.org/abs/2009.00393}{{\ttfamily 2009.00393}}].

\bibitem{Harris:2025cxx}
S.~Harris, \emph{{Sparsity in the numerical six-point bootstrap}},  \href{https://arxiv.org/abs/2507.00124}{{\ttfamily 2507.00124}}.

\bibitem{Rosenhaus:2018zqn}
V.~Rosenhaus, \emph{{Multipoint Conformal Blocks in the Comb Channel}}, \href{https://doi.org/10.1007/JHEP02(2019)142}{\emph{JHEP} {\bfseries 02} (2019) 142} [\href{https://arxiv.org/abs/1810.03244}{{\ttfamily 1810.03244}}].

\bibitem{Petkou:2018ynm}
A.C.~Petkou and A.~Stergiou, \emph{{Dynamics of Finite-Temperature Conformal Field Theories from Operator Product Expansion Inversion Formulas}}, \href{https://doi.org/10.1103/PhysRevLett.121.071602}{\emph{Phys. Rev. Lett.} {\bfseries 121} (2018) 071602} [\href{https://arxiv.org/abs/1806.02340}{{\ttfamily 1806.02340}}].

\bibitem{Iliesiu:2018fao}
L.~Iliesiu, M.~Kolo{\v{g}}lu, R.~Mahajan, E.~Perlmutter and D.~Simmons-Duffin, \emph{{The Conformal Bootstrap at Finite Temperature}}, \href{https://doi.org/10.1007/JHEP10(2018)070}{\emph{JHEP} {\bfseries 10} (2018) 070} [\href{https://arxiv.org/abs/1802.10266}{{\ttfamily 1802.10266}}].

\bibitem{Iliesiu:2018zlz}
L.~Iliesiu, M.~Kolo{\u{g}}lu and D.~Simmons-Duffin, \emph{{Bootstrapping the 3d Ising model at finite temperature}}, \href{https://doi.org/10.1007/JHEP12(2019)072}{\emph{JHEP} {\bfseries 12} (2019) 072} [\href{https://arxiv.org/abs/1811.05451}{{\ttfamily 1811.05451}}].

\bibitem{Karlsson:2022osn}
R.~Karlsson, A.~Parnachev, V.~Prilepina and S.~Valach, \emph{{Thermal stress tensor correlators, OPE and holography}}, \href{https://doi.org/10.1007/JHEP09(2022)234}{\emph{JHEP} {\bfseries 09} (2022) 234} [\href{https://arxiv.org/abs/2206.05544}{{\ttfamily 2206.05544}}].

\bibitem{Alday:2020eua}
L.F.~Alday, M.~Kologlu and A.~Zhiboedov, \emph{{Holographic correlators at finite temperature}}, \href{https://doi.org/10.1007/JHEP06(2021)082}{\emph{JHEP} {\bfseries 06} (2021) 082} [\href{https://arxiv.org/abs/2009.10062}{{\ttfamily 2009.10062}}].

\bibitem{Dodelson:2022yvn}
M.~Dodelson, A.~Grassi, C.~Iossa, D.~Panea~Lichtig and A.~Zhiboedov, \emph{{Holographic thermal correlators from supersymmetric instantons}}, \href{https://doi.org/10.21468/SciPostPhys.14.5.116}{\emph{SciPost Phys.} {\bfseries 14} (2023) 116} [\href{https://arxiv.org/abs/2206.07720}{{\ttfamily 2206.07720}}].

\bibitem{Huang:2022vet}
K.-W.~Huang, R.~Karlsson, A.~Parnachev and S.~Valach, \emph{{Freedom near lightcone and ANEC saturation}}, \href{https://doi.org/10.1007/JHEP05(2023)065}{\emph{JHEP} {\bfseries 05} (2023) 065} [\href{https://arxiv.org/abs/2210.16274}{{\ttfamily 2210.16274}}].

\bibitem{Dodelson:2023vrw}
M.~Dodelson, C.~Iossa, R.~Karlsson and A.~Zhiboedov, \emph{{A thermal product formula}}, \href{https://doi.org/10.1007/JHEP01(2024)036}{\emph{JHEP} {\bfseries 01} (2024) 036} [\href{https://arxiv.org/abs/2304.12339}{{\ttfamily 2304.12339}}].

\bibitem{Esper:2023jeq}
C.~Esper, K.-W.~Huang, R.~Karlsson, A.~Parnachev and S.~Valach, \emph{{Thermal stress tensor correlators near lightcone and holography}}, \href{https://doi.org/10.1007/JHEP11(2023)107}{\emph{JHEP} {\bfseries 11} (2023) 107} [\href{https://arxiv.org/abs/2306.00787}{{\ttfamily 2306.00787}}].

\bibitem{Marchetto:2023xap}
E.~Marchetto, A.~Miscioscia and E.~Pomoni, \emph{{Sum rules {\&} Tauberian theorems at finite temperature}}, \href{https://doi.org/10.1007/JHEP09(2024)044}{\emph{JHEP} {\bfseries 09} (2024) 044} [\href{https://arxiv.org/abs/2312.13030}{{\ttfamily 2312.13030}}].

\bibitem{Barrat:2024aoa}
J.~Barrat, B.~Fiol, E.~Marchetto, A.~Miscioscia and E.~Pomoni, \emph{{Conformal line defects at finite temperature}}, \href{https://doi.org/10.21468/SciPostPhys.18.1.018}{\emph{SciPost Phys.} {\bfseries 18} (2025) 018} [\href{https://arxiv.org/abs/2407.14600}{{\ttfamily 2407.14600}}].

\bibitem{Buric:2025anb}
I.~Buri{\'c}, I.~Gusev and A.~Parnachev, \emph{{Thermal holographic correlators and KMS condition}},  \href{https://arxiv.org/abs/2505.10277}{{\ttfamily 2505.10277}}.

\bibitem{Barrat:2025wbi}
J.~Barrat, E.~Marchetto, A.~Miscioscia and E.~Pomoni, \emph{{Thermal Bootstrap for the Critical O(N) Model}}, \href{https://doi.org/10.1103/PhysRevLett.134.211604}{\emph{Phys. Rev. Lett.} {\bfseries 134} (2025) 211604} [\href{https://arxiv.org/abs/2411.00978}{{\ttfamily 2411.00978}}].

\bibitem{Barrat:2025nvu}
J.~Barrat, D.N.~Bozkurt, E.~Marchetto, A.~Miscioscia and E.~Pomoni, \emph{{The analytic bootstrap at finite temperature}},  \href{https://arxiv.org/abs/2506.06422}{{\ttfamily 2506.06422}}.

\bibitem{Buric:2024kxo}
I.~Buric, F.~Russo, V.~Schomerus and A.~Vichi, \emph{{Thermal one-point functions and their partial wave decomposition}}, \href{https://doi.org/10.1007/JHEP12(2024)021}{\emph{JHEP} {\bfseries 12} (2024) 021} [\href{https://arxiv.org/abs/2408.02747}{{\ttfamily 2408.02747}}].

\bibitem{Buric:2025uqt}
I.~Buri{\'c}, F.~Mangialardi, F.~Russo, V.~Schomerus and A.~Vichi, \emph{{Heavy-Heavy-Light Asymptotics from Thermal Correlators}},  \href{https://arxiv.org/abs/2506.21671}{{\ttfamily 2506.21671}}.

\bibitem{Gobeil:2018fzy}
Y.~Gobeil, A.~Maloney, G.S.~Ng and J.-q.~Wu, \emph{{Thermal Conformal Blocks}}, \href{https://doi.org/10.21468/SciPostPhys.7.2.015}{\emph{SciPost Phys.} {\bfseries 7} (2019) 015} [\href{https://arxiv.org/abs/1802.10537}{{\ttfamily 1802.10537}}].

\bibitem{Alkalaev:2024jxh}
K.~Alkalaev and S.~Mandrygin, \emph{{One-point thermal conformal blocks from four-point conformal integrals}}, \href{https://doi.org/10.1007/JHEP10(2024)241}{\emph{JHEP} {\bfseries 10} (2024) 241} [\href{https://arxiv.org/abs/2407.01741}{{\ttfamily 2407.01741}}].

\bibitem{Hadasz2010}
L.~Hadasz, Z.~Jaskolski and P.~Suchanek, \emph{{Recursive representation of the torus 1-point conformal block}}, \href{https://doi.org/10.1007/JHEP01(2010)063}{\emph{JHEP} {\bfseries 01} (2010) 063} [\href{https://arxiv.org/abs/0911.2353}{{\ttfamily 0911.2353}}].

\bibitem{Kraus2017}
P.~Kraus, A.~Maloney, H.~Maxfield, G.S.~Ng and J.-q.~Wu, \emph{{Witten Diagrams for Torus Conformal Blocks}}, \href{https://doi.org/10.1007/JHEP09(2017)149}{\emph{JHEP} {\bfseries 09} (2017) 149} [\href{https://arxiv.org/abs/1706.00047}{{\ttfamily 1706.00047}}].

\bibitem{Cho:2017oxl}
M.~Cho, S.~Collier and X.~Yin, \emph{{Recursive Representations of Arbitrary Virasoro Conformal Blocks}}, \href{https://doi.org/10.1007/JHEP04(2019)018}{\emph{JHEP} {\bfseries 04} (2019) 018} [\href{https://arxiv.org/abs/1703.09805}{{\ttfamily 1703.09805}}].

\bibitem{Alkalaev:2017bzx}
K.B.~Alkalaev and V.A.~Belavin, \emph{{Holographic duals of large-c torus conformal blocks}}, \href{https://doi.org/10.1007/JHEP10(2017)140}{\emph{JHEP} {\bfseries 10} (2017) 140} [\href{https://arxiv.org/abs/1707.09311}{{\ttfamily 1707.09311}}].

\bibitem{Alkalaev:2020yvq}
K.~Alkalaev and V.~Belavin, \emph{{More on Wilson toroidal networks and torus blocks}}, \href{https://doi.org/10.1007/JHEP11(2020)121}{\emph{JHEP} {\bfseries 11} (2020) 121} [\href{https://arxiv.org/abs/2007.10494}{{\ttfamily 2007.10494}}].

\bibitem{Alkalaev:2022kal}
K.~Alkalaev, S.~Mandrygin and M.~Pavlov, \emph{{Torus conformal blocks and Casimir equations in the necklace channel}}, \href{https://doi.org/10.1007/JHEP10(2022)091}{\emph{JHEP} {\bfseries 10} (2022) 091} [\href{https://arxiv.org/abs/2205.05038}{{\ttfamily 2205.05038}}].

\bibitem{Alkalaev:2023evp}
K.~Alkalaev and S.~Mandrygin, \emph{{Torus shadow formalism and exact global conformal blocks}}, \href{https://doi.org/10.1007/JHEP11(2023)157}{\emph{JHEP} {\bfseries 11} (2023) 157} [\href{https://arxiv.org/abs/2307.12061}{{\ttfamily 2307.12061}}].

\bibitem{Pavlov:2023asi}
M.~Pavlov, \emph{{Global torus blocks in the necklace channel}}, \href{https://doi.org/10.1140/epjc/s10052-023-12166-7}{\emph{Eur. Phys. J. C} {\bfseries 83} (2023) 1026} [\href{https://arxiv.org/abs/2302.10153}{{\ttfamily 2302.10153}}].

\bibitem{Dolan:2000ut}
F.A.~Dolan and H.~Osborn, \emph{{Conformal four point functions and the operator product expansion}}, \href{https://doi.org/10.1016/S0550-3213(01)00013-X}{\emph{Nucl. Phys. B} {\bfseries 599} (2001) 459} [\href{https://arxiv.org/abs/hep-th/0011040}{{\ttfamily hep-th/0011040}}].

\bibitem{Dolan:2003hv}
F.A.~Dolan and H.~Osborn, \emph{{Conformal partial waves and the operator product expansion}}, \href{https://doi.org/10.1016/j.nuclphysb.2003.11.016}{\emph{Nucl. Phys. B} {\bfseries 678} (2004) 491} [\href{https://arxiv.org/abs/hep-th/0309180}{{\ttfamily hep-th/0309180}}].

\bibitem{Dolan:2011dv}
F.A.~Dolan and H.~Osborn, \emph{{Conformal Partial Waves: Further Mathematical Results}},  \href{https://arxiv.org/abs/1108.6194}{{\ttfamily 1108.6194}}.

\bibitem{Buric:2020dyz}
I.~Buric, S.~Lacroix, J.A.~Mann, L.~Quintavalle and V.~Schomerus, \emph{{From Gaudin Integrable Models to $d$-dimensional Multipoint Conformal Blocks}}, \href{https://doi.org/10.1103/PhysRevLett.126.021602}{\emph{Phys. Rev. Lett.} {\bfseries 126} (2021) 021602} [\href{https://arxiv.org/abs/2009.11882}{{\ttfamily 2009.11882}}].

\bibitem{Buric:2021ywo}
I.~Buric, S.~Lacroix, J.A.~Mann, L.~Quintavalle and V.~Schomerus, \emph{{Gaudin models and multipoint conformal blocks: general theory}}, \href{https://doi.org/10.1007/JHEP10(2021)139}{\emph{JHEP} {\bfseries 10} (2021) 139} [\href{https://arxiv.org/abs/2105.00021}{{\ttfamily 2105.00021}}].

\bibitem{Buric:2021ttm}
I.~Buric, S.~Lacroix, J.A.~Mann, L.~Quintavalle and V.~Schomerus, \emph{{Gaudin models and multipoint conformal blocks. Part II. Comb channel vertices in 3D and 4D}}, \href{https://doi.org/10.1007/JHEP11(2021)182}{\emph{JHEP} {\bfseries 11} (2021) 182} [\href{https://arxiv.org/abs/2108.00023}{{\ttfamily 2108.00023}}].

\bibitem{Buric:2021kgy}
I.~Buric, S.~Lacroix, J.A.~Mann, L.~Quintavalle and V.~Schomerus, \emph{{Gaudin models and multipoint conformal blocks III: comb channel coordinates and OPE factorisation}}, \href{https://doi.org/10.1007/JHEP06(2022)144}{\emph{JHEP} {\bfseries 06} (2022) 144} [\href{https://arxiv.org/abs/2112.10827}{{\ttfamily 2112.10827}}].

\bibitem{Ammon:2024axd}
M.~Ammon, J.~Hollweck, T.~H\"ossel and K.~W\"olfl, \emph{{Conformal Blocks in Two and Four Dimensions from Oscillator Representations}},  \href{https://arxiv.org/abs/2406.19436}{{\ttfamily 2406.19436}}.

\bibitem{Kraus2020}
M.~Besken, S.~Datta and P.~Kraus, \emph{{Quantum thermalization and Virasoro symmetry}}, \href{https://doi.org/10.1088/1742-5468/ab900b}{\emph{J. Stat. Mech.} {\bfseries 2006} (2020) 063104} [\href{https://arxiv.org/abs/1907.06661}{{\ttfamily 1907.06661}}].

\bibitem{Luscher:1974ez}
M.~Luscher and G.~Mack, \emph{{Global Conformal Invariance in Quantum Field Theory}}, \href{https://doi.org/10.1007/BF01608988}{\emph{Commun. Math. Phys.} {\bfseries 41} (1975) 203}.

\bibitem{Calixto2010}
M.~{Calixto} and E.~{Perez-Romero}, \emph{{Extended MacMahon-Schwinger's Master Theorem and Conformal Wavelets in Complex Minkowski Space}}, \href{https://doi.org/10.48550/arXiv.1002.3498}{\emph{arXiv e-prints} (2010) arXiv:1002.3498} [\href{https://arxiv.org/abs/1002.3498}{{\ttfamily 1002.3498}}].

\bibitem{Fortin:2019zkm}
J.-F.~Fortin, W.~Ma and W.~Skiba, \emph{{Higher-Point Conformal Blocks in the Comb Channel}}, \href{https://doi.org/10.1007/JHEP07(2020)213}{\emph{JHEP} {\bfseries 07} (2020) 213} [\href{https://arxiv.org/abs/1911.11046}{{\ttfamily 1911.11046}}].

\bibitem{Parikh:2019dvm}
S.~Parikh, \emph{{A multipoint conformal block chain in $d$ dimensions}}, \href{https://doi.org/10.1007/JHEP05(2020)120}{\emph{JHEP} {\bfseries 05} (2020) 120} [\href{https://arxiv.org/abs/1911.09190}{{\ttfamily 1911.09190}}].

\bibitem{Dolan:2005wy}
F.A.~Dolan, \emph{{Character formulae and partition functions in higher dimensional conformal field theory}}, \href{https://doi.org/10.1063/1.2196241}{\emph{J. Math. Phys.} {\bfseries 47} (2006) 062303} [\href{https://arxiv.org/abs/hep-th/0508031}{{\ttfamily hep-th/0508031}}].

\bibitem{Gozzini:2021kbt}
F.~Gozzini, \emph{{A high-performance code for EPRL spin foam amplitudes}}, \href{https://doi.org/10.1088/1361-6382/ac2b0b}{\emph{Class. Quant. Grav.} {\bfseries 38} (2021) 225010} [\href{https://arxiv.org/abs/2107.13952}{{\ttfamily 2107.13952}}].

\bibitem{Dona:2018nev}
P.~Dona and G.~Sarno, \emph{{Numerical methods for EPRL spin foam transition amplitudes and Lorentzian recoupling theory}}, \href{https://doi.org/10.1007/s10714-018-2452-7}{\emph{Gen. Rel. Grav.} {\bfseries 50} (2018) 127} [\href{https://arxiv.org/abs/1807.03066}{{\ttfamily 1807.03066}}].

\bibitem{Dona:2022dxs}
P.~Dona and P.~Frisoni, \emph{{How-to Compute EPRL Spin Foam Amplitudes}}, \href{https://doi.org/10.3390/universe8040208}{\emph{Universe} {\bfseries 8} (2022) 208} [\href{https://arxiv.org/abs/2202.04360}{{\ttfamily 2202.04360}}].

\bibitem{TensorOperationsjl}
{Devos, Lukas, and Van Damme, Maarten and Haegeman, Jutho}, \emph{Tensoroperations.jl},  (10, 2023).
\newblock \url{https://github.com/Jutho/TensorOperations.jl}, \href{https://doi.org/10.5281/zenodo.3245496}{10.5281/zenodo.3245496}.

\bibitem{Perlmutter:2015iya}
E.~Perlmutter, \emph{{Virasoro conformal blocks in closed form}}, \href{https://doi.org/10.1007/JHEP08(2015)088}{\emph{JHEP} {\bfseries 08} (2015) 088} [\href{https://arxiv.org/abs/1502.07742}{{\ttfamily 1502.07742}}].

\bibitem{varsh1988}
D.A.~Varshalovich, A.N.~Moskalev and V.K.~Khersonskii, \emph{Quantum Theory of Angular Momentum}, WORLD SCIENTIFIC (1988), \href{https://doi.org/10.1142/0270}{10.1142/0270}, [\href{https://arxiv.org/abs/https://www.worldscientific.com/doi/pdf/10.1142/0270}{{\ttfamily https://www.worldscientific.com/doi/pdf/10.1142/0270}}].

\bibitem{Fortin:2020bfq}
J.-F.~Fortin, W.-J.~Ma and W.~Skiba, \emph{{Seven-point conformal blocks in the extended snowflake channel and beyond}}, \href{https://doi.org/10.1103/PhysRevD.102.125007}{\emph{Phys. Rev. D} {\bfseries 102} (2020) 125007} [\href{https://arxiv.org/abs/2006.13964}{{\ttfamily 2006.13964}}].

\bibitem{Fortin:2020yjz}
J.-F.~Fortin, W.-J.~Ma and W.~Skiba, \emph{{Six-point conformal blocks in the snowflake channel}}, \href{https://doi.org/10.1007/JHEP11(2020)147}{\emph{JHEP} {\bfseries 11} (2020) 147} [\href{https://arxiv.org/abs/2004.02824}{{\ttfamily 2004.02824}}].

\bibitem{Benjamin:2023qsc}
N.~Benjamin, J.~Lee, H.~Ooguri and D.~Simmons-Duffin, \emph{{Universal asymptotics for high energy CFT data}}, \href{https://doi.org/10.1007/JHEP03(2024)115}{\emph{JHEP} {\bfseries 03} (2024) 115} [\href{https://arxiv.org/abs/2306.08031}{{\ttfamily 2306.08031}}].

\bibitem{Shaghoulian:2015kta}
E.~Shaghoulian, \emph{{Modular forms and a generalized Cardy formula in higher dimensions}}, \href{https://doi.org/10.1103/PhysRevD.93.126005}{\emph{Phys. Rev. D} {\bfseries 93} (2016) 126005} [\href{https://arxiv.org/abs/1508.02728}{{\ttfamily 1508.02728}}].

\bibitem{Belin:2016yll}
A.~Belin, J.~de~Boer, J.~Kruthoff, B.~Michel, E.~Shaghoulian and M.~Shyani, \emph{{Universality of sparse $d > 2$ conformal field theory at large $N$}}, \href{https://doi.org/10.1007/JHEP03(2017)067}{\emph{JHEP} {\bfseries 03} (2017) 067} [\href{https://arxiv.org/abs/1610.06186}{{\ttfamily 1610.06186}}].

\bibitem{ruehl_1970}
W.~Ruehl, \emph{Lorentz group and harmonic analysis}, W. A. Benjamin, Inc. (1970).

\bibitem{Makinen:2019rou}
I.~M\"akinen, \emph{{Introduction to SU(2) Recoupling Theory and Graphical Methods for Loop Quantum Gravity}},  \href{https://arxiv.org/abs/1910.06821}{{\ttfamily 1910.06821}}.

\bibitem{Calixto_2011}
M.~Calixto and E.~P{\'{e}}rez-Romero, \emph{Extended {MacMahon}{\textendash}schwinger's master theorem and conformal wavelets in complex minkowski space}, \href{https://doi.org/https://doi.org/10.1016/j.acha.2010.11.004}{\emph{Applied and Computational Harmonic Analysis} {\bfseries 31} (2011) 143}.

\bibitem{Ruehl:1972jy}
W.~Rühl, \emph{{Distributions on minkowski space and their connection with analytic representations of the conformal group}}, \href{https://doi.org/10.1007/BF01649659}{\emph{Commun. Math. Phys.} {\bfseries 27} (1972) 53}.

\bibitem{Martin-Dussaud:2019ypf}
P.~Martin-Dussaud, \emph{{A Primer of Group Theory for Loop Quantum Gravity and Spin-foams}}, \href{https://doi.org/10.1007/s10714-019-2583-5}{\emph{Gen. Rel. Grav.} {\bfseries 51} (2019) 110} [\href{https://arxiv.org/abs/1902.08439}{{\ttfamily 1902.08439}}].

\end{thebibliography}\endgroup
\bibliographystyle{JHEP}

\end{document}